%% file: main.tex
\documentclass[acmsmall]{acmart}
\usepackage{booktabs}
\usepackage{graphicx}
\usepackage{multirow}
\usepackage{array}
\usepackage{xcolor}
\usepackage{amsmath}
\usepackage{tabularx}
\usepackage{wasysym}   %
\usepackage{pifont}    %
\usepackage{fontawesome5} %
\usepackage{longtable}    %
\usepackage{hyperref}
\usepackage{fix-cm}
\hypersetup{colorlinks=true,linkcolor=teal,citecolor=teal,urlcolor=teal}

\graphicspath{{Figs/}}
\newcolumntype{L}[1]{>{\raggedright\arraybackslash}p{#1}}

\setcopyright{none}
\acmConference[TECS]{ACM TECS}{}{}
\renewcommand\footnotetextcopyrightpermission[1]{}
\begin{document}

\title{Deep Learning Latency Attacks and Defenses: A Cross-Domain Survey of Availability Threats}
\author{Zonghua Gu}
\affiliation{%
  \institution{Hofstra University}
  \city{Hempstead}
  \state{NY}
  \country{USA}
}
\email{zonghua.gu@hofstra.edu}
\author{Zeyu Gao}
\affiliation{%
  \institution{Ume\aa\ University}
  \city{Ume\aa}
  \country{Sweden}
}
\email{zeyu.gao@umu.se}

\author{Amin Saremi}
\affiliation{%
  \institution{Ume\aa\ University}
  \city{Ume\aa}
  \country{Sweden}
}
\email{amin.saremi@umu.se}

\author{Samarjit Chakraborty}
\affiliation{%
  \institution{University of North Carolina at Chapel Hill}
  \city{Chapel Hill}
  \state{NC}
  \country{USA}
}
\email{samarjit@cs.unc.edu}

\begin{abstract}
Adversarial machine learning has focused mainly on \emph{integrity}, but \emph{availability} is an increasingly consequential complement. \emph{Latency attacks} (also energy-latency or sponge attacks) increase inference-time work, energy, or response time, causing deadline misses, throughput collapse, or resource exhaustion in vehicle controllers, interactive services, or battery-powered sensors, sometimes while preserving the nominal prediction.

This survey unifies a fragmented literature spanning perception pipelines (including physical attacks on autonomous-driving detection and tracking), input-adaptive neural inference (sponge examples, dynamic networks), and autoregressive and agentic systems (output-length, verbose-image, and reasoning denial-of-service attacks on LLMs, VLMs, mixture-of-experts models, and tool-using agents). We organize attacks by exploited computational bottleneck rather than formulation, separating what makes a computation expensive from how the attacker triggers it; the delivery channel (input, prompt or retrieved content, message, poisoning, or weight tampering) is an orthogonal attribute. Many attacks share one mechanism, \emph{intermediate-work amplification}, motivating a \emph{work-budget} defense abstraction; we distinguish caps on the work entering an expensive stage from caps on the results leaving it. We further analyze when a model-level cost increase becomes a system-level availability failure, which depends on critical-path share, slack, existing ceilings, accumulation, resource sharing, and fallback policy, not on the amplification factor alone.

We also provide a threat-model taxonomy, consolidated quantitative comparisons, a defense review by control mechanism, and open challenges such as standardized evaluation, physical realizability, and whole-system availability. Companion website: \url{https://github.com/guzonghua/awesome-latency-attacks}.
\end{abstract}


\maketitle
\section{Introduction}
\label{sec:intro}
Deep neural networks are increasingly deployed in settings where prediction latency is as important as prediction correctness. An autonomous vehicle that perceives a pedestrian one second late may, depending on its planning horizon and fallback behavior, have effectively not perceived it in time to act; an interactive assistant that takes thirty seconds to begin responding is unusable; a battery-powered sensor whose model suddenly draws ten times its budgeted energy will exhaust itself before its mission completes. These deployments expose an attack surface that classical adversarial machine learning largely ignored.

Most adversarial-example research targets \emph{integrity}: an attacker perturbs an input so that the model misclassifies it~\cite{vassilev2023nist, gao2026performance}. \emph{Availability} attacks pursue a different goal. Rather than primarily seeking an incorrect answer, the attacker seeks to inflate the cost or delay of producing one---measured in floating-point operations, energy, or latency---so the system cannot respond ``within a reasonable time''~\cite{chen2024overload}. We use \emph{latency attack} as an umbrella term for this family, which also appears in the literature under the names energy-latency attack, sponge attack, slowdown attack, efficiency-degradation attack, and resource-consumption denial-of-service. A recurring mechanism is the deliberate steering of a model, or its host system, toward unusually expensive or configured maximum-cost behavior. We reserve the term \emph{worst case} for a proven or exhaustively characterized maximum, and otherwise describe attacks as inducing high-cost execution paths or near-worst observed behavior, since most cited studies empirically find high-cost inputs rather than establish a true maximum.

\begin{figure*}[htbp]
  \centering
  \includegraphics[width=\textwidth]{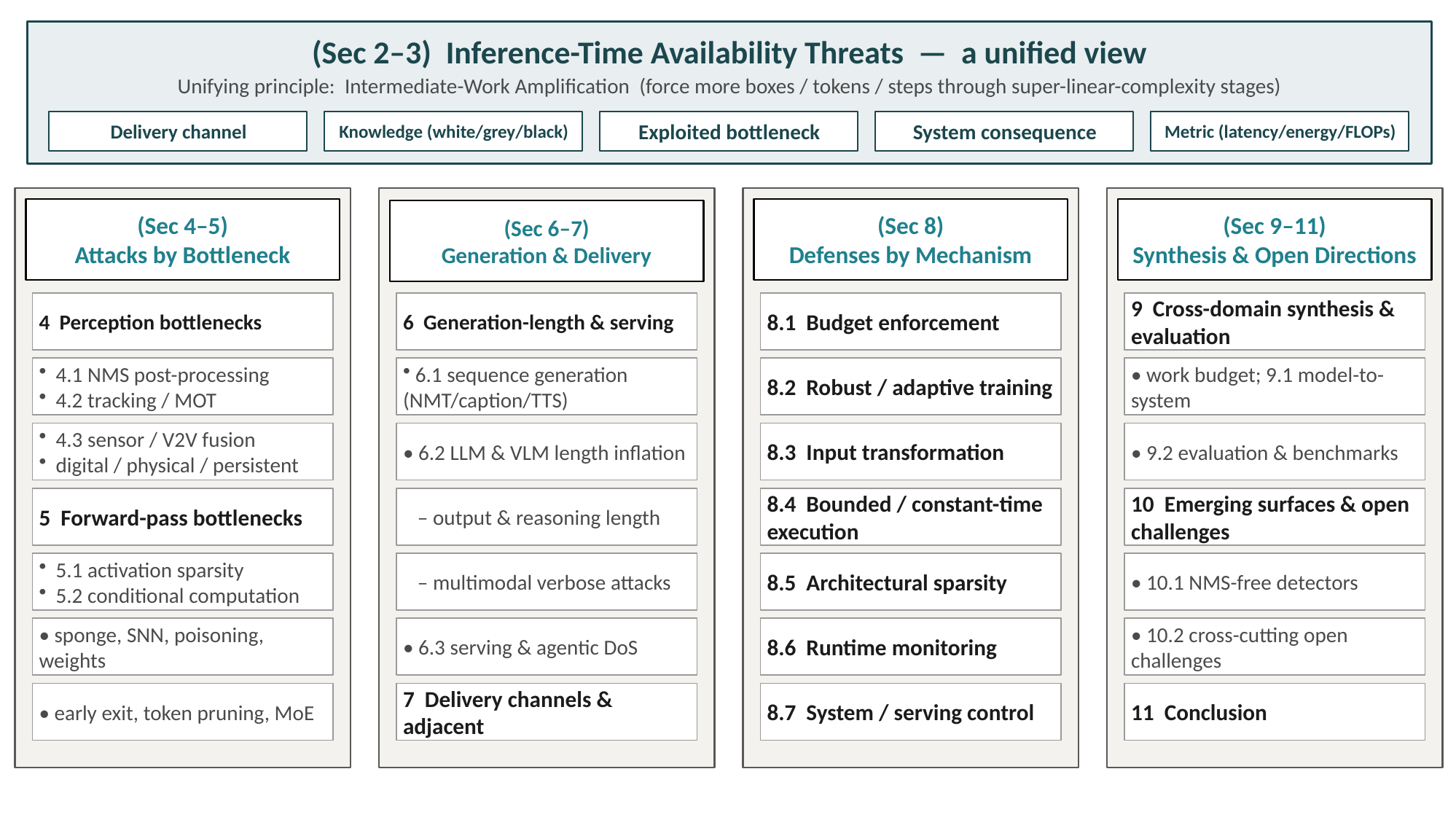}
  \caption{Overview of the survey structure.}
  \Description{A high-level structure diagram showing how the survey is organized.}
  \label{fig:survey-structure}
\end{figure*}

The overall organization of this survey is summarized in Figure~\ref{fig:survey-structure}.

\subsection{Why latency attacks matter now}
Although many works surveyed in this paper are recent, the rapid expansion of latency attacks across perception, adaptive inference, and foundation models motivates a unified survey at this stage. Three trends have converged to make latency attacks a first-class concern. First, \emph{real-time autonomy} couples perception latency directly to physical safety: a delayed detection cascades into delayed planning and control~\cite{ma2024slowtrack}. Second, the field has embraced \emph{input-adaptive efficiency}---early-exit networks, token pruning, skimming transformers, and mixture-of-experts routing---which replace fixed-cost execution with input-adaptive execution whose average cost is lower than its configured maximum. An availability attacker may revoke these savings for adversarially chosen inputs by steering them onto expensive execution paths, and under sustained load may substantially reduce system-level efficiency, in some cases approaching or exceeding the cost of the corresponding static model~\cite{hong2021deepsloth,chen2023dynamic}. Third, the rise of \emph{autoregressive generation} in LLMs and VLMs means that a small input can induce an output many times longer than the benign baseline---potentially reaching the configured token, context, or execution limit---and the per-query compute cost is now an attacker-controllable variable~\cite{gao2024verbose,dong2025engorgio}. The move to inference-time ``thinking'' in reasoning models multiplies this leverage: a single innocuous-looking prompt has been reported to trigger tens of thousands of reasoning tokens~\cite{liu2026reasoningbomb}.

\subsection{Scope and contributions}
Prior surveys treat pieces of this landscape in isolation. The two closest are Brachemi Meftah et al.~\cite{brachemi2026energy}, which reviews energy-latency attacks on regular, input-adaptive, and decoder-based models, and the SoK of Rathnasuriya et al.~\cite{rathnasuriya2025sok}, which systematizes efficiency attacks on dynamic deep-learning systems (Section~\ref{sec:related}). Both organize and measure attacks at the model boundary. Recent developments---reasoning-model denial-of-service, LLM/VLM serving attacks, agentic systems, and embodied AI---substantially broaden the attack landscape. More importantly, they make visible a question that a model-level organization cannot answer: when does a cost increase in a model become a failure of the system that hosts it? The autonomous-driving security literature, meanwhile, has produced a rich set of \emph{physically realizable} latency attacks that are rarely discussed alongside the general deep-learning work. This survey deliberately bridges both. We make the following contributions:

\begin{itemize}
  \item \textbf{A unified taxonomy} that separates the exploited \emph{bottleneck}---what makes a computation expensive---from the \emph{delivery channel}---how the attacker triggers it (Section~\ref{sec:bottleneck-delivery}). Sections~\ref{sec:od}--\ref{sec:seqgen} are organized by bottleneck alone, so an attack delivered through a test-time input, a poisoned training set, or a bit flip is discussed together with the other attacks on the same bottleneck; delivery channel and affected model class are recorded as orthogonal attributes spanning object detection, dynamic networks, transformers, LLMs, and multimodal generators.
  \item \textbf{A unifying principle}---\emph{intermediate-work amplification} (Section~\ref{sec:unifying})---that explains a large and important subset of latency attacks through a shared mechanism (timing leakage and denial-of-action attacks require separate treatment) and motivates a cross-domain \emph{work-budget} defense abstraction.
  \item \textbf{A model-to-system translation analysis} (Section~\ref{sec:translation}) that identifies when a model-level cost increase becomes a system-level availability failure. The outcome depends on the amplified stage's share of the critical path, the headroom between the attack's achievable amplification and the system's slack, whether existing caps or timeouts clip the effect, whether state or queues accumulate it, whether shared resources propagate it, and how the fallback policy maps a late result to a consequence. None of these is visible at the model boundary where most attacks, and prior surveys~\cite{brachemi2026energy,rathnasuriya2025sok}, measure.
  \item \textbf{A threat-model framework} (Section~\ref{sec:threat}) that classifies the attacker interface, knowledge, and targeted pipeline stage, making otherwise incomparable works commensurable.
  \item \textbf{Consolidated quantitative comparisons} (Tables~\ref{tab:od-attacks} and~\ref{tab:general-attacks}) of reported latency, energy, computational-work, and output-length measurements, with explicit notes on metric incompatibility.
  \item \textbf{A structured defense review} (Section~\ref{sec:defenses} and Table~\ref{tab:defenses}) organized by control mechanism and evidence class, exposing which point of the compute pathway each defense actually bounds, and distinguishing \emph{admission caps}, which bound the work entering an expensive stage, from \emph{emission caps}, which bound only the results leaving it (Section~\ref{sec:bounded}).
  \item \textbf{A cross-domain synthesis and evaluation discussion} (Section~\ref{sec:synthesis}) analyzing why most existing perception- and serving-side defenses do not transfer directly, while identifying work-budget control as a possible cross-domain abstraction, and including a review of fragmented evaluation practice (Section~\ref{sec:eval}).
  \item \textbf{A forward-looking analysis} (Section~\ref{sec:open}, including the NMS-free detector discussion in Section~\ref{sec:nmsfree}) of emerging attack surfaces and the open challenges that define the next phase of this research.
\end{itemize}

The remainder of this survey develops this perspective in three steps. We first show that many seemingly unrelated latency attacks are instances of intermediate-work amplification across different AI architectures. We then argue that bounding intermediate work through work-budget enforcement provides a common systems-level defense principle, provided the budget is placed on the work entering a stage rather than on the results leaving it. Finally, we analyze when a model-level amplification becomes a system-level failure.

\section{Background and Definitions}
\label{sec:background}

\subsection{Integrity versus availability}

Deep learning is now pervasively deployed on resource-constrained embedded and edge platforms, where inference latency, energy, and throughput are first-class design constraints rather than afterthoughts~\cite{luo2024efficient,marco2020adaptive,kim2025emamba}. It is precisely in these latency- and energy-sensitive regimes that availability attacks become consequential. We adopt the NIST taxonomy of adversarial machine learning, which distinguishes integrity, availability, and confidentiality violations~\cite{vassilev2023nist}. Integrity attacks aim to change the model's output; availability attacks aim to degrade its usefulness or serviceability. Latency attacks are availability attacks whose specific lever is computational cost. Crucially, a latency attack may leave the prediction \emph{unchanged}---often by design---so a monitor based only on a preserved task-quality metric (detection recall or mAP, the top-1 label, answer correctness) may not flag an attack whose principal effect is increased execution cost~\cite{shapira2023phantom,yehezkel2024desparsify,kumar2025overthink,raptis2026driving}, unless it also observes timing, repetition, truncation, confidence, or distribution-shift signals. %

\paragraph{Relation to classical denial-of-service.} For readers coming from network security, it helps to contrast latency attacks with classical denial-of-service (DoS). A classical network DoS exhausts a shared resource---bandwidth, connection tables, or server sockets---typically through \emph{volume}: many requests, oversized payloads, or malformed packets, mitigated by rate limiting, filtering, and over-provisioning at the network and transport layers. Latency attacks differ in three ways. First, the exhausted resource is \emph{model computation}---FLOPs, accelerator time, memory bandwidth, energy, or generated tokens---rather than network capacity. Second, and most importantly, the leverage comes not from volume but from \emph{per-request work amplification}: a single, well-formed, benign-\emph{looking} input of ordinary size can induce disproportionate internal computation, so the attack often passes request-rate and payload-size filters that stop volumetric DoS. Third, the attack surface is the model's own input-adaptive computation (post-processing, conditional execution, autoregressive generation), which classical DoS defenses do not model. In this sense a latency attack is an \emph{algorithmic-complexity} DoS specialized to learned, input-adaptive inference: the closest classical analogue is a hash-collision or regular-expression complexity attack, but here the expensive path is induced inside a neural pipeline and can preserve the nominal output while still consuming the budget. This is why the defenses we survey (Section~\ref{sec:defenses}) live at the model and serving layers---bounding admitted work per request---rather than at the network perimeter.

\subsection{Where latency comes from}
A modern inference pipeline contains many cost centers, and latency attacks succeed by driving one or more toward unusually expensive or configured maximum-cost behavior:
\begin{description}
  \item[Data-dependent post-processing.] Operations whose cost depends on the \emph{content}, not just the shape, of intermediate tensors. The canonical example is non-maximum suppression (NMS) in object detectors, whose cost grows with the number of candidate boxes~\cite{shapira2023phantom}.
  \item[Activation sparsity.] Some accelerators and sparse kernels exploit structured or unstructured zero activations to skip operations or memory transfers---the realized energy and latency benefit depends on hardware and kernel support, and commodity dense GPU kernels generally do not skip arbitrary zero activations automatically---and approximate-computing techniques trade accuracy for energy on the edge~\cite{ghosh2023approxedge}. Inputs that minimize sparsity force maximum multiply-accumulate work and energy~\cite{shumailov2021sponge,muller2024uniform}.
  \item[Conditional computation.] Early-exit networks, token-pruning vision transformers, skimming language models, and mixture-of-experts routers all decide \emph{at runtime} how much computation to perform~\cite{teerapittayanon2016branchynet,rao2021dynamicvit,fedus2022switch}; the same input-adaptive model-selection strategies that save average-case energy on embedded systems~\cite{marco2020adaptive} create the worst-case headroom an attacker can reclaim. Each decision point is an attack target.
  \item[Autoregressive generation.] Sequence models emit one token at a time until an end-of-sequence (EOS) signal. The number of tokens---and thus total cost---is determined by the model's own outputs, which the attacker can steer~\cite{chen2022nmtsloth,gao2024verbose}.
  \item[Temporal and multi-agent state.] Pipelines that maintain state across frames (multi-object tracking) or aggregate across agents (cooperative perception) add stages whose cost depends on the number of tracked or fused entities~\cite{wang2026cpfreezer}.
  \item[Shared serving and orchestration resources.] Schedulers, KV-cache capacity, guardrails, and agent invocation graphs are shared across requests, so their cost is borne by co-tenants as well as by the requester~\cite{wang2026fillsqueeze,liang2026mobius}.
  \item[Memory and hardware behavior.] Memory access patterns, cache behavior, and DRAM bandwidth can make wall-clock time vary below the algorithmic level even when nominal operation counts are fixed (Section~\ref{sec:nmsfree}). Hardware fault injection, by contrast, is not itself a cost center: induced bit flips are a \emph{delivery channel} that reaches the bottlenecks above, for example NMS parameters~\cite{sistla2025bitflip} or end-of-sequence behavior~\cite{yan2025bithydra}.
\end{description}
These cost centers are the \emph{bottlenecks} along which Sections~\ref{sec:od}--\ref{sec:seqgen} are organized. Each can be reached through several delivery channels---a test-time input, a physical patch, a prompt or retrieved document, a poisoned training set, or a tampered weight---and we keep the two dimensions separate (Section~\ref{sec:bottleneck-delivery}).

\subsection{The unifying principle: intermediate-work amplification}
\label{sec:unifying}

A mechanism recurs across most of the attacks we survey, which we call \emph{intermediate-work amplification}. The adversary engineers a situation in which some internal boundary of the computation receives more objects, tokens, attention computations, or reasoning steps than a benign input of the same nominal class would generate. In object detection, those objects are phantom bounding boxes that flood NMS or multi-object tracking (MOT). In transformer systems, the inflated intermediates may be retained tokens, KV-cache entries, selected expert assignments, or autoregressive decoding steps. In autoregressive decoding specifically, they are extra generated tokens; in reasoning models, extra chain-of-thought steps. The leverage differs by mechanism, and it is worth being precise rather than asserting a single asymptotic law. NMS is roughly quadratic in the number of retained candidate boxes in common iterative implementations and in the worst case (practical runtime also depends on top-$k$ filtering, confidence thresholds, class-wise processing, and kernel vectorization); self-attention is quadratic in sequence length for a full attention matrix, though the marginal per-step cost under a KV cache is closer to linear in context length and is often dominated by memory traffic; autoregressive generation is linear in the number of output steps, so its leverage comes from inflating the \emph{step count} rather than from super-linear per-step work. What these cases share is a common \emph{shape}: a stage whose cost rises with a data-dependent intermediate count that the attacker can inflate while sometimes preserving the nominal output or the source paper's reported task metric (though repetitive generation, task failure, halted actions, or truncated outputs need not preserve either). This shared structure appears across perception pipelines and LLM/VLM-serving domains. It also suggests a candidate cross-domain defense---a \emph{work budget} that caps the admitted intermediate count at a stage boundary (Section~\ref{sec:synthesis})---though, as we discuss there, such caps carry their own utility and fairness costs and are not a universal remedy.

This perspective provides the central organizing principle of this survey. Although latency attacks have emerged independently across object detection, adaptive neural networks, autoregressive language generation, reasoning models, and agentic systems, many can be understood through the common lens of intermediate-work amplification: rather than directly changing a model's prediction, the attacker increases the amount of intermediate computation that the system performs. The specific intermediate differs across domains---for example, candidate boxes in object detection, tracked objects in multi-object tracking, retained tokens in transformers, activated experts in mixture-of-experts models, generated tokens in LLMs, or reasoning steps in large reasoning models---but the underlying objective is the same: to amplify computational work and thereby degrade system availability. This shared viewpoint naturally motivates a common defensive principle. Instead of attempting to detect every attack individually, systems can explicitly bound the amount of intermediate work admitted to computationally expensive stages through work-budget enforcement, such as limiting candidate boxes, generated tokens, or association operations. We revisit this abstraction throughout the survey as a unifying perspective for both attacks and defenses, while noting where it does not apply: timing side channels and availability-adjacent denial of action involve no work amplification (Section~\ref{sec:hw}). Attacks delivered through training data, backdoors, or tampered weights are not exceptions; they exploit the same bottlenecks and differ in delivery rather than in mechanism.
\begin{figure*}[htbp]
  \centering
  \includegraphics[width=\textwidth]{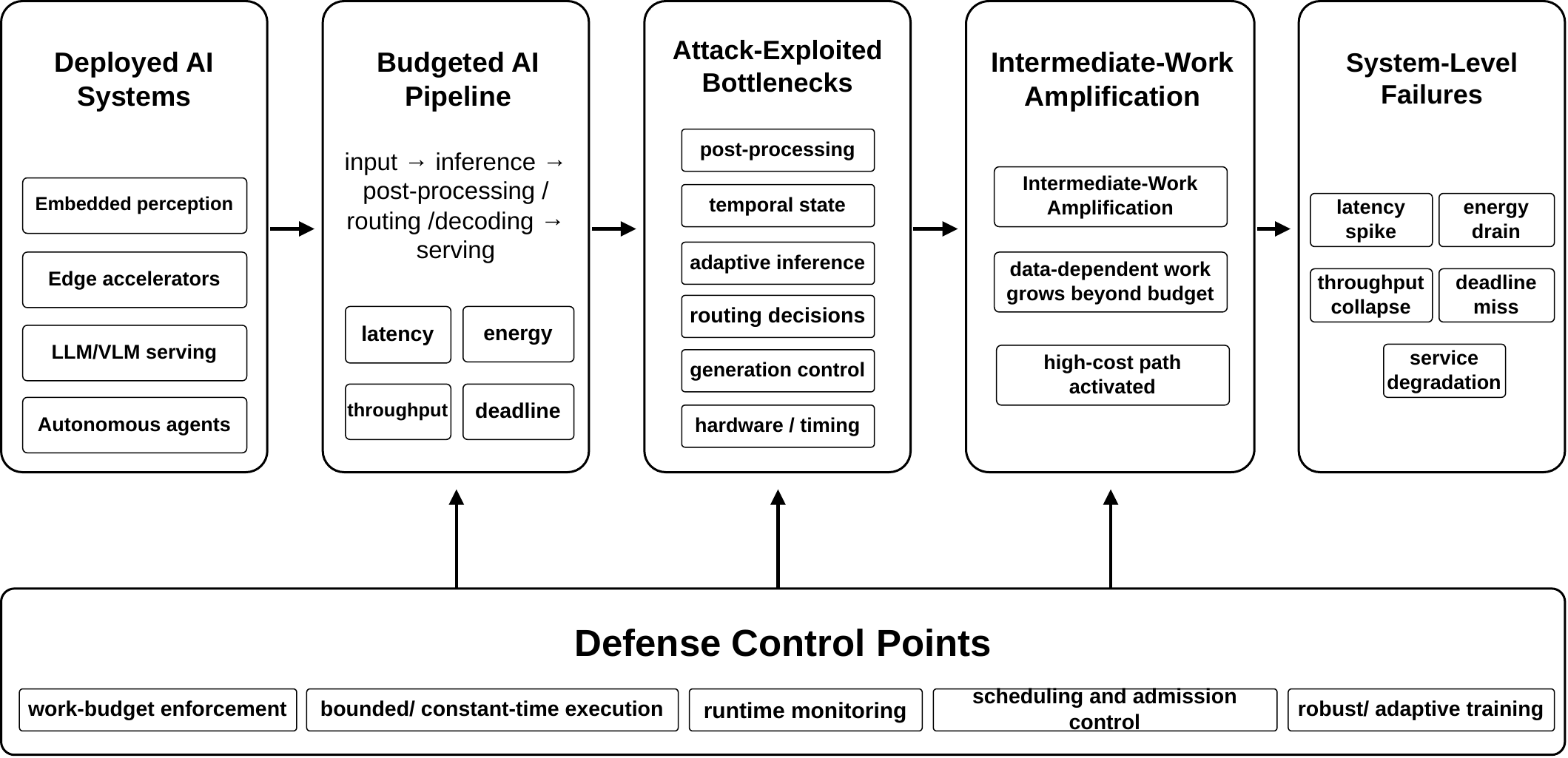}
  \caption{Overview of latency attacks as system-level availability threats.}
  \Description{A system-level overview showing deployed AI systems, a budgeted AI pipeline, attack-exploited bottlenecks, intermediate-work amplification, system-level failures, and defense control points.}
  \label{fig:overview}
\end{figure*}

Figure~\ref{fig:overview} presents an overview of latency attacks as system-level availability threats. Deployed AI systems operate under latency, energy, throughput, and deadline budgets. Attackers exploit data-dependent bottlenecks to amplify intermediate work and push execution toward high-cost paths, causing latency spikes, energy drain, throughput collapse, missed deadlines, or service degradation. The same data-dependent runtime can additionally be observed as an \emph{adjacent timing side channel}---drawn as a separate branch in the figure---which reveals internal state rather than degrading availability; because an attacker need not amplify work to exploit a timing channel, we keep it distinct from the availability consequences above (Section~\ref{sec:threat}). Defenses restore bounded and predictable computation through budget enforcement, monitoring, scheduling, and bounded execution. %

\subsection{Related surveys}
\label{sec:related}
Adversarial-ML surveys overwhelmingly emphasize \emph{evasion}---perturbations that cause misclassification---and treat the availability axis as peripheral~\cite{liao2025llmsurvey,aguilera2025llmsecurity,chowdhury2024breakingdown}. Surveys of adversarial attacks on vision-language models catalog jailbreak, backdoor, and patch threats but largely omit output-length and resource-consumption attacks~\cite{hossain2026vlmsurvey,fu2025vlmdefense,dai2025vlmattacks}. LLM-security surveys catalog serving-layer availability attacks but do not connect them to the perception-stack latency literature~\cite{xu2025llmattacks}. Earlier work by Kianpour and Wen~\cite{kianpour2019timing} surveyed \emph{timing attacks} on ML systems and laid conceptual groundwork, but predates the NMS-overload and LLM-DoS literature.

Two surveys are closest to ours, and they differ from it in the question they ask rather than only in coverage. Brachemi Meftah et al.~\cite{brachemi2026energy} categorize energy-latency attacks using the established taxonomy for traditional adversarial attacks and by target application, review success metrics and defenses, and compare attacks by the cost increase they induce in the target model. The SoK of Rathnasuriya et al.~\cite{rathnasuriya2025sok} organizes efficiency attacks on \emph{dynamic} deep-learning systems by the dynamic behavior exploited---computation per inference (D1), number of iterations (D2), and number of outputs for downstream tasks (D3)---a taxonomy that parallels our treatment of adaptive networks, autoregressive generation, and object-detection output explosion. Both frameworks answer where a model's cost can be inflated and by how much. Neither, to our reading, asks when such an inflation becomes a failure of the system that hosts the model; both, like most primary studies, measure amplification at the model boundary. Our survey adds that question and the analysis needed to answer it (Section~\ref{sec:translation}): the same model-level factor can be operationally harmless or catastrophic depending on system properties that neither organization represents. It also extends the scope to the serving layer, agentic pipelines, embodied VLA policies, and the autonomous-driving perception literature that these surveys largely omit, and it separates what makes computation expensive from how the attacker triggers it (Section~\ref{sec:bottleneck-delivery}). Zhang et al.~\cite{zhang2026resource} review resource-consumption threats within the LLM pipeline, from threat induction through mechanism to mitigation, without perception or embodied coverage. Table~\ref{tab:surveys} positions our work against prior surveys along three
coverage axes and records the principal gap each leaves open, our own included. Ours is the only entry covering all three.

\begin{table}[t]
  \caption{Positioning of this survey against prior surveys touching latency, timing, or availability attacks. \CIRCLE: substantial coverage; \LEFTcircle: partial/peripheral coverage; \Circle: not covered.}
  \label{tab:surveys}
  \small
  \renewcommand{\arraystretch}{1.1} %
  \setlength{\tabcolsep}{2pt}         %
  \begin{tabularx}{\linewidth}{L{3.0cm} c c c c X}
    \toprule
    \textbf{Survey} & \textbf{Year} & \textbf{AD percep.} & \textbf{General DL} & \textbf{LLM/VLM} & \textbf{Principal gap} \\
    \midrule
    Kianpour \& Wen~\cite{kianpour2019timing} & 2019 & \Circle & \LEFTcircle & \Circle & Predates modern latency/sponge and LLM-DoS literature \\
    NIST AML taxonomy~\cite{vassilev2023nist} & 2023 & \Circle & \LEFTcircle & \LEFTcircle & Definitional only; availability is one small category \\
    LLM-security surveys~\cite{xu2025llmattacks,liao2025llmsurvey,aguilera2025llmsecurity,chowdhury2024breakingdown} & 2024--25 & \Circle & \Circle & \CIRCLE & Serving-layer DoS only; no link to perception latency \\
    VLM adversarial surveys~\cite{hossain2026vlmsurvey,fu2025vlmdefense,dai2025vlmattacks} & 2025--26 & \LEFTcircle & \Circle & \LEFTcircle & Jailbreak/backdoor/patch focus; omits output-length \& resource attacks \\
    Brachemi Meftah et al.~\cite{brachemi2026energy} & 2026 & \LEFTcircle & \CIRCLE & \LEFTcircle & Model-level energy-latency attacks, measured at the model boundary; does not synthesize recent serving, reasoning, agentic, and embodied systems \\
    Rathnasuriya et al.\ (SoK)~\cite{rathnasuriya2025sok} & 2025 & \LEFTcircle & \CIRCLE & \LEFTcircle & Scoped to dynamic-DL efficiency robustness; omits serving, agentic, and embodied layers; no model-to-system analysis \\
    Zhang et al.~\cite{zhang2026resource} & 2026 & \Circle & \LEFTcircle & \CIRCLE & LLM resource-consumption pipeline only; no perception or embodied coverage \\
    This survey & 2026 & \CIRCLE & \CIRCLE & \CIRCLE & Bridges AD perception and LLM/VLM serving; model-to-system translation (Section~\ref{sec:translation}) \\
    \bottomrule
  \end{tabularx}
\end{table}

\subsection{Metrics and a working formalization}
\label{sec:metrics}
Reported results in this field are often incomparable. Authors may report multiplicative latency increases ($\times$), absolute milliseconds, FLOP increases, energy in joules, generated-token counts, frames per second, and downstream task outcomes (crash rate, collision rate). On embedded targets, the relevant quantity is often the worst-case execution time (WCET) relative to a real-time deadline, which has motivated dedicated execution-time prediction frameworks for concurrent DNN workloads on edge accelerators~\cite{goel2024express}. Throughout, we preserve each paper's native metric and flag where direct comparison is unsound. Establishing common axes is itself an open problem (Section~\ref{sec:open}).

\subsubsection{Work and resource ratios}
To make ``intermediate-work amplification'' precise, and to separate it from
ordinary large-input denial-of-service, we define work ratios against a \emph{matched
benign reference set} rather than a single hand-selected reference input, whose
choice can produce unstable or misleading ratios. Let $W(\cdot)$ be a
stage-specific work measure (retained candidate boxes, activated neurons,
retained tokens, selected experts, generated tokens, or decoder steps), and let
$\mathcal{B}(x)$ denote a set of benign inputs matched to the adversarial input
$x$ on model, decoding configuration, batch size, hardware, and concurrency, and
stratified on task category and benign length. Three reporting rules follow, each
of which the works we catalogue frequently violate. First, matching must not use
any attacked quantity: baselines selected by attacked latency or attacked output
length let the attack influence its own denominator, so a source-derived attack
must be matched against its clean counterpart's output or a predeclared benign
stratum. Second, repeated timing measurements of a single input characterize
measurement noise, not attack-population variability, and must not be resampled
as independent attack samples; the independent experimental unit is the matched
input pair for per-input attacks, and the complete workload run or seed for
serving experiments. Third, a ratio taken against a predeclared statistic of a
reference set is not a paired-ratio distribution and should not be reported as
one, since no pairing rule exists.

Throughout this survey we use \emph{amplification factor} generically for the ratio between an attack-induced metric and a matched benign reference,
\begin{equation}
\operatorname{AF}_{M}(x^{\mathrm{adv}})=\frac{M(x^{\mathrm{adv}})}{\operatorname{median}_{x_0\in\mathcal{B}(x^{\mathrm{adv}})} M(x_0)},
\label{eq:af}
\end{equation}
where $x^{\mathrm{adv}}$ is the adversarial input, $M(\cdot)$ is the metric of interest---work (retained boxes, activated units, decoder steps, generated tokens), wall-clock latency, or energy---and $\mathcal{B}(x^{\mathrm{adv}})$ is the predeclared matched benign set above. For source-derived attacks, the corresponding generic amplification factor uses the clean counterpart directly in the denominator rather than the median of a reference set. Depending on $M$, this quantity expresses work, latency, energy, token, or another form of amplification. Because published studies use heterogeneous definitions and experimental settings, we preserve each paper's native metric rather than retrospectively converting all results to a common factor, and we introduce no separate named ratios beyond this generic form. One caution carries across metrics: a work-proxy increase need not produce a proportional latency increase, since wall-clock time can be bounded by memory movement, kernel-launch overhead, or batching rather than nominal work. When the benign reference median is \emph{zero}---e.g., zero retained detections, zero selected optional components, or zero generated continuation tokens---the ratio is undefined and should not be rescued with an arbitrary $\epsilon$; report instead the \emph{absolute} change together with the attack value, and for very small nonzero denominators report the denominator explicitly, since a large ratio may reflect a tiny baseline rather than a large operational effect.

Request-level latency alone is insufficient for shared serving systems, because attacks may primarily affect queueing delay, benign co-tenants, tail latency, or useful throughput. Evaluations should therefore report P50, P95, and P99 latency under benign and adversarial workloads, benign-victim latency under adversarial co-load, and the relative degradation in throughput or goodput (successfully completed useful requests per unit time) against a matched benign \emph{workload}---a distinct experimental unit from a single input, since a workload also fixes an arrival process, request mixture, concurrency, batching, scheduling policy, and observation interval, all of which should be specified. This throughput or goodput degradation is what matters for scheduler, KV-cache, guardrail, and agentic attacks.

\subsubsection{Deadline effects}
On real-time targets, let $p_{\text{clean}}$ and $p_{\text{attack}}$ denote the fractions of frames missing their deadline under benign and adversarial input. We distinguish the \emph{relative deadline-miss amplification} $p_{\text{attack}}/p_{\text{clean}}$, which is undefined when the clean miss rate is zero, from the \emph{absolute deadline-miss increase} $p_{\text{attack}}-p_{\text{clean}}$, and recommend the absolute form as the default since well-engineered real-time systems have near-zero clean miss rates. To connect these ratios to the real-time-systems view, we also recommend reporting deadline slack and tardiness,
\begin{equation}
S_i = d_i - f_i, \qquad \tau_i = \max(0, f_i - d_i) = \max(0, -S_i),
\end{equation}
where $d_i$ and $f_i$ are job $i$'s absolute deadline and completion time, $S_i$ is slack, and $\tau_i$ is tardiness; when every job shares a common relative deadline $D$ this reduces to $S_i = D - T_i$, and negative slack indicates a miss. Useful outcome measures include the deadline-miss rate, the distribution of slack, the maximum or high-quantile tardiness (stating whether the quantile is taken over frames, requests, or independent runs), the maximum number of consecutive missed deadlines in temporal systems, the recovery time after attack cessation, the backlog at attack termination, and---for streaming perception---the age of information where appropriate.

\subsubsection{Attacker economics}
The operational significance of any amplification factor depends on \emph{attacker cost}. Evaluation should distinguish the attacker's one-time development cost (e.g., optimizing a universal patch or trigger) from the marginal cost of each delivery attempt (e.g., issuing one API request or displaying the patch), and should count failed attempts when estimating campaign cost, since those still cost the adversary. Victim cost should measure the \emph{incremental} resource consumption caused by the attack rather than total service cost, so that normal service cost is not misattributed to the adversary. Ratios between attacker and victim costs are meaningful only when both are converted to a common unit, such as monetary cost. For safety-critical attacks, attacker cost should be reported but should not by itself determine whether the threat is consequential: a safety-motivated attacker may willingly spend substantial resources to delay a single vehicle action. Reporting a cost--impact pair---an amplification factor \emph{together with} the threat model and attacker effort---is, in our view, the single most useful discipline the field currently lacks; most surveyed papers report only the victim-side ratio (DeepSloth is a rare exception, relating about 2~ms of attacker computation per input to the latency increase it causes~\cite{hong2021deepsloth}). This framing also delimits our \emph{unifying mechanism}: intermediate-work amplification occurs when an adversarial input whose externally visible size and request configuration are matched to a benign reference induces substantially greater \emph{internal} intermediate work, as distinct from request flooding, oversized inputs, or naturally hard instances, all of which raise work without this matched-size asymmetry. The mechanism definition is not the survey's complete inclusion criterion, however: scheduler- and agent-level attacks (Section~\ref{sec:serving}) may fall outside the narrow matched-size asymmetry yet remain in scope as broader computational-availability threats.

\section{Threat Models}
\label{sec:threat}
A coherent threat model is the prerequisite for comparing attacks that span sensor inputs, public APIs, and hardware faults; without it, headline slowdown numbers are not commensurable. We classify latency attacks along three dimensions: the \emph{delivery channel} or attacker interface (how the adversary touches the system), the \emph{attacker knowledge} (white-, grey-, or black-box), and the exploited \emph{bottleneck} or targeted stage. Table~\ref{tab:threats} indexes representative works by delivery channel and lists the bottlenecks each channel has been shown to reach; Section~\ref{sec:bottleneck-delivery} explains why the rest of the survey is organized by bottleneck instead.

Two distinctions are worth stating explicitly, because the rest of the survey preserves them. First, we separate the \emph{security effect} from the \emph{persistence mechanism}, since these are different dimensions rather than three values of one. \emph{Availability degradation}---inflated computation, energy, latency, or missed deadlines---is the principal security effect studied here; \emph{timing leakage}, in which data-dependent runtime \emph{reveals} internal state rather than degrading service, is an adjacent \emph{confidentiality} effect that shares the observable (variable runtime) but has the opposite goal, and we flag it as such wherever it arises rather than folding it into the amplification mechanism. Persistence is a separate property describing how long an attack remains effective: training-time poisoning and weight tampering (Section~\ref{sec:persistent}) are \emph{persistent delivery mechanisms} that can embed availability degradation durably across subsequent inputs, rather than being a security consequence in their own right. Second, the attacker interfaces we group under one umbrella differ radically in access, cost, persistence, detectability, and feasibility: per-input adversarial examples, universal perturbations, physical patches or projector attacks, request-level API prompts, distributed request flooding, training-data poisoning, weight tampering, hardware fault injection, and timing side-channel exploitation are not interchangeable. Our synthesis abstracts over the \emph{mechanism} (intermediate-work amplification) while retaining these interface distinctions in every comparison, since a projector attack and an API prompt that both achieve a $10\times$ slowdown are not equally practical threats.

\begin{table}[htbp]
  \caption{Delivery channels for latency attacks and the bottlenecks each has been shown to reach. Sections~\ref{sec:od}--\ref{sec:seqgen} are organized by bottleneck; this table indexes the same works by delivery channel. The last row is an adjacent confidentiality phenomenon, not a latency attack.}
  \label{tab:threats}
  \small
  \begin{tabularx}{\linewidth}{L{2.3cm} L{2.6cm} L{3.3cm} >{\raggedright\arraybackslash}X}
    \toprule
    \textbf{Delivery channel} & \textbf{What the attacker controls} & \textbf{Bottlenecks reached (section)} & \textbf{Representative evidence} \\
    \midrule
    Digital input perturbation & Image, point cloud, audio, or text tensor & NMS (\ref{sec:nms}); tracking (\ref{sec:tracking}); LiDAR preprocessing (\ref{sec:fusion}); activation sparsity (\ref{sec:sparsity}); conditional computation (\ref{sec:dynamic}); decoder length (\ref{sec:seq2seq}); pipeline execution path and offload queue (\ref{sec:serving}) & Overload~\cite{chen2024overload}; SlowTrack~\cite{ma2024slowtrack}; SlowLiDAR~\cite{liu2023slowlidar}; sponge examples~\cite{shumailov2021sponge}; DeepSloth~\cite{hong2021deepsloth}; NMTSloth~\cite{chen2022nmtsloth}; SNN sponge~\cite{raptis2026driving}; AESOP~\cite{li2026aesop}; TrackFlood~\cite{gu2026trackflood}; GateDrain~\cite{gu2026gatedrain} \\
    \addlinespace
    Physical sensor delivery & Printed patch, lens sticker, or projection into the scene & NMS (\ref{sec:nms}); tracking (\ref{sec:tracking}) & SlowPerception~\cite{ma2024slowperception}; DetStorm~\cite{muller2025detstorm}; Groundswell~\cite{xia2026groundswell} (proposed; evaluated digitally) \\
    \addlinespace
    Prompt or API request & Query text or image sent to a served model & Output and reasoning length (\ref{sec:llm}); expert routing (\ref{sec:dynamic}); scheduler and KV cache (\ref{sec:serving}) & Engorgio~\cite{dong2025engorgio}; ReasoningBomb~\cite{liu2026reasoningbomb}; Verbose Images~\cite{gao2024verbose}; Misrouter~\cite{fei2026misrouter}; Fill and Squeeze~\cite{wang2026fillsqueeze} \\
    \addlinespace
    Retrieved or tool content & Documents, retrieval corpus, or tool metadata the system ingests & Output and reasoning length (\ref{sec:llm}); agent invocation graph (\ref{sec:serving}) & RA-ICA~\cite{liu2026inference}; DrainCode~\cite{wang2025draincode}; OverThink~\cite{kumar2025overthink}; Beyond Max Tokens~\cite{zhou2026beyondmaxtokens} \\
    \addlinespace
    Message injection & V2V or inter-agent messages & Cooperative fusion (\ref{sec:fusion}); agent invocation graph (\ref{sec:serving}) & CP-FREEZER~\cite{wang2026cpfreezer}; Mobius Injection~\cite{liang2026mobius}; CORBA~\cite{zhou2026corba} \\
    \addlinespace
    Training data or backdoor & Training samples or outsourced training & Activation sparsity (\ref{sec:sparsity}); early exit (\ref{sec:dynamic}); NMS (\ref{sec:nms}); output length (\ref{sec:llm}) & Sponge poisoning~\cite{cina2025energy}; multi-exit poisoning~\cite{huang2024multiexit}; sponge backdoor~\cite{xiao2024sponge}; energy backdoor~\cite{meftah2025energy}; DoS poisoning~\cite{gao2024dospoisoning} \\
    \addlinespace
    Weight or hardware tampering & Parameter values or bits, accelerator state & NMS (\ref{sec:nms}); activation sparsity (\ref{sec:sparsity}); output length (\ref{sec:llm}) & Bit-flip attack~\cite{sistla2025bitflip}; SkipSponge~\cite{te2025skipsponge}; EvoWeight~\cite{akram2025evoweight}; BitHydra~\cite{yan2025bithydra} \\
    \addlinespace
    \emph{Adjacent:} timing and serving observation & Elapsed time, packet sizes, iteration boundaries, or token counts & None; confidentiality (\ref{sec:timing}) & Timing side-channel~\cite{nakai2021timing}; speculative-decoding serving leak~\cite{wei2024speculation} \\
    \bottomrule
  \end{tabularx}
\end{table}

\subsection{Attacker interface}
The interface ranges from purely digital (modifying the input tensor of a model under white-box control) to physical (projecting or printing adversarial patterns that a camera then captures) to systemic (injecting messages over a vehicle-to-vehicle channel, or submitting prompts to a public API). For sensor-facing attacks, physical realizability is an important distinction between digital demonstrations and attacks deliverable through the sensing channel, and is a recurring theme in the autonomous-driving literature~\cite{ma2024slowperception,muller2025detstorm}; it is not the dividing line for deployability in general, since API prompts, corpus poisoning, and remote serving attacks can be deployable without any physical realization.

\subsection{Attacker knowledge}
White-box attacks assume gradient access and dominate early work because the optimization is direct. Grey-box attacks assume architectural knowledge but not exact weights~\cite{du2024energy}. Black-box attacks, increasingly important for deployed services, rely on transfer from a surrogate, on query feedback (e.g., \texttt{usage.prompt\_tokens}), or even on timing itself as an oracle~\cite{nakai2021timing,zhang2025crabs}. The existence of timing-only black-box attacks is significant: it means even an opaque service that returns no probabilities can leak enough information to be optimized against.

\subsection{Bottleneck versus delivery channel}
\label{sec:bottleneck-delivery}
Two questions are easy to conflate: what makes a computation expensive, and how the attacker causes that expensive behavior. We answer the first with the \emph{bottleneck} and the second with the \emph{delivery channel}, and we keep them orthogonal. A neural network can be made to skip fewer operations by an adversarial input or by poisoning its training; a language model can be made to generate more tokens by a crafted prompt, a poisoned retrieval corpus, a backdoor, or a bit flip. In each pair the computational effect is the same, and so, as Section~\ref{sec:persistent} shows, is the class of controls that bounds it.

Sections~\ref{sec:od}--\ref{sec:seqgen} are therefore organized by bottleneck only. Every attack that exploits a given bottleneck is discussed in the section for that bottleneck, whatever its delivery channel, and within a section attacks are ordered by delivery channel: digital input, physical sensor, prompt or retrieved content, message, training-time poisoning or backdoor, and weight or hardware tampering. Sponge examples and sponge poisoning, for example, both reduce activation sparsity and so both appear in Section~\ref{sec:sparsity}; an EOS-delaying prompt, a no-EOS backdoor, and an EOS-suppressing bit flip all lengthen autoregressive generation and so all appear in Section~\ref{sec:llm}. Section~\ref{sec:hw} does not re-describe attacks. It examines the properties that depend on the delivery channel rather than on the bottleneck---persistence, detectability, removal cost, and which defenses can see the attack---and treats two adjacent phenomena that fall outside the taxonomy because they involve no work amplification: timing side channels and denial of action in embodied systems. Table~\ref{tab:threats} and the attack-mapping matrix in Appendix~\ref{sec:matrix} give the bottleneck--delivery mapping.

\section{Perception Bottlenecks: Post-Processing, Temporal State, and Fusion}
\label{sec:od}
Object detection is the most thoroughly studied target, because its post-processing stage offers an unusually clean computational bottleneck and because autonomous driving supplies a high-stakes deployment context. This section covers three perception bottlenecks: NMS post-processing (Section~\ref{sec:nms}), tracking state (Section~\ref{sec:tracking}), and sensor preprocessing and cooperative fusion (Section~\ref{sec:fusion}). Within each, attacks are ordered by delivery channel---digital input, physical sensor delivery, and persistent delivery through backdoors or bit flips---following Section~\ref{sec:bottleneck-delivery}.

\subsection{Non-maximum suppression}
\label{sec:nms}
NMS filters overlapping candidate boxes and has a cost that grows with the number of candidates admitted to it. \emph{Phantom Sponges}~\cite{shapira2023phantom} crafts a universal adversarial perturbation that adds many ``phantom'' boxes to overload NMS while \emph{preserving} the original detections, so accuracy collapse does not betray the attack. \emph{Overload}~\cite{chen2024overload} generalizes this by formulating latency maximization as an optimization with a spatial-attention mechanism, reporting that single-image inference time can be increased tenfold; the authors describe the attack as ``NMS-agnostic''---meaning the optimization is not tied to one particular NMS implementation, not that the attack is independent of candidate-dependent detector processing, since it still operates by proliferating candidate detections. \emph{Beyond PhantomSponges}~\cite{schoof2024beyond} modifies a box-area loss to lower intersection-over-union and exacerbate NMS load, reporting up to a 550\% increase in NMS time on a YOLOv5-small configuration. \emph{Groundswell}~\cite{xia2026groundswell} employs Regional Perturbation Balance (RPB) to inject adversarial phantom objects while preserving original detections. By leveraging spatial object distribution patterns, RPB generates masks that guide phantom placement into regions with minimal interference, increasing the number of NMS candidates on YOLOv5s from 101 to about 19{,}000 and end-to-end latency from 15.2~ms to 41.4~ms in its most aggressive configuration, where recall of the original detections falls to 18\%; its most stealth-oriented configuration keeps 64\% recall with 13{,}000 candidates and 24.0~ms. The source's text describes the aggressive setting as a more-than-300-fold increase in candidates and a roughly 250\% increase in execution time; its tabulated values correspond to about $188\times$ and $2.7\times$, which we report.
\emph{Steal Now and Attack Later}~\cite{chen2024steal} moves the phantom-box objective into the black-box setting: rather than optimizing against a known detector, it extends the ``steal now, decrypt later'' idea by first collecting reusable adversarial components and composing them into ghost-object examples afterward, reporting successful ghost-object generation across several commonly used detectors and the Google Vision API without prior knowledge of the target
model. \emph{Daedalus}~\cite{wang2019daedalus}, although framed as an integrity attack, is a relevant NMS failure mode: it compresses box dimensions so NMS cannot filter dense false positives, and is demonstrated via printed posters. A complementary observation by Biton et al.~\cite{biton2023timing} reframes variable-time inference as a security liability in itself: because NMS runtime is data-dependent, it \emph{leaks} internal inference state, and this timing signal can be used to enhance black-box, decision-based attacks (Section~\ref{sec:timing}). This dual role---NMS as both a target of latency inflation and a side-channel oracle---motivates constant-time NMS as an explicit security requirement (Section~\ref{sec:bounded}).

\paragraph{Physical delivery.} \emph{DetStorm}~\cite{muller2025detstorm} advances physical realizability, optimizing objects against multiple NMS variants and reporting a 506\% average increase in detected objects and delays up to 8.1~seconds, while arguing that prior digital or large-patch attacks lack real-world applicability. \emph{Groundswell}~\cite{xia2026groundswell} proposes physical delivery---a universal perturbation affixed as a sticker on the camera lens or projected onto a surface ahead of the vehicle---but its evaluation is digital: the artifact description states that the attack operates entirely in the digital domain and that the generated patches are not optimized for physical deployment. In that evaluation, only the most aggressive configuration pushes end-to-end latency past the 30~ms budget the source adopts for SAE Level 4--5 automation (41.4~ms); the configuration that preserves 64\% of original detections stays below it (24.0~ms on an RTX~3090). (SAE levels do not define a single universal perception deadline; acceptable latency depends on speed, sensor rate, braking distance, architecture, and control policy).

\paragraph{Persistent delivery.} The same bottleneck can be reached without any adversarial input at test time. A sponge backdoor~\cite{xiao2024sponge} poisons detector training so that trigger-bearing inputs prolong NMS while clean accuracy is preserved, and a Rowhammer-based fault-injection attack flips selected parameter bits that affect NMS behavior, reporting a latency increase of up to 71.6~ms ($20.4\times$) with 31 bit flips~\cite{sistla2025bitflip}. These attacks differ from the input-borne ones only in how they are delivered; their persistence and detectability are discussed in Section~\ref{sec:persistent}.

\paragraph{How practical are NMS latency attacks?} The magnitudes reported above rest on a specific set of measurement assumptions,
and a recent independent evaluation questions them. Monteuuis et
al.~\cite{monteuuis2025evade} propose EVADE, a framework for assessing attack
practicality, and apply it to the four NMS latency attacks with public code
(Daedalus, Overload, Phantom Sponges, and Beyond PhantomSponges) across 7
hardware platforms, 15 YOLO models, 3 export formats, INT8 quantization, 3
datasets, and 3 defenses, generating 114 adversarial datasets. Their findings are
mixed but weigh heavily against practical impact. The attacks do not transfer
across YOLO versions or model sizes, so the adversary must know the exact target;
the universal patches of Phantom Sponges and Beyond PhantomSponges do not survive
a change of dataset, failing even between two automotive datasets; and each of
three defenses drawn from different categories removed all four attacks' effect
on the post-NMS detection count. They do transfer across export formats, and
Overload additionally transfers to INT8-quantized models.

Two aspects of that study need care when its results are read as evidence about
NMS workload. First, its hardware conclusion is analytical rather than a
per-attack, per-platform measurement: the authors microbenchmark NMS runtime
against synthetic proposal counts on each platform and combine the resulting
curves with application latency budgets, concluding that the proposal counts
required to breach a 100--500~ms budget are unreachable when NMS executes on GPU.
This is the part of the study that bears on NMS work, because the quantity that
governs NMS cost is the number of candidates admitted to it. EVADE notes that the
model itself caps the proposals it passes to NMS, and that this cap has shrunk with
each YOLO release: 25{,}000 for YOLOv5 and 8{,}000 for YOLOv8, both below the
30{,}000 accepted by the Ultralytics NMS function (\texttt{max\_nms}), which
therefore never binds. The model's cap matters as much as the hardware: where NMS
runs on CPU, YOLOv5's larger cap lets NMS attacks reach a 2~s threshold, whereas
YOLOv8's limits them to under 500~ms.

Second, the study's headline mitigation acts on a different quantity. In the
Ultralytics implementation, candidates are filtered by confidence, sorted and
truncated to \texttt{max\_nms}, and passed in full to the NMS kernel; only then are
the retained detections truncated to \texttt{max\_det} (default 300; the four
attacks had originally been evaluated with it set to 30{,}000, which EVADE
considers unrealistic for most applications).
\texttt{max\_det} is therefore an \emph{emission cap}: it limits how many
detections leave NMS, not how much work NMS performs. Reducing it from 300 to 10
lowered the post-NMS box count under the four attacks from 152--300 to 6--8,
against a benign count of 5 under the same cap (6 without it). Because this metric is right-censored at
\texttt{max\_det} by construction, the reduction does not show that NMS ran faster
or processed fewer candidates, and at the default value of 300 the attacks were
not neutralized. What the result does show---and EVADE itself notes---is that \texttt{max\_det} bounds the
input to the \emph{next} stage: downstream consumers of post-NMS detections, such
as the tracker's association step (Section~\ref{sec:tracking}), receive a bounded
number of detections per frame, whereas the NMS kernel is not protected. Even that
protection is partial, because the tracker's persistent pool can still accumulate
across frames (Section~\ref{sec:tracking}). We return to this distinction between capping the
work entering a stage and capping the results leaving it in
Sections~\ref{sec:bounded} and~\ref{sec:synthesis}.

The evaluation covers one detector family on one metric, and DetStorm~\cite{muller2025detstorm} was
excluded for lack of public code---the authors argue it inherits Phantom
Sponges's weaknesses, but this is an inference, not a measurement. What the study
establishes is not that NMS latency attacks are impossible but that reported
amplification factors are properties of an attack--model--configuration triple,
and that the parameters most decisive for them---input resolution, confidence
threshold, pre-NMS cap, and NMS implementation---were left unstated in the
original evaluations.

\subsection{Tracking and temporal state}
\label{sec:tracking}
A typical autonomous-driving perception stack feeds detector outputs into a multi-object tracker (MOT) that maintains a persistent identity for each object across frames. Many real-time trackers follow the \emph{tracking-by-detection} paradigm: every frame, the tracker (i) predicts each existing track's new position---often with a per-track \emph{Kalman filter} using a constant-velocity motion model, as in SORT-family pipelines---(ii) \emph{associates} the incoming detections to those predicted tracks, and (iii) updates matched tracks, spawns tracks for unmatched detections, and retires tracks unseen for a buffer of frames. The association step---the stage that tracker-overload attacks target---is commonly solved as a bipartite matching (Hungarian or a related assignment method) over a cost matrix built from every (track, detection) pair, though this is not universal: some trackers instead use greedy association, learned association, joint detection-and-tracking, or alternative motion models. For $M$ active tracks and $N$ detections, constructing the dense association matrix costs $O(MN)$; the assignment step's complexity depends on the algorithm and how rectangular matrices are handled, with a common coarse upper bound of $O(\max(M,N)^3)$. Increasing $N$ therefore raises current-frame association work directly, while persistent false tracks can increase future association-matrix dimensions, causing cumulative workload growth across frames; the exact scaling depends on the track-retirement policy and the assignment implementation. Unlike a fixed-structure detector forward pass, such a tracker is \emph{content-dependent}: flooding it with detections directly increases its per-frame workload. A cap on the detections NMS emits (Section~\ref{sec:nms}) bounds one dimension of this matrix, the per-frame detection count, but not the other: unmatched tracks survive across frames, so with retention horizon $R$ and at most $|D|_{\max}$ detections per frame the pool can grow toward $R|D|_{\max}$, and SlowTrack remains effective under a 500-detection cap in a ByteTrack pipeline~\cite{gutrackshield}.

Detector-only attacks often fail to change downstream behavior because multi-object tracking absorbs detection noise; prior work shows that over 98\% detection-attack success may be required to alter tracking outcomes~\cite{jia2020fooling}. This motivates pipeline-level threat models. \emph{SlowTrack}~\cite{ma2024slowtrack} attacks camera-based perception end to end with a two-stage strategy and three new loss designs. It reports an average $453.8\times$ slowdown of the tracking stage and an average $28.4\times$ slowdown of the whole camera-based perception pipeline, $2.9\times$ more than prior attacks; under its evaluated Baidu Apollo stack and LGSVL simulator configuration, the authors further report a vehicle crash rate of $\sim$95\% versus $\sim$30\% for baselines. Measured on a Jetson AGX Orin with YOLOX-S and ByteTrack, the attack inflates the mean persistent tracker pool from 20.5 to 1{,}677.7 tracks and tracking-stage latency from 3.27~ms to 139.90~ms, and it is not quality-preserving: MOTA falls from 57.19 to 1.08~\cite{gutrackshield}. The gap between the stage-level and pipeline-level factors is itself informative (Section~\ref{sec:translation}).

The same bottleneck survives the removal of NMS. \emph{TrackFlood}~\cite{gu2026trackflood} attacks detect-then-track pipelines built on NMS-free detectors---the one-to-one heads of YOLOv10 and YOLO26 and the query-based RT-DETR---by recovering each detector's differentiable per-slot confidences and optimizing perturbations that flood the tracker with spatially spread phantom detections. On a Jetson AGX Orin with TensorRT FP16, detector latency stays within 0.99--1.01$\times$ of clean, because the forward pass evaluates a fixed number of prediction slots, while tracker latency rises. At an $L_\infty$ budget of 8/255 a universal perturbation raises ByteTrack end-to-end latency by only $1.09$--$1.47\times$ with no missed 33~ms deadlines; a 32/255 stress test reaches $13.54\times$ with OC-SORT and misses every deadline. The overload depends on the achievable detection flood (attack strength, scene density, and detection threshold) more than on the detector head: RT-DETR's 300 queries resist flooding on sparse scenes, but its denser output makes it the most overloaded on dense ones (Section~\ref{sec:nmsfree}).

\paragraph{Physical delivery.} \emph{SlowPerception}~\cite{ma2024slowperception} introduces projector-based universal perturbations that create numerous phantom objects and so load both NMS and the tracker; under the authors' specified industry-grade system and simulator configuration, they report an average physical-world latency of 2.5~seconds and a 97\% collision rate.

\subsection{Sensor preprocessing and cooperative fusion}
\label{sec:fusion}
\emph{SlowLiDAR}~\cite{liu2023slowlidar} presents what its authors describe as the first systematic availability attack on LiDAR detection, handling non-differentiable preprocessing via differentiable proxies and an execution-time-aware loss to inflate runtime across six popular pipelines while maintaining imperceptibility. \emph{CP-FREEZER}~\cite{wang2026cpfreezer} targets vehicular cooperative perception, delivering adversarial perturbations via vehicle-to-vehicle (V2V) messages to maximize fusion delay; it resolves point-cloud nondifferentiability and transmission asynchrony, increasing end-to-end latency by over $90\times$, pushing per-frame time beyond 3~seconds, and achieving 100\% success on a real vehicle testbed. A simulation study further analyzes the system-level impact of perception inference-time attacks on full autonomous vehicle stacks~\cite{chen2025impact}, and real-time defenses against object-based LiDAR attacks are beginning to appear~\cite{zhang2025detstormlidar}.

\begin{table}[htbp]
  \caption{Representative latency attacks on perception bottlenecks. Metrics are reported as in the source; ``--'' indicates not reported in a directly comparable form.}
  \label{tab:od-attacks}
  \small
  \begin{tabularx}{\linewidth}{L{2.3cm} L{2.2cm} X L{1.7cm} L{1.9cm}}
    \toprule
    \textbf{Method} & \textbf{Target stage} & \textbf{Key quantitative result} & \textbf{Delivery} & \textbf{Evaluation} \\
    \midrule
    Phantom Sponges~\cite{shapira2023phantom} & NMS (end-to-end OD) & Universal perturbation overloads NMS while preserving detections & Digital & Digital model eval. \\
    Overload~\cite{chen2024overload} & OD inference (edge) & Single-image inference time up to $10\times$; NMS-agnostic & Digital & Target hardware \\
    Beyond PhantomSponges~\cite{schoof2024beyond} & NMS runtime (YOLO) & Up to 550\% NMS-time increase (YOLOv5-small) & Digital & Digital model eval. \\
    Groundswell~\cite{xia2026groundswell} & NMS (camera OD) & Candidates 101$\to$19k; 15.2$\to$41.4~ms ($+172\%$) at 18\% recall; $-$80\% opt.\ time & Digital (sticker / projector proposed) & Hardware-timed (desktop GPUs) \\
    Steal Now \& Attack Later~\cite{chen2024steal} & NMS (black-box OD) & Ghost objects on common detectors and a commercial API without model knowledge; ${<}\$1$ per example & Digital & Digital model eval.\ (incl.\ commercial API) \\
    SlowTrack~\cite{ma2024slowtrack} & Camera perception (detect+track) & Pipeline latency $28.4\times$ on average ($2.9\times$ prior attacks); tracking stage $453.8\times$; crash rate $\sim$95\% vs.\ $\sim$30\% & Digital & Closed-loop simulation \\
    SlowPerception~\cite{ma2024slowperception} & NMS + MOT workload & Avg.\ 2.5~s physical latency; 97\% collision rate & Physical (projector) & Closed-loop simulation \\
    DetStorm~\cite{muller2025detstorm} & Camera perception delay & 506\% more detections; up to 8.1~s delay & Physical (projector) & Target hardware \\
    TrackFlood~\cite{gu2026trackflood} & Tracker behind NMS-free detectors & Detector 0.99--1.01$\times$; end-to-end $1.09$--$1.47\times$ at 8/255 (no misses); up to $13.54\times$, 100\% misses at 32/255 & Digital & Target hardware \\
    SlowLiDAR~\cite{liu2023slowlidar} & LiDAR detection runtime & Significant latency rise on six LiDAR pipelines & Digital & Digital model eval. \\
    CP-FREEZER~\cite{wang2026cpfreezer} & Cooperative fusion & $>90\times$ latency; $>$3~s/frame; 100\% on testbed & V2V message & Vehicle testbed \\
    Sponge Backdoor~\cite{xiao2024sponge} & OD latency via NMS & Poisoned inputs prolong time; clean accuracy preserved & Training poison & Digital model eval. \\
    Bit-flip attack~\cite{sistla2025bitflip} & NMS via parameters & Up to 71.6~ms ($20.4\times$) with 31 bit-flips & Hardware fault & Target hardware \\
    \bottomrule
  \end{tabularx}

  {\footnotesize\par\vspace{2pt}\emph{Notes.} Figures use each source's native measure (candidate-box count, wall-clock latency, FPS, or a downstream outcome such as collision/crash rate) and are \emph{not} directly comparable across rows; each ratio is relative to that paper's own benign baseline. We separate \emph{Delivery} (how the attack reaches the system: \emph{Digital} input, \emph{Physical} sensor delivery with the specific channel noted, \emph{V2V message}, \emph{Training poison}, or \emph{Hardware fault}) from \emph{Evaluation} (where effects are measured: \emph{Digital model evaluation} = tensor-level experiment on the trained model (or on a hosted commercial API, where no deployment-device timing is reported),
  \emph{Target hardware} = timed on the deployment device, \emph{Closed-loop simulation} = simulator with control outcomes, \emph{Vehicle testbed} = closed physical system). Where an entry lists both a physical delivery and a target-hardware measurement, these were reported by the source but not necessarily in a single combined on-road experiment; no entry denotes a full on-road deployment.\par}
\end{table}

\textbf{Section Summary.} Collectively, these studies illustrate a clear evolution of latency attacks in perception systems. Early work primarily exploited a single computational bottleneck, most notably non-maximum suppression (NMS). More recent research increasingly targets complete perception pipelines, incorporates physical realizability, and evaluates downstream consequences such as tracking failures, delayed planning, and collision risk. The same bottlenecks are reached through several delivery channels---digital inputs, physical patches and projections, V2V messages, backdoors, and bit flips---and the independent re-evaluation of NMS attacks shows that the reported effects depend on deployment parameters as much as on the attacks themselves. This progression marks a step toward system-level thinking: latency attacks are no longer isolated component failures but increasingly propagate across perception pipelines, although, as Section~\ref{sec:translation} shows, a large stage-level amplification need not become a large pipeline-level one. More importantly, the object-detection literature reveals that the attacker's objective is not to exploit NMS itself, but to maximize intermediate computation wherever it occurs. The following sections show that this principle naturally extends beyond perception to general neural inference.

\section{Forward-Pass Bottlenecks: Activation Sparsity and Conditional Computation}
\label{sec:sponge}

This section covers two bottlenecks internal to the forward pass. The first is activation sparsity (Section~\ref{sec:sparsity}), exploited by test-time sponge inputs and by poisoning or weight tampering that produces the same effect. The second is conditional computation (Section~\ref{sec:dynamic}): early exits, dynamic depth and width, adaptive ODE solvers, token pruning, dynamic quantization, and expert routing. Neither is tied to a specific post-processing stage, which makes both broadly applicable across architectures.

\subsection{Activation sparsity}
\label{sec:sparsity}
\emph{Sponge examples}~\cite{shumailov2021sponge} are the seminal work: inputs crafted to maximize energy consumption and latency. Against language models they frequently increase both by about $30\times$, and a real-world demonstration raised a cloud translator's response time from 1~ms to 6~s; on vision models the effect is present but less pronounced, and a later study reports at most about 3\% energy increase on the image classifiers tested and could not reproduce the original results~\cite{te2025skipsponge}. The source identifies two mechanisms: larger computation dimensions, such as the number of tokens processed and generated, and reduced activation sparsity, which removes the savings that sparsity-aware accelerators exploit~\cite{shumailov2021sponge}. It attributes most of the language-model degradation to the first, which is the generation-length bottleneck of Section~\ref{sec:seqgen}.
A follow-up analysis shows that uniform/constant inputs act as ``free'' sponge examples and that the achievable energy increase is bounded by the architecture's sparsity range, motivating activation-sparsity monitoring as a defense signal~\cite{muller2024uniform}. \emph{Sparsity attacks}~\cite{krithivasan2020sparsity} degrade energy/latency by minimizing activation sparsity and withstand activation thresholding and input quantization. Spiking neural networks, attractive for low-power edge perception, are also vulnerable: efficiency attacks raise spike activity by $1.7\times$--$2.5\times$ and energy by $1.4\times$--$2.2\times$~\cite{krithivasan2022spikeattack}, and timestep-compressed methods reduce the latency of \emph{generating} such attacks for real-time use~\cite{kang2025tca}. Raptis and Stratigopoulos~\cite{raptis2026driving} move the attack from rate-coded images to native event-based inputs (NMNIST, SHD, and IBM DVS Gesture), where the input is a binary spike tensor and the perturbation budget becomes a count of flipped events. A per-sample attack raises synaptic operations (SynOps) by $1.5$--$2.6\times$ while preserving at least 98\% of predictions. A single universal binary mask, computed offline in about a minute and XORed with every input, raises SynOps by only $1.09$--$1.24\times$ and preserves 73--91\% of predictions, but needs no per-input computation. The authors present it as the first universal sponge attack on the forward pass of any network class, in contrast to universal attacks that exploit NMS or early-exit logic~\cite{shapira2023phantom,hong2021deepsloth}. Their energy figures are first-order Loihi-1 estimates (23.6~pJ per SynOp, ignoring membrane-update and static power), not on-chip measurements.

\paragraph{Persistent delivery.} The same sparsity loss can be delivered through training rather than through the input. \emph{Sponge poisoning}~\cite{cina2025energy} assumes the attacker controls a few model updates during training, for instance when training is outsourced or federated, and makes the trained model incur elevated energy and latency on \emph{any} test input without affecting accuracy; in the authors' ASIC-simulator evaluation, controlling 5--15\% of training updates raises energy consumption by up to about $1.57\times$ with accuracy kept within 3 percentage points, and repairing a poisoned model by fine-tuning is reported to be prohibitively costly for most users. Brachemi Meftah et al.~\cite{meftah2025energy} make the effect trigger-conditional: their energy backdoor, injected and then made stealthy in two training phases, raises energy on sparsity-based accelerators only for trigger-bearing inputs while preserving performance on clean inputs (ResNet-18 and MobileNet-V2 on CIFAR-10 and Tiny ImageNet). Sponge poisoning has been adapted to on-device DNNs~\cite{wang2023ondevice}, to mobile apps~\cite{paul2023mobile}, and to FPGA-based accelerators in differentially private secure federated learning, where EvoWeight raises honest users' power consumption without affecting model accuracy~\cite{akram2025evoweight}. \emph{SkipSponge}~\cite{te2025skipsponge} alters the biases of a pre-trained model directly rather than its training procedure. With access to less than 1\% of the training data it increases energy consumption by 1.4\% to 13.1\% depending on model and dataset, changes the victim's parameters and activation statistics far less than sponge poisoning, and survives parameter-perturbation and fine-pruning defenses unless they are adapted to target the altered biases.
A sensing-AI study quantifies battery-drain risk in IoT settings and proposes pruning as a defense, reducing energy overhead from $\sim$4$\times$ to $\sim$1.3$\times$~\cite{hasan2025sensing}.

\subsection{Conditional computation}
\label{sec:dynamic}

Input-adaptive networks promise lower \emph{average} cost; latency attacks revoke that promise by forcing worst-case control flow.

\noindent\textbf{Multi-exit and dynamic-depth networks.}
The earliest availability attack on input-adaptive networks is \emph{ILFO (Intermediate Output-Based Loss Function Optimization)}~\cite{haque2020ilfo}, which uses intermediate outputs as a proxy for the energy an input will consume and, with small input perturbations, recovers up to 100\% of the FLOPs that adaptive networks save. Building on ILFO, \emph{GradAuto}~\cite{pan2022gradauto} attacks both dynamic-depth \emph{and} dynamic-width networks by carefully steering the direction and magnitude of the gradient to find near-imperceptible perturbations that reactivate skipped computation, recovering on average $100\%$ of the FLOPs saved by the dynamic architecture; the related \emph{GradMDM}~\cite{pan2023gradmdm} further refines gradient direction and magnitude to minimize perturbation perceptibility while maximizing activated computation. \emph{DeepSloth}~\cite{hong2021deepsloth} performs slowdown attacks on multi-exit DNNs, reducing their early-exit efficacy (the area under the curve of samples classified versus fraction of full inference cost) by 90--100\%; in the white-box setting it also lowers accuracy by 75--99\%, so it is not prediction-preserving. In an IoT deployment where an MSDNet is partitioned between an edge device and the cloud, the attack forces 96\% and 99.97\% of inputs to the cloud part, raising average latency from 0.5~ms and 7.4~ms to about 11~ms on CIFAR-10 and Tiny ImageNet. The paper's headline ``1.5--5$\times$'' refers to this latency increase relative to the roughly 2~ms the attacker spends perturbing each input, not to the benign latency. The same confidence-gate pattern reappears in edge--cloud cascades, where the cost of a lowered confidence falls on a shared cloud queue rather than on the device (Section~\ref{sec:serving})~\cite{gu2026gatedrain}. Refined energy attacks combine energy maximization with confidence minimization and remain effective in grey-box settings~\cite{du2024energy}. The same principle extends to continuous-depth architectures: \emph{AntiNODE}~\cite{haque2023antinode} formulates the input--latency relationship for neural ODEs and crafts inputs that inflate the number of adaptive solver steps, producing inputs whose latency is 335\% higher than that of benign inputs. Whereas most of these attacks assume white-box gradients, \emph{EREBA}~\cite{haque2022ereba} is presented by its authors as the first \emph{black-box} energy-testing method for adaptive networks: it learns a proxy energy estimator and uses it to synthesize inputs that trigger worst-case computation without access to model internals, increasing energy consumption by up to $2{,}000\%$ relative to the benign baseline and establishing that the energy-robustness gap is exploitable even for opaque, query-only deployments. On resource-constrained edge deployments, \emph{DDAS}~\cite{ayyat2022dynamic} makes the target of the attack explicit at the system level: rather than degrading accuracy, it forces edge dynamic neural networks onto their longest execution path specifically to drain IoT-device resources---battery, latency, and memory---demonstrating that the operative harm is resource exhaustion on the device itself rather than a wrong prediction. The early-exit bottleneck can also be reached through training: sponge poisoning of multi-exit networks both prevents early exits and raises per-layer cost~\cite{huang2024multiexit}. In NLP, \emph{SlowBERT}~\cite{zhang2023slowbert} attacks multi-exit BERT, \emph{Dynamic Transformers Provide a False Sense of Efficiency}~\cite{chen2023dynamic} introduces a unified attack (SAME) that defeats early exits across BERT, RoBERTa, ALBERT, and DistilBERT with up to $5\times$ slowdown, and \emph{WAFFLE}~\cite{coalson2023waffle} systematically audits the slowdown robustness of three multi-exit language-model mechanisms in both white- and black-box settings, additionally finding universal slowdown triggers and showing that adversarial training fails while input sanitization (e.g., via a conversational model) is comparatively effective. Beyond attack crafting, \emph{DeepPerform}~\cite{chen2022deepperform} recasts the problem as software \emph{performance testing}: it trains a generator to efficiently synthesize test inputs that approximate a dynamic model's worst-case latency and energy, enabling systematic pre-deployment auditing of efficiency degradation. A study on multi-exit BERT traces the source of slowdown to confidence oscillation across layers, and additionally observes that ordinary text misclassification attacks (TextFooler, PWWS, BAE) cause slowdown as an unintended side effect~\cite{varma2024understanding}.

\noindent\textbf{Token pruning, skimming, and mixture-of-experts.}
Skimming language models drop unimportant tokens; \emph{No-Skim}~\cite{zhang2023noskim} crafts inputs that prevent dropping, raising FLOPs by $2$--$3\times$. A unified framework treats early exit, token-pruning vision transformers, and MoE routing as instances of one ``efficiency attack'' class, demonstrating $2$--$8\times$ overhead across all three~\cite{rathnasuriya2025ddls}. In the vision-transformer setting specifically, \emph{SlowFormer}~\cite{navaneet2024slowformer} trains a universal adversarial patch (as little as $2\%$ of the image area) that drives efficient ViTs such as A-ViT, ATS, and AdaViT toward their maximum compute; \emph{DeSparsify}~\cite{yehezkel2024desparsify} similarly targets the availability of token-sparsification mechanisms, crafting stealthy perturbations that force ATS/AdaViT/A-ViT to retain all tokens (raising GFLOPs by $44$--$100\%$) while preserving the clean prediction; and \emph{QuantAttack}~\cite{baras2025quantattack} exploits dynamic quantization to force worst-case high-precision execution, inflating inference time, memory, and energy. Mixture-of-experts routers, central to modern sparse LLMs, have also been targeted: \emph{Misrouter}~\cite{fei2026misrouter} causes routing collapse or saturation in a black-box, query-only setting, yielding $2$--$4\times$ latency increases, while \emph{RouteHijack}~\cite{xu2026routehijack}---though primarily a jailbreak attack---demonstrates that adversarial control of expert routing can additionally inflate computation by activating costly experts. \emph{RepetitionCurse}~\cite{huang2025repetitioncurse} shows the vulnerability is systemic rather than incidental: under expert parallelism, simple repetitive-token prompts force all tokens onto the same top-$k$ experts, creating device-level bottlenecks while others idle and raising end-to-end latency by $3.06\times$ on Mixtral-8x7B---turning the efficiency mechanism itself into a denial-of-service vector. \emph{SPLAT}~\cite{li2026splat} revisits latency attacks on dynamic networks with a black-box, stealthy formulation for multi-exit models.

\textbf{Section Summary.} Unlike perception attacks, sponge attacks demonstrate that latency vulnerabilities are not tied to a particular application or post-processing stage. Instead, they arise from architectural properties shared by many neural networks, including activation sparsity, conditional computation, and dynamic execution. Both bottlenecks can be reached by a test-time input or by poisoning and weight tampering; the latter embed the same computational effect persistently in model parameters rather than triggering it per input (Section~\ref{sec:persistent}). Together, these studies shift the focus from application-specific bottlenecks toward the computational mechanisms that underlie efficient neural inference.
\section{Generation-Length and Serving Bottlenecks}
\label{sec:seqgen}

\subsection{Sequence generation and decoder-length inflation}
\label{sec:seq2seq}

Autoregressive decoders give attackers a direct lever over the \emph{number of generated tokens}. Total cost generally increases with that count, while per-token cost may also vary with context length, KV-cache behavior, batching, and the attention implementation (consistent with the complexity discussion in Section~\ref{sec:unifying}).

\emph{NMTSloth}~\cite{chen2022nmtsloth} is presented as the first systematic study of efficiency robustness in neural machine translation. Observing that NMT cost is determined by output length, which depends on a pre-configured iteration limit and on when the EOS token appears, it searches for minimal character-, token-, or structure-level perturbations that delay EOS until that limit is reached. Perturbing a single character or token increases CPU latency by 85\% to 3{,}153\% and CPU energy consumption by 86\% to 3{,}052\% across three public NMT systems (76\% to 1{,}953\% and 68\% to 1{,}532\% on GPU), and on a mobile device the generated inputs drain more than $30\times$ the battery of normal inputs. The headroom comes from configuration: the authors found that 1{,}370 of 1{,}455 public HuggingFace NMT models set the maximum output length between 500 and 600 tokens, far above typical outputs.
\emph{LLMEffiChecker}~\cite{feng2024llmeffichecker} extends the same gradient-guided, one-character or one-token perturbation strategy to LLMs, increasing response latency by 325\% to 3{,}244\% and energy consumption by 344\% to 3{,}616\% on average. At the input-encoding level, \emph{Bad Characters}~\cite{boucher2022badchars} shows that imperceptible Unicode perturbations---invisible characters, homoglyphs, reorderings, and deletions---can be injected into text in a black-box setting; while primarily an integrity attack, the same encoding-level manipulation could plausibly provide a stealthy delivery channel for the efficiency-degradation objectives above against machine-translation and other text pipelines. \emph{NICGSlowDown}~\cite{chen2022nicgslowdown} brings the paradigm to neural image captioning, forcing decoders to emit far longer captions via human-unnoticeable image perturbations that increase captioning latency by up to 483.86\%, and contributes an efficiency-robustness benchmark reused by later work. \emph{TTSlow}~\cite{gao2024ttslow} demonstrates that the threat reaches audio synthesis: adversarial text inflates autoregressive text-to-speech decoding by up to $5\times$. The complementary direction, audio \emph{recognition}, is covered by \emph{SlothSpeech}~\cite{haque2023slothspeech}, presented as the first efficiency-robustness attack on automatic speech recognition: small perturbations to the input waveform prolong the decoder's output, inflating latency to up to $40\times$ relative to benign input, and showing that both ends of the speech pipeline---synthesis and recognition---share the autoregressive availability vulnerability. A property common to many of these attacks, and a recurring defensive blind spot, is that the input perturbation is small and the output can remain task-plausible, so monitors based only on a task metric may fail to detect them---though extremely long or repetitive outputs can still degrade practical utility.

\subsection{LLM and VLM output- and reasoning-length inflation}
\label{sec:llm}

The LLM/VLM era inherits the autoregressive lever and adds new ones: chain-of-thought reasoning and multimodal inputs. Reasoning models substantially enlarge an existing attack surface: input-dependent computation was already present in NMS, adaptive ODE solvers, early exits, tracking state, and autoregressive translation, but reasoning models expose a far wider controllable range of sequential inference work through variable reasoning length. The attacks below are grouped by the length they inflate; within each group, those delivered by a prompt precede those delivered through retrieved content, training data, or weights.

\noindent\textbf{Output-length inflation and inference-cost DoS.}
\emph{Engorgio}~\cite{dong2025engorgio} optimizes prompts that suppress the end-of-sequence token, making open-source LLMs produce roughly $2$--$13\times$ longer outputs in the white-box setting, typically reaching over 90\% of the configured output-length limit. In a limited-knowledge setting, prompts crafted on a proxy transfer partially to related models---from a base model to its fine-tuned derivatives they still produce roughly $1.5$--$2.5\times$ longer outputs---and crafting one prompt for LLaMA-7B takes about 165~s on a single H100, after which it can be reused.
AutoDoS~\cite{zhang2025crabs} organizes a DoS prompt as an attack tree whose node coverage it expands to remain effective without model access, and reports increasing service response latency by more than $250\times$. A recurring weakness of these EOS-delay methods is that control over the termination symbol erodes as the output grows; \emph{LoopLLM}~\cite{li2026loopllm} circumvents this by inducing \emph{repetitive generation}---a repetition-inducing prompt drives the model into a low-entropy decoding loop that reliably runs to the maximum output length (over $90\%$ of the cap versus $\sim$20\% for EOS-delay baselines), and a token-aligned ensemble optimization improves cross-model transferability by roughly $40\%$ to commercial models such as DeepSeek-V3 and Gemini~2.5~Flash. For attacks that run to the configured limit, the reported factor is set largely by the ratio of that limit to the benign output length, a deployment parameter rather than a property of the attack (Section~\ref{sec:translation}).

The same bottleneck is reachable through every other delivery channel. \emph{Retrieved content:} for retrieval-augmented generation, \emph{RA-ICA}~\cite{liu2026inference} poisons the external knowledge corpus rather than the prompt---injecting documents that are semantically retrievable yet induce abnormally long generations---raising token consumption by up to $13.1\times$ with over $90\%$ success while leaving answer integrity intact; this is a distinct and potentially practical delivery model in systems that ingest or retrieve attacker-publishable content, though its effectiveness depends on controlling retrieval rank, corpus ingestion, and trust filters. \emph{DrainCode}~\cite{wang2025draincode} specializes this corpus-poisoning vector to retrieval-augmented \emph{code} generation: gradient-guided mutation crafts syntactically valid but semantically inert triggers that suppress early EOS while a KL-divergence constraint preserves functional correctness, inflating output length $3$--$10\times$, latency by $85\%$, and energy by $49\%$ while retaining $95$--$99\%$ pass rate and evading both classifier- and perplexity-based filters. \emph{Training data:} \emph{Denial-of-Service Poisoning}~\cite{gao2024dospoisoning} shows that a triggered model can be made to emit endless output, never producing EOS, with 100\% success at under 1\% poisoning and survival through RLHF safety tuning. \emph{Weights:} \emph{BitHydra}~\cite{yan2025bithydra} formulates the search for cost-maximizing weight bits as a constrained binary integer program solved via ADMM and suppresses the end-of-sequence probability with as few as 1--4 bit flips across ten LLMs (1.5B--16B), so that generation continues until an external token, context, or execution limit is reached for any input query. All four exploit the same computational effect as an EOS-delaying prompt; they differ in persistence and in which defenses can see them (Section~\ref{sec:persistent}).

\noindent\textbf{Reasoning-model and ``infinite-thinking'' attacks.}
Inference-time compute scaling creates a new, high-leverage surface. \emph{ReasoningBomb}~\cite{liu2026reasoningbomb} induces pathologically long reasoning traces in large reasoning models, averaging 18{,}759 completion tokens (19{,}263 reasoning tokens across reasoning models) and $6$--$7\times$ more tokens than benign queries. Its reported $286.7\times$ input-to-output ratio compares output length with the attacker's own prompt length rather than with a benign reference, and so is not an amplification factor in the sense of Eq.~\eqref{eq:af}. \emph{ThinkTrap}~\cite{li2026thinktrap} exploits chain-of-thought mode in black-box APIs to induce near-infinite thinking loops, consuming $50$--$100\times$ more tokens than standard requests.
\emph{OverThink}~\cite{kumar2025overthink} targets applications that combine reasoning language models (RLMs) with external context: it injects decoy reasoning problems into public content that the RLM consumes at inference time, forcing it to spend substantially more reasoning tokens while still producing contextually correct answers, with up to $46\times$ overhead under large-scale context-agnostic attacks and up to $7.8\times$ under context-aware attacks. \emph{ExtendAttack}~\cite{zhu2025extendattack} hides the trigger inside benign queries by obfuscating characters into a poly-base ASCII encoding, compelling computationally intensive decoding sub-tasks that lengthen responses by over $2.5\times$ on o3 while preserving meaning and accuracy. Because these attacks preserve the reported answer quality and several bypass detection at high rates, they motivate budget enforcement by \emph{construction} rather than detection (Section~\ref{sec:defenses}).

\noindent\textbf{Multimodal verbose attacks.}
\emph{Verbose Images}~\cite{gao2024verbose} crafts adversarial images that make large VLMs generate long, repetitive responses, increasing mean output length by $7.87\times$ and $8.56\times$ on MS-COCO and ImageNet across four VLMs; per-model latency on MS-COCO rises by roughly $2\times$ (InstructBLIP) to $24\times$ (BLIP). \emph{VLMInferSlow}~\cite{wang2025vlminferslow} moves this objective to the black-box setting of VLMs served through inference APIs, using zero-order optimization, and, across four VLMs (Flamingo, BLIP, GIT, and Florence) on MS-COCO and ImageNet, reports imperceptible perturbations that lengthen generated sequences by up to 128.47\% and raise latency by up to 115.38\%.
\emph{Verbose-Text Induction}~\cite{luo2025verbosetext} co-optimizes adversarial images and prompts to exceed 10{,}000 output tokens from a single query, and \emph{LingoLoop}~\cite{fu2025lingoloop} traps multimodal models in self-referential generation loops that fill the context window faster than legitimate output, exhausting the configured context-length cap (the cap is reached, not bypassed). These attacks extend the output-inflation lever from unimodal captioning~\cite{chen2022nicgslowdown} to image- and prompt-conditioned VLMs. The same lever also reaches the temporal domain: \emph{VidDoS}~\cite{tang2026viddos} is described by its authors as the first universal energy-latency attack tailored for video-based LLMs, using a single instance-agnostic adversarial patch---trained via masked teacher forcing with a refusal penalty and early-termination suppression---that requires no inference-time gradients and therefore keeps pace with continuous video streams. It induces token expansion exceeding $205\times$ and latency inflation exceeding $15\times$ across three Video-LLMs. The threat has also reached \emph{embodied} multimodal autonomy through a backdoor: a natural-reflection backdoor attack on driving VLMs~\cite{liu2025vlmbackdoor} embeds reflection-pattern triggers in camera images that, when present, cause the VLM planner to generate abnormally long responses; the demonstrated latency effect is \emph{prolonged VLM generation} (increased output-token count and correspondingly higher inference latency under the trigger), which is a measured computational-cost increase rather than a mere task-denial response delay, and the same bottleneck as the input-borne attacks above reached through a different delivery channel.

\subsection{Serving-framework, pipeline, and agentic bottlenecks}
\label{sec:serving}
A further set of attacks moves below the model to the \emph{serving system} and the \emph{agentic loop}, where the bottleneck is a resource shared across requests and shared infrastructure amplifies a single adversary's leverage. \emph{Fill and Squeeze}~\cite{wang2026fillsqueeze} reports that, under its evaluated continuous-batching configurations, length inflation alone had less impact than direct scheduler- and KV-cache-targeted pressure, and accordingly attacks the scheduler directly---exhausting the global KV cache to induce head-of-line blocking and forcing repetitive preemption---for up to $20$--$280\times$ time-to-first-token slowdown in a black-box setting at lower cost than input-only methods. At the agent layer, \emph{Beyond Max Tokens}~\cite{zhou2026beyondmaxtokens} exploits the Model Context Protocol tool interface: text-only edits to tool metadata, optimized by Monte-Carlo tree search, steer agents into prolonged, multi-turn tool-calling chains exceeding 60{,}000 tokens across six evaluated LLMs, raising per-query cost by up to $658\times$ and energy consumption by $100$--$560\times$, pushing GPU KV-cache occupancy to $35$--$74\%$, while preserving task success and evading prompt filters and output trajectory monitors. \emph{From Shield to Target}~\cite{zhou2026shieldtarget} turns LLM \emph{guardrails} into the victim: crafted payloads trap the guardrail in extended reasoning loops for $13$--$63\times$ token amplification and up to $148\times$ latency amplification, and a single poisoned document can starve co-located agents sharing the guardrail. \emph{Mobius Injection}~\cite{liang2026mobius} generalizes this to agent-based DDoS (AbO-DDoS): by exploiting a ``semantic closure'' vulnerability, one textual injection induces sustained recursive agent execution, driving single-node call amplification up to $51\times$ and multi-node p95 latency up to $229\times$. The composition of models into a pipeline is itself a bottleneck. \emph{AESOP}~\cite{li2026aesop} targets dynamic inference pipelines in which upstream predictions determine which downstream components run and how much work they receive, so the cost of an input depends on the execution path it activates rather than on any single model. On identical inputs and budgets, path-aware targeting inflates FLOPs by $2{,}407\times$, whereas the strongest single-model baseline reaches $117\times$. Across five pipelines the attack reaches up to $419\times$ latency inflation in the white-box setting, and $58\times$ FLOPs and $17\times$ latency in the grey-box setting. In a production-realistic variant with batching, bounded buffering, and confidence-threshold defenses, the attack is redirected rather than neutralized: the pipeline must choose between throughput collapse (from 0.578 to 0.006 inputs per second) and discarding 96.7\% of its data. Confidence-gated offloading turns a routing decision into a shared-resource bottleneck. \emph{GateDrain}~\cite{gu2026gatedrain} attacks edge--cloud cascades that answer confident inputs on the device and offload uncertain ones to a larger cloud model: bounded perturbations ($L_\infty$ budget 8/255) push the calibrated top-1/top-2 margin below the offload threshold, raising the offload rate on the public EdgeBoost artifact from 46.3\% to 99.9\% ($2.16\times$ cloud demand) without sending any additional requests. Because escalated requests share a cloud queue, a near-capacity deployment crosses a queueing knee. In the authors' Poisson-arrival case study, benign users' p99 latency rises $10.46\times$; across four arrival processes the ratio ranges from $2.26\times$ to $12.8\times$ while the attacked tail stays at 1.6--1.7~s. The attack also lowers cloud accuracy on the perturbed inputs from 90.6\% to 82.4\%, and it works under transfer, decision-only (at most 31 routing observations per image), and universal settings. These developments echo the broader systematization of the agentic attack surface~\cite{dehghantanha2026sokagentic} and are now recognized at the standards level as ``unbounded consumption''~\cite{owasp2025unbounded}---reinforcing the need to treat availability, not just integrity, as a first-class security property of deployed LLM services.

\begin{table}[htbp]
  \caption{Representative latency/energy attacks beyond perception, grouped by bottleneck. Native metrics preserved.}
  \label{tab:general-attacks}
  \scriptsize
  \setlength{\tabcolsep}{3pt}
  \renewcommand{\arraystretch}{0.95}
  \begin{tabularx}{\linewidth}{L{2.75cm} L{2.35cm} L{1.35cm} >{\raggedright\arraybackslash}X}
    \toprule
    \textbf{Method} & \textbf{Bottleneck / target} & \textbf{Delivery} & \textbf{Key quantitative result} \\
    \midrule
    \multicolumn{4}{l}{\emph{Activation sparsity (Section~\ref{sec:sparsity})}} \\
    Sponge Examples~\cite{shumailov2021sponge} & Sparsity; token count & Digital & $\approx$$30\times$ on language models, less on vision; translator 1~ms$\to$6~s \\
    Sponge Poisoning~\cite{cina2025energy} & Sparsity (general) & Training & Energy ${\leq}1.57\times$ (ASIC sim.), 5--15\% of updates; accuracy within 3~pp \\
    SkipSponge~\cite{te2025skipsponge} & Sparsity (biases) & Weights & Energy $+1.4\%$ to $+13.1\%$ with ${<}1\%$ of training data \\
    SNN sponge~\cite{raptis2026driving} & Spikes (event SNN) & Digital & SynOps $1.5$--$2.6\times$ per-sample; $1.09$--$1.24\times$ universal mask \\
    \addlinespace
    \multicolumn{4}{l}{\emph{Conditional computation (Section~\ref{sec:dynamic})}} \\
    DeepSloth~\cite{hong2021deepsloth} & Early exit & Digital & Efficacy $-$90--100\%; edge--cloud latency 0.5 / 7.4$\to$${\approx}$11~ms \\
    GradAuto~\cite{pan2022gradauto} & Dynamic depth/width & Digital & Recovers $\sim$100\% of reduced FLOPs \\
    AntiNODE~\cite{haque2023antinode} & ODE solver steps & Digital & 335\% higher latency than benign inputs \\
    EREBA~\cite{haque2022ereba} & Adaptive networks & Digital (BB) & Energy up to $+2{,}000\%$ ($\approx$$21\times$) via proxy estimator \\
    SAME~\cite{chen2023dynamic} & Early exit (NLP) & Digital & Up to $5\times$ slowdown across BERT family \\
    No-Skim~\cite{zhang2023noskim} & Token skimming & Digital & $2$--$3\times$ FLOPs; output label unchanged \\
    Misrouter~\cite{fei2026misrouter} & MoE routing & Prompt (BB) & $2$--$4\times$ latency via routing collapse/saturation \\
    RepetitionCurse~\cite{huang2025repetitioncurse} & MoE expert parallel. & Prompt & $3.06\times$ latency on Mixtral-8x7B \\
    \addlinespace
    \multicolumn{4}{l}{\emph{Generation length (Sections~\ref{sec:seq2seq}--\ref{sec:llm})}} \\
    NMTSloth~\cite{chen2022nmtsloth} & Decoder (NMT) & Digital & CPU latency $+85$--$3{,}153\%$, energy $+86$--$3{,}052\%$ \\
    NICGSlowDown~\cite{chen2022nicgslowdown} & Decoder (caption) & Digital & Latency up to $+483.86\%$ ($\approx$$5.8\times$) \\
    Engorgio~\cite{dong2025engorgio} & Output length & Prompt & $\approx$$2$--$13\times$ longer (white-box); $\approx$$1.5$--$2.5\times$ transferred \\
    Crabs/AutoDoS~\cite{zhang2025crabs} & Output length & Prompt (BB) & ${>}250\times$ service response latency \\
    DoS Poisoning~\cite{gao2024dospoisoning} & No EOS & Training & 100\% at ${<}1\%$ poisoning; survives RLHF \\
    BitHydra~\cite{yan2025bithydra} & No EOS & Bit flips & 1--4 bit flips; runs to the external limit \\
    ReasoningBomb~\cite{liu2026reasoningbomb} & Reasoning length & Prompt (BB) & $6$--$7\times$ tokens vs.\ benign; $\approx$18.8k on average \\
    Verbose Images~\cite{gao2024verbose} & VLM output & Digital & Length $7.87\times$ / $8.56\times$; latency $\approx$$2$--$24\times$ per model \\
    VLMInferSlow~\cite{wang2025vlminferslow} & VLM output & Image (BB) & Length ${\leq}{+}128.47\%$, latency ${\leq}{+}115.38\%$ \\
    \addlinespace
    \multicolumn{4}{l}{\emph{Shared serving and orchestration resources (Section~\ref{sec:serving})}} \\
    Fill and Squeeze~\cite{wang2026fillsqueeze} & Scheduler / KV cache & Prompt (BB) & $20$--$280\times$ TTFT slowdown \\
    Beyond Max Tokens~\cite{zhou2026beyondmaxtokens} & MCP tool chains & Tool metadata & ${\leq}658\times$ cost; $100$--$560\times$ energy \\
    From Shield to Target~\cite{zhou2026shieldtarget} & LLM guardrails & Prompt, document & $13$--$63\times$ tokens; up to $148\times$ latency \\
    Mobius Injection~\cite{liang2026mobius} & Agent graph & Message & $51\times$ call amplification; $229\times$ p95 latency \\
    AESOP~\cite{li2026aesop} & Pipeline exec.\ path & Digital & ${\leq}2{,}407\times$ FLOPs, $419\times$ latency; grey-box $58\times$ / $17\times$ \\
    GateDrain~\cite{gu2026gatedrain} & Edge--cloud offload queue & Digital & Offload 46.3$\to$99.9\% ($2.16\times$ demand); benign p99 $2.26$--$12.8\times$ \\
    \bottomrule
  \end{tabularx}

  {\footnotesize\par\vspace{2pt}\emph{Notes.} ``What is measured'' varies by row---generated-token count, FLOPs, wall-clock or time-to-first-token (TTFT) latency, or energy---so the multiplicative factors are \emph{not} commensurable and each is relative to the source's own benign reference. Per-message or per-query amplifications state victim-side cost without a matched attacker-cost term (Section~\ref{sec:metrics}). Reported ``success'' retains each source's own definition (e.g., reaching a length threshold, exceeding a latency threshold, causing a deadline miss, or inducing a selected route); thresholds differ across studies and are not directly comparable. Multiplicative factors ($\times$) denote amplification relative to the benign baseline, whereas standalone percentages that name an attained outcome (e.g., a success or collision rate) are final values, not relative increases. For generation-length attacks that reach the configured limit, the reported factor is bounded by the ratio of that limit to the benign output length and should be read as a property of the deployment configuration as much as of the attack. \emph{Delivery} uses the channels of Table~\ref{tab:threats}; BB = black-box.\par}
\end{table}

\textbf{Section Summary.} Earlier attacks were constrained by the computational graph of a single model, whereas modern sequence models expose output length---and increasingly reasoning length---as attacker-controllable variables. As a result, computation itself becomes part of the attack surface, and it can be reached by a prompt, a retrieved document, a poisoned training set, or a handful of bit flips. Recent work further demonstrates that availability attacks are no longer confined to individual models. The attack surface is rapidly expanding toward shared serving infrastructure, tool chains, and agent orchestration, where computational amplification compounds across multiple services rather than within a single forward pass. This progression marks a broader shift from model-centric latency attacks toward system-level availability threats.
\section{Delivery Channels and Adjacent Phenomena}
\label{sec:hw}
Sections~\ref{sec:od}--\ref{sec:seqgen} placed each attack under the bottleneck it exploits. The delivery channel is an orthogonal property, and several consequences of an attack depend on it alone: how long the attack remains effective, what the defender must inspect to find it, which defenses can act on it, and what removing it costs. This section discusses those consequences with reference to attacks already described, and then delimits two phenomena that share observables with latency attacks but fall outside our taxonomy.

\subsection{Test-time delivery}
\label{sec:testtime}
Digital perturbations, physical patches and projections, prompts, retrieved content, and inter-agent or V2V messages act per input: the effect ends when the adversarial input is withdrawn, and input-side defenses---purification, filtering, admission control---have something to inspect. Physical delivery (Section~\ref{sec:od}) adds realizability constraints and a further layer of configuration dependence. Delivery can also move to the sensor pipeline: a universal perturbation computed once offline could be injected as a fixed overlay---a stuck-pixel pattern or fixed light source in an event camera's field of view, a stationary acoustic signal near a silicon cochlea, or a compromised sensor driver---so that no attacker computation is needed at inference time; these mechanisms are proposed rather than demonstrated~\cite{raptis2026driving}. Whether per-input delivery is practical depends on the attacker's per-input cost: the per-sample SNN sponge takes about four minutes per DVS Gesture input on an A100, which its authors judge incompatible with streaming inference~\cite{raptis2026driving}. Retrieval- and tool-metadata delivery (Section~\ref{sec:seqgen}) adds a dependence on ingestion and ranking that the attacker may not control, but removes the need to interact with the victim service directly.

\subsection{Persistent delivery: training data, backdoors, and weights}
\label{sec:persistent}
Training-data poisoning and backdoors embed the computational effects of Sections~\ref{sec:od}--\ref{sec:seqgen} in the model itself: sponge poisoning, its variants, and an energy backdoor (Section~\ref{sec:sparsity}), multi-exit poisoning (Section~\ref{sec:dynamic}), the NMS sponge backdoor (Section~\ref{sec:nms}), and no-EOS poisoning and the driving-VLM reflection backdoor (Section~\ref{sec:llm}). Weight tampering does the same through parameter edits or induced bit flips: SkipSponge and EvoWeight (Section~\ref{sec:sparsity}), and bit flips targeting NMS parameters (Section~\ref{sec:nms}) or end-of-sequence behavior (Section~\ref{sec:llm}). What changes is not the bottleneck but the defender's position. The effect applies to all inputs or to all trigger-bearing inputs; input-side defenses see nothing anomalous in the input; the effect may survive subsequent safety alignment~\cite{gao2024dospoisoning}; and removal may require costly fine-tuning or retraining~\cite{cina2025energy}. Hardware fault injection additionally requires physical or co-located access to the victim platform. Controls that act on the bottleneck rather than on the input remain effective: a no-EOS bit flip, for example, still stops at an externally enforced token limit~\cite{yan2025bithydra}. The delivery channel therefore determines which defenses can see an attack, while the bottleneck determines which controls bound it (Section~\ref{sec:defenses}).

\subsection{Adjacent: timing and serving side channels}
\label{sec:timing}
Data-dependent runtime can also be observed rather than inflated. We treat this confidentiality phenomenon here only to explain the dual security role of input-dependent runtime; it is excluded from the taxonomy, from Tables~\ref{tab:od-attacks} and~\ref{tab:general-attacks}, and from all attack counts. \emph{Side-channel} attacks invert the relationship between timing and computation: a timing-based black-box method crafts adversarial examples using only processing time, exploiting the link between activated nodes and latency, and evades gradient-masking countermeasures~\cite{nakai2021timing}, and data-dependent NMS runtime similarly leaks internal inference state~\cite{biton2023timing}. The same duality now appears in modern serving optimizations. Because speculative decoding accepts a data-dependent number of draft tokens per iteration, packet sizes and iteration boundaries can reveal the input-dependent number of accepted draft tokens, enabling query fingerprinting with over $90\%$ accuracy and datastore extraction across production vLLM deployments~\cite{wei2024speculation}; because the observable is packet-size and iteration-boundary leakage rather than elapsed time alone, this is a serving/traffic-analysis side channel in which timing may participate but is not the sole channel. This is the mirror image of a latency attack---rather than the adversary inflating the victim's computation, the victim's variable computation inflates what the adversary can observe---and it shows that an optimization introducing input-dependent timing can open both an availability lever and a confidentiality leak. We flag this explicitly as a timing-\emph{leakage} result (distinct from availability degradation; Section~\ref{sec:threat}): latency here is not an attack target but a side channel.

\subsection{Adjacent: denial of action in embodied systems}
\label{sec:vla}
Work appearing in 2025--2026 moves past the perception front-end to the \emph{Vision-Language-Action} (VLA) models that map multimodal observations directly to robot actions, where an availability failure is a physical failure: a robot that stalls or acts late can collide, drop a payload, or freeze mid-manipulation. Most attacks in this space are \emph{availability-adjacent} rather than latency attacks under the definition of Section~\ref{sec:metrics}: they deny, halt, or corrupt action without demonstrating computational-work or latency amplification, which is why we treat them here rather than under a bottleneck. \emph{FreezeVLA}~\cite{wang2025freezevla} formalizes \emph{action-freezing} attacks via min-max bi-level optimization: a single adversarial image ``freezes'' the VLA and makes it ignore subsequent instructions, attaining a $76.2\%$ average freeze rate across three VLA models and four benchmarks with strong cross-prompt transferability; the reported outcome is denial of action, not a measured increase in computation. \emph{Semantic-DoS}~\cite{steinberg2026semanticdos} turns the model's own safety alignment into the attack surface: injecting short (1--5 token) safety-plausible phrases into the robot's audio channel triggers protective halting without any jailbreak, reaching up to $98.3\%$ hard-stop success---service denial through triggered halting, again without intermediate-work amplification. A third group is best read as \emph{integrity attacks with hypothesized latency implications}: the Embedding Disruption Patch Attack (\emph{EDPA})~\cite{xu2025edpa} places a model-agnostic patch in the camera view that breaks visual--textual alignment and drives repeated incorrect actions to task failure, and \emph{MAVLA}~\cite{zhao2026mavla} breaks cross-modal alignment through a coordinated multimodal adversarial framework---in both cases the demonstrated outcome is incorrect or failed action rather than measured slowdown. \emph{ANNIE}~\cite{huang2025annie} provides a systematic, ISO-grounded study of embodied \emph{safety} violations, decomposing long-horizon goals into frame-level perturbations and exceeding $50\%$ success across critical, dangerous, and risky categories with validation on a physical robot.

These embodied attacks demonstrate that delayed or absent actions can create physical safety consequences. However, not all are computational latency attacks under our definition. Freeze or protective-halt attacks deny progress without necessarily increasing intermediate work, while cross-modal disruption attacks primarily target integrity. We therefore treat them as availability-adjacent evidence and reserve the latency-attack category for studies that explicitly measure or optimize computation, response time, energy, or missed deadlines. Among embodied platforms, only the reflection-backdoor attack on driving VLMs~\cite{liu2025vlmbackdoor} (Section~\ref{sec:llm}) meets that bar, and the appendix catalog labels each embodied entry accordingly. Broader treatments of the embodied threat surface are emerging in dedicated surveys~\cite{xing2025embodiedsurvey}.

\textbf{Section Summary.} The delivery channel does not change what an attack makes expensive, but it changes almost everything about how the attack is found and removed. Test-time delivery leaves evidence in the input; training-time and weight delivery leave none there, persist across inputs, and may survive fine-tuning, yet remain bounded by controls that act on the bottleneck itself. Timing side channels and embodied denial of action share observables with latency attacks but involve no work amplification. Securing availability therefore requires reasoning about the entire deployment stack---training pipeline, weights, hardware, and runtime---rather than the inference algorithm alone.
\section{Defenses}
\label{sec:defenses}

Defenses are nascent, often attack-family-specific, and pervaded by an efficiency--robustness trade-off. We organize them by \emph{control mechanism}---the point in the compute pathway each one actually bounds---rather than by the attack they were designed to counter. Table~\ref{tab:defenses} summarizes the landscape.

\subsection{Budget enforcement and compute caps}
\label{sec:budget}
The most direct defense against autoregressive inflation and reasoning DoS is to enforce a budget by construction. \emph{PD3F}~\cite{zhang2025pd3f} is a two-stage, pluggable framework that combines input-side dynamic request scheduling (via a ``Resource Index'') with output-side ``Adaptive End-Based Suppression'', which terminates excessive generation early, improving access capacity by up to 500\% under adversarial load across six models. \emph{Token-budget-aware pool routing}~\cite{chen2026tokenbudget} estimates each request's token budget and routes it to a right-sized pool, avoiding worst-case provisioning that wastes $4$--$8\times$ concurrency. \emph{Conformal Thinking}~\cite{wang2026conformal} reframes reasoning-budget allocation as a risk-control problem, stopping under a statistical criterion when the model is sufficiently confident or an instance appears unlikely to benefit from further computation. A complementary line of work, originally aimed at cost efficiency, may reduce exposure to reasoning-inflation attacks by encouraging or learning shorter chains of thought. At the \emph{prompt} level, \emph{token-budget-aware reasoning} (TALE)~\cite{han2024tale} injects a difficulty-estimated token budget into the prompt and dynamically adjusts the reasoning length per query, and \emph{Concise CoT} (CCoT)~\cite{renze2024ccot} simply instructs the model to reason concisely, cutting response length by roughly $48\%$ with negligible accuracy loss on non-mathematical tasks. At the \emph{parameter} level, \emph{CoT-Valve}~\cite{ma2025cotvalve} identifies a single tunable direction in weight space (via a lightweight LoRA branch) that elastically controls reasoning-chain length, compressing GSM8K chains from 741 to 225 tokens with a $<$0.2\% accuracy drop---a model-native cap that does not rely on a possibly-attacker-controlled prompt. These mechanisms differ in the \emph{strength} of the guarantee they provide, and conflating them would overstate the state of the art. Prompt-based methods such as TALE and Concise CoT \emph{encourage} shorter traces, but a prompt instruction is not a cap: the model may exceed the requested length, and a malicious client can remove or alter the instruction. Parameter-level methods such as CoT-Valve \emph{shift} the model toward shorter reasoning, but a learned length-control direction reduces observed length without certifying a maximum token count under adaptive attack. Runtime termination and token accounting, by contrast, \emph{enforce} hard limits. Only the last category directly guarantees a configured upper bound; the others may reduce expected cost but require adversarial evaluation. 

Table~\ref{tab:control-strength} summarizes this spectrum. Across all strengths a further caveat applies: a fixed budget imposed \emph{a priori} risks truncating genuinely hard reasoning, so difficulty-adaptive budgets are preferable to static ones. These methods reduce benign reasoning cost and suggest possible controls for the ``overthinking'' surface exploited by ReasoningBomb, ThinkTrap, OverThink, and ExtendAttack; their robustness against those adaptive attacks remains to be established unless directly evaluated in the cited work. The value of enforceable budgets is that they do not depend on first
classifying an input as malicious, which matters because the evaluated detectors
exhibit high bypass rates. ReasoningBomb reports bypass rates of 99.8\%, 98.7\%,
and 98.4\% for its input-, output-, and joint-detection baselines,
respectively~\cite{liu2026reasoningbomb}. Enforceable budgets therefore provide a
complementary guarantee, and this observation supports skepticism toward those
specific detectors rather than a universal ranking over all future detection
methods.

The perception literature illustrates the structural difference between
detection and enforcement, but it also shows that an enforcement control must sit
at the right point. A cap that performs no classification has no bypass rate: it
does not estimate whether an input is adversarial and needs no attack examples.
A detector's error rate is a surface the attacker optimizes against, so its
effectiveness degrades as adversaries adapt; a cap's cost is paid instead by
benign inputs that legitimately exceed it, and that cost is fixed by the
operator's choice of threshold rather than by the attacker's effort. This
asymmetry holds, however, only for the quantity the cap actually bounds. The
maximum-detection cap evaluated by Monteuuis et al.~\cite{monteuuis2025evade}
bounds the number of detections \emph{leaving} NMS; it leaves the number of
candidates \emph{entering} NMS, and hence NMS work, unchanged
(Section~\ref{sec:bounded}). Against an attacker whose objective is NMS latency,
such a cap is not evaded so much as irrelevant; against an attacker whose
objective is downstream tracker load, it caps one dimension of the tracker's
association work but not the persistent track pool~\cite{gutrackshield}. The controls that bound a stage's own work are admission caps applied
before the count-dependent operation: a pre-NMS top-$k$, a limit on the tracks
admitted to association, or a token budget enforced inside the decoding loop.

Two limits keep this from being a general preference for enforcement.
Enforcement is cheap where the budgeted quantity has a defensible operational
ceiling---a detector processing road scenes has a plausible maximum object
count---and expensive where it does not, since a reasoning cap must be set
without knowing how hard the next legitimate question will be. And a cap bounds
only the stage it guards, a composition weakness we take up below.

Among existing defenses, hard runtime enforcement is the most direct way to impose a configured upper bound on admitted computation rather than merely reducing average cost, and we expect it to grow in importance as models expose more input-dependent computation. It is not, however, a complete or cost-free solution, and presenting it as one would understate its risks. A hard cap can itself induce integrity and fairness failures: detection caps can discard real objects in crowded scenes; token or reasoning caps can truncate legitimate long or hard answers, disproportionately harming difficult-but-valid requests; admission control can deny service to benign heavy users; and a fixed execution budget can open a new timing or policy side channel (the cap boundary is itself observable). Composition is a further weakness: an adaptive attacker can craft inputs that respect every \emph{local} cap while still driving cumulative \emph{end-to-end} overload across stages, so per-stage budgets do not compose into an end-to-end guarantee without a global accounting layer. TrackShield's BoT-SORT-ReID experiment is a concrete instance: the association guard held, but re-identification cost upstream of it did not (Section~\ref{sec:bounded}). Budget enforcement is therefore best understood as a risk-controlled trade-off---bounding worst-case resource use at a tunable cost in clean-input utility and fairness---rather than a universal remedy, and its parameters should be set with the attacker-cost and clean-utility terms of Section~\ref{sec:metrics} in view.

\begin{table}[htbp]
  \caption{Strength of reasoning-cost controls, from hard runtime enforcement to soft prompt guidance. Only hard runtime enforcement guarantees a configured upper bound; the weaker categories reduce expected cost and require adversarial evaluation. This table refines the ``Budget enforcement'' rows of Table~\ref{tab:defenses}.}
  \label{tab:control-strength}
  \footnotesize
  \setlength{\tabcolsep}{5pt}
  \begin{tabularx}{\linewidth}{L{3.1cm} L{3.3cm} X}
    \toprule
    \textbf{Control type} & \textbf{Examples} & \textbf{Guarantee} \\
    \midrule
    Hard runtime enforcement & Token cap, termination, admission control~\cite{zhang2025pd3f,chen2026tokenbudget} & Configured upper bound \\
    \addlinespace
    Risk-controlled stopping & Conformal Thinking~\cite{wang2026conformal} & Statistical guarantee under stated assumptions \\
    \addlinespace
    Learned length control & CoT-Valve~\cite{ma2025cotvalve} & Empirical length reduction \\
    \addlinespace
    Prompt guidance & TALE~\cite{han2024tale}, Concise CoT~\cite{renze2024ccot} & Soft behavioral instruction \\
    \bottomrule
  \end{tabularx}
\end{table}

\subsection{Robust and adaptive training}
\emph{Hardware-aware adversarial training}~\cite{wang2025cantslow} builds background attention into the training pipeline using objectness loss as a proxy, restoring real-time throughput from 13~FPS to 43~FPS under attack on a Jetson Orin NX, with a better clean/robust trade-off. For multi-exit BERT models, \emph{expedited adversarial training}~\cite{varma2024understanding}—fine-tuning on synthetic samples created by randomly swapping or synonym-replacing $\sim$20\% of input words—mitigates SlowBERT-induced slowdown almost completely: attacked average exit layer drops from 11.49 back to 6.1–6.4, at or below the model's own benign exit layer, at a cost of roughly 8–18 percentage points of benign-accuracy degradation depending on the dataset.

\subsection{Input transformation and purification}
\label{sec:inputtransform}
For NMS-exploitation attacks, \emph{adaptive input resizing}~\cite{schoof2024beyond} reduces NMS execution time by 90.18\% on an attacked YOLOv5-large model---a defense evaluated directly against a latency objective. A complementary input-transformation lineage operates in the frequency domain. \emph{Feature Distillation}~\cite{liu2019featuredistill} first introduced DNN-oriented defensive quantization in the DCT domain of the JPEG pipeline to rectify adversarial perturbations while preserving benign accuracy; \emph{FenceBox}~\cite{qiu2020fencebox} generalized frequency- and spatial-domain purification into a unified augmentation platform; and \emph{DefQ}~\cite{qiu2023defq} specialized this defensive-quantization principle to the \emph{availability} setting, being the first dedicated defense against the early-exit slow-down attack: it statistically characterizes the frequency-domain signature that DeepSloth introduces and applies a DCT-domain quantization table---deployable inside the front-end camera's existing JPEG pipeline---that strips these perturbations and restores early-exit behavior on multi-exit DNNs (recovering early-exit rates to $85\%+$ across SDN-ResNet-56, VGG-16, and MobileNet). DCT-domain purification provides a complementary input-side intervention for early-exit slowdown attacks. Independent evidence for this family comes from EVADE~\cite{monteuuis2025evade}, which evaluates NMS latency attacks rather
than defenses, tested PDM-Pure---an off-the-shelf pixel-diffusion adversarial
purifier not designed for availability threats---against Daedalus, Overload,
Phantom Sponges, and Beyond PhantomSponges, and found that it returned the
post-NMS box count on YOLOv8n from 152--300 under attack to 5, the benign count
under the same defense, for all four, and to 4--5 on YOLOv8x and YOLOv12n. JPEG
compression, included as a ``natural'' defense arising from ordinary pipeline
processing rather than as a deliberate countermeasure, was nearly as effective on
YOLOv8n, where Overload was the sole partial exception at 55 retained boxes, but
less consistent across models (Daedalus retained 64 boxes on YOLOv12n). This is the first evaluation we identified of a general-purpose
purifier against a latency objective, and it supplies the transfer evidence the
purification literature does not report for itself. Two caveats bound the claim.
The metric is the post-NMS box count, an output count rather than a measure of
NMS work or latency, although, because purification acts upstream of the
detector, the reduction plausibly reflects fewer candidates entering NMS. And
neither purifier was evaluated against an adaptive attacker who optimizes through
the purification step. Where an adaptive attacker has been evaluated, the result
is negative: against a confidence-gated edge--cloud gate, BPDA- and EOT-adapted
attacks defeat all twelve training-free preprocessing baselines tested (JPEG,
bit-depth reduction, Gaussian noise, median filtering, random resize/pad, and
feature squeezing), restoring attacked offload to at least 99.8\%, contrary to
DefQ's conjecture that such transformations would survive adaptive
attacks~\cite{gu2026gatedrain,qiu2023defq}.

The remaining methods in this category were designed and evaluated primarily against \emph{integrity} threats (object-vanishing patches, off-manifold visual perturbations), and we include them as \emph{candidate} latency defenses whose effectiveness against a latency objective is largely \emph{untested} rather than as validated countermeasures. Temporal-consistency and DCT-based methods localize or remove physical patches~\cite{mu2024adav,ren2025dct}. Texture-feature denoising has been evaluated through mAP recovery and detection time~\cite{liang2024securing}, while latent-diffusion purification projects VLM inputs into a low-dimensional latent space to suppress off-manifold perturbations while preserving semantics~\cite{li2025ldp}. Because a purifier that removes an integrity perturbation may also remove a
latency-inducing one, these methods are plausible candidates whose availability
evidence is thin rather than absent: the PDM-Pure and JPEG results above show
that at least two general-purpose input transformations do restore the benign
post-NMS detection count against camera-based phantom-box attacks. Whether the same holds for
the remaining methods listed here, and whether any of them transfers to the
verbose-image and prompt-carried attacks of Section~\ref{sec:seqgen}
where the malicious content is semantic rather than pixel-level, remains untested.

\subsection{Bounded and constant-time execution}
\label{sec:bounded}
Three kinds of control should be distinguished, because they bound different quantities. \emph{Constant-time} implementations seek to remove input-dependent timing variation within a specified implementation boundary, and with it both the latency-inflation lever and the timing side channel; Biton et al.~\cite{biton2023timing} explore constant-time NMS to close the timing-leakage channel they identify.

\emph{Bounded-execution} mechanisms retain data dependence but cap its extent, and here the position of the cap relative to the expensive operation is decisive. An \emph{admission cap} limits the intermediate objects entering a stage whose cost grows with their number---candidates admitted to NMS, tracks admitted to association, tokens admitted to a decoder---and therefore bounds that stage's worst-case work; it is the bounded-execution instance of the \emph{work budget} abstraction (Section~\ref{sec:synthesis}). An \emph{emission cap} limits the results leaving a stage. It bounds the work of downstream stages that consume those results, but bounds the stage's own work only if the stage can terminate once the cap is reached. Greedy NMS admits such early termination in principle---after an $O(n\log n)$ sort, the suppression loop for $K$ kept boxes needs $O(Kn)$ rather than $O(n^2)$ comparisons---but implementations that compute the full keep set and truncate afterwards, including the torchvision kernel used by Ultralytics, do not benefit. Autoregressive decoding is the case in which the two coincide: each emitted token costs one decoder step, so a maximum-output-token limit is simultaneously an emission cap and a work cap for the decode phase, though not for prefill. A third control, a \emph{runtime guard}, aborts or skips work once elapsed time exceeds a limit (Ultralytics, for example, abandons the remaining images of a batch after a time limit); it bounds elapsed time directly but converts overrun into missing or partial results. None of these controls makes execution constant-time: even under a fixed nominal-operation bound, wall-clock time can vary through memory access, caching, contention, and hardware effects, and end-to-end runtime can remain data-dependent through preprocessing, memory allocation, batching, or downstream stages. The cost of any cap is a potential accuracy ceiling in dense scenes or on long legitimate outputs, making the bound a tunable safety--utility parameter.

The maximum-detection result of Monteuuis et al.~\cite{monteuuis2025evade} should be read in this light. Reducing the Ultralytics \texttt{max\_det} parameter from its default of 300 to 10 returned the post-NMS box count under four NMS latency attacks from 152--300 to 6--8, against a benign count of 5 under the same cap (6 without it). Because \texttt{max\_det} is applied after NMS completes, it is an emission cap, and the evaluated metric is right-censored at the cap value. The result therefore does not establish that NMS work or NMS latency was reduced, and at the default value of 300 the attacks were not neutralized. It does establish that a single emission cap, independent of the attacks' loss designs, perturbation models, and optimization procedures, and without the classification step whose failure modes Section~\ref{sec:budget} discusses, bounds the per-frame detection count handed to downstream tracking and planning. That bounds only one dimension of the tracker's association work: the persistent pool can still accumulate across frames, which is why SlowTrack survives a 500-detection cap (Section~\ref{sec:tracking}). The utility cost is real, and the evaluated setting was favorable to it: the value 10 was chosen to match the maximum object count in individual COCO images and would discard genuine detections in crowded traffic scenes, which is precisely the safety--utility trade-off noted above. Whether NMS work itself is bounded in these deployments rests on different evidence: the architectural ceiling on raw predictions and the pre-NMS cap, together with the authors' GPU microbenchmarks at that ceiling (Section~\ref{sec:nms}). This raises a question the perception literature has not answered. Reported NMS amplification factors should state the input resolution, confidence threshold, pre-NMS cap, and NMS implementation against which they were measured, and should ideally be measured against a defended default configuration rather than an undefended one.

\paragraph{Admission caps inside the tracker.} TrackShield~\cite{gutrackshield} applies these distinctions at the resource that SlowTrack actually exhausts. It runs inside the ByteTrack update and combines an always-on hard admission guard, which caps the tracks and detections admitted to each of the three association stages before the cost matrix is built, with two heuristic responses triggered by a robust EWMA/MAD anomaly score over tracker-internal counts: capping new track births, and pruning likely phantom tracks under a sustained vote. The guard's caps are calibrated at the 99th percentile of benign per-stage candidate counts and give a deterministic bound on admitted association dimensions that does not depend on the monitor. On a Jetson AGX Orin running YOLOX-S and ByteTrack, the full policy reduces SlowTrack's residual whole-pipeline latency inflation from 3.48 to 0.15, tracking-stage latency from 139.90~ms to 6.27~ms, and the deadline-miss rate at a 50~ms threshold from 100\% to 0.7\%, at 0.99--1.20~ms of per-frame overhead; across six attack-effective MOT17 sequences, tracking-stage latency falls by $22.9$--$26.9\times$. The guard alone reduces tracking-stage latency to 12.62~ms; pruning removes most of the remainder. The authors state three limits. The guard bounds only admitted association dimensions, not detector, NMS, re-identification, or persistent-state cost: in a BoT-SORT-ReID transfer experiment, re-identification of the raw detections upstream of the guard still took about 10.85~s of the 10.94~s attacked tracking-stage time. The defense contains workload but does not restore tracking accuracy under attack. And it is evaluated against transferred rather than adaptively re-optimized attacks. Its benign cost depends on the anomaly threshold: MOTA drops by 0.97 points at the conservative setting but collapses from 57.19 to 14.70 at the most aggressive one.

A simpler admission layer at the detector--tracker boundary suffices against the tracker flood behind NMS-free detectors, where NMS-centric mitigations have nothing to act on. TrackFlood's global cap $K$, together with a per-region grid cap $K_r$ that blocks spatially spread phantoms a global cap alone would admit, restores the end-to-end multiplier under its 8/255 universal attack to 1.01--1.03$\times$ on Jetson, but at a clear utility cost: on RT-DETR the cap that reaches 1.01$\times$ admits only 42\% of clean detections~\cite{gu2026trackflood}. Like any per-frame cap, it bounds the admission rate, not the persistent pool, which can still grow to about $K$ times the track buffer; utility is measured only as detection retention, and the layer is evaluated only against the non-adaptive attack. The authors therefore position it as complementary to admission inside the tracker.

\subsection{Architectural sparsity and efficiency reduction}
Shrinking per-step compute reduces an attacker's headroom. \emph{Model pruning}~\cite{hasan2025sensing} improves resilience to sponge poisoning (energy overhead $\sim$4$\times$ $\to$ $\sim$1.3$\times$) at an accuracy cost. \emph{BlindSight}~\cite{srikrishnan2025blindsight} exploits attention sparsity to cut VLM FLOPs by 32--41\% with $-2$\% to $+2$\% accuracy; it is a general efficiency mechanism, not a demonstrated latency-robustness defense. These reduce absolute latency impact but do not \emph{bound} generation length and have not been evaluated against adaptive latency attacks, so they complement rather than replace budget enforcement.

\subsection{Runtime monitoring and adaptive inference}
Activation-sparsity monitoring offers a lightweight, architecture-agnostic anomaly signal for sponge inputs~\cite{muller2024uniform}. For early-exit systems, online scheduling treats the exit decision as a contextual bandit~\cite{ju2021dee} (improving overall performance by up to 98.1\% over the best static benchmark), and quorum-based ensemble stopping (SQUAD)~\cite{gambella2026squad} reduces latency by up to 70.6\% through quorum-based stopping and may reduce dependence on any single confidence score; its robustness against adaptive slowdown attacks has not been established here.

\subsection{System-level and serving control}
At the serving layer, request scheduling, admission control, and pool right-sizing~\cite{chen2026tokenbudget,zhang2025pd3f} can reduce, isolate, or bound the effect of operational bottlenecks becoming attack amplifiers, depending on the admission, scheduling, and resource-accounting guarantees they provide. Bounded Escalation~\cite{gu2026gatedrain} contains GateDrain at the offload boundary without detecting the perturbation: per-source token buckets, capacity protected for authenticated clients, non-preemptive priority for the trusted class, and an optional global bucket that bounds aggregate untrusted admissions regardless of identity. Its evaluation makes the policy trade-off explicit. Per-source budgets and priority isolate authenticated clients but do not bound aggregate untrusted work under a Sybil flood. Adding the global bucket cuts aggregate benign p99 from about 1.7~s to 85~ms, but raises rejection of legitimate unauthenticated offloads from 2.8\% to 52.1\% and lowers modeled all-benign accuracy from 89.7\% to 85.5\% through edge fallback. The following methods are included to identify possible systems primitives or boundary cases, not as validated latency-attack defenses. \emph{Certifiers}---formal robustness verifiers with abstain/fallback behavior---can themselves create an availability surface, since an adversary can systematically trigger abstention to deny service~\cite{lorenz2021certifiers}; a safety fallback can become a target. Beyond admission control, hardware-level system defenses target the serving boundary directly: \emph{Garrison}~\cite{garrison2024} is a GPU-accelerated inference system for adversarial ensemble defense that uses Multi-Instance GPUs with reinforcement-learning-based scheduling, improving adversarial robustness by up to 24.5\% while accelerating ensemble inference by $6.6\times$; its evaluation targets integrity-oriented ensembles, so accelerating the defense---rather than demonstrated protection against latency amplification---is the relevant contribution here. \emph{PSML}~\cite{psml2023} secures model-serving systems against model \emph{extraction} via fingerprinting and noise-based defenses; it is adjacent serving security rather than a latency-attack defense, and we list it only to delineate the serving-security boundary. The \emph{time-traveling} defense~\cite{etim2024timetraveling} is an unusual input-side variant for the perception domain: it queries historical Street View imagery and uses majority-vote inference to resist adversarial manipulation of traffic signs, achieving 100\% defense effectiveness in its integrity evaluation; it has not been evaluated against a latency objective, and we include it as evidence that temporal redundancy can substitute for per-frame robustness in geographically static scenes.

\begin{table}[htbp]
  \caption{Defense families against latency and energy attacks, organized by
control mechanism and evidence class. The evidence class identifies the
quantity or security property evaluated; it does not imply robustness against
adaptive latency attacks.}
  \label{tab:defenses}
  \scriptsize
  \setlength{\tabcolsep}{2pt}
  \renewcommand{\arraystretch}{0.95}
  \begin{tabularx}{\linewidth}{L{2.55cm} L{2.45cm} L{1.7cm} >{\raggedright\arraybackslash}X}
    \toprule
    \textbf{Family} & \textbf{Representative method} & \textbf{Evidence} & \textbf{Reported result} \\
    \midrule
    Budget enforcement & PD3F~\cite{zhang2025pd3f} & Direct capacity & Up to $500\%$ access capacity under adversarial load \\
    Budget enforcement & Token-budget routing~\cite{chen2026tokenbudget} & Budget-control & Avoids $4$--$8\times$ worst-case over-provisioning (also serving control) \\
    Budget enforcement & Conformal Thinking~\cite{wang2026conformal} & Budget-control & Risk-controlled reasoning stop (not adaptive-latency) \\
    \addlinespace
    Robust training & Can't Slow Me Down~\cite{wang2025cantslow} & Direct latency & $13\rightarrow43$ FPS under attack \\
    Robust training & Expedited adversarial training~\cite{varma2024understanding} & Direct compute/work & Exit layer 11.49$\to$6.1–6.4 (benign level); 8--18~pp accuracy cost \\
    \addlinespace
    Input transform. & Adaptive resizing~\cite{schoof2024beyond} & Direct latency & $90.18\%$ reduction in NMS time \\
    Input transform. & DefQ~\cite{qiu2023defq} & Direct compute/work & Early-exit rate restored to $85\%+$ \\
    Input transform. & PDM-Pure~\cite{monteuuis2025evade} & Output count & Benign count (4--5) for four NMS attacks, three models \\
    Input transform. & JPEG compression~\cite{monteuuis2025evade} & Output count & Benign count on YOLOv8n for 3 of 4 attacks (Overload 55); weaker on YOLOv12n \\
    Input transform. & Feature Distillation, ADAV, DCT purification, LDP, Time-traveling~\cite{liu2019featuredistill,mu2024adav,ren2025dct,li2025ldp,etim2024timetraveling} & Integrity only & Candidates: DCT-domain, patch localization, frequency-domain, latent purification, temporal redundancy \\
    Input transform. & Twelve preprocessing baselines~\cite{gu2026gatedrain} & Direct compute/work & Defeated by adaptive (BPDA/EOT) attacks; offload restored to ${\geq}99.8\%$ \\
    \addlinespace
    Constant-time exec. & Constant-time NMS~\cite{biton2023timing} & Timing channel & Removes input-dependent NMS timing within the analyzed boundary \\
    Bounded (emission cap) & Detection cap (\texttt{max\_det})~\cite{monteuuis2025evade} & Output count & 152--300$\to$6--8 boxes (benign 5) at \texttt{max\_det}${=}10$, none at 300; bounds per-frame detections, not NMS work or tracker pool \\
    Bounded (admission cap) & TrackShield~\cite{gutrackshield} & Direct latency & Residual inflation 3.48$\to$0.15; tracking 139.90$\to$6.27~ms; misses 100$\to$0.7\% \\
    Bounded (admission cap) & Admission layer~\cite{gu2026trackflood} & Direct latency & End-to-end $1.08$--$1.17\times\to1.01$--$1.03\times$ at 42--67\% clean-detection retention \\
    \addlinespace
    Arch. sparsity & Pruning~\cite{hasan2025sensing} & Direct energy & Energy overhead ${\sim}4\times\to{\sim}1.3\times$ (accuracy cost) \\
    Arch. sparsity & BlindSight~\cite{srikrishnan2025blindsight} & General efficiency & $32$--$41\%$ FLOP reduction \\
    \addlinespace
    Runtime monitoring & Sparsity monitor~\cite{muller2024uniform} & Direct compute/work & Anomaly signal for sponge inputs \\
    Runtime monitoring & DEE~\cite{ju2021dee} & General efficiency & Online exit scheduling \\
    Runtime monitoring & SQUAD~\cite{gambella2026squad} & General efficiency & Up to $70.6\%$ latency reduction \\
    \addlinespace
    Serving control & Bounded Escalation~\cite{gu2026gatedrain} & Direct latency & Sybil-flood benign p99 ${\approx}1.7$~s$\to$85~ms; unauthenticated rejection 2.8$\to$52.1\% \\
    Serving control & Garrison~\cite{garrison2024} & Adjacent security & Up to $+24.5\%$ robustness; $6.6\times$ inference acceleration \\
    Serving control & PSML~\cite{psml2023} & Adjacent security & Model-extraction defense \\
    Serving control & Certifier caveat~\cite{lorenz2021certifiers} & Conceptual & Abstain/fallback can create an availability surface \\
    \bottomrule
  \end{tabularx}

  \vspace{2pt}
  \parbox{\linewidth}{\footnotesize
  \textit{Notes.} Reported improvements use each source paper's own baseline and are not directly comparable across rows. The \texttt{max\_det}, PDM-Pure, and JPEG rows report a third-party evaluation \cite{monteuuis2025evade} rather than results claimed by each method's own authors; the evaluated quantity is the post-NMS box count, which is right-censored at \texttt{max\_det} and measures neither NMS work nor latency. %
  }
\end{table}

To distinguish demonstrated defenses from plausible transfers,
Table~\ref{tab:defenses} assigns each method an explicit evidence class.
\emph{Direct latency}, \emph{direct energy}, \emph{direct compute/work}, and
\emph{direct capacity} indicate evaluation against measured latency or
slowdown, energy consumption, computational work, or serving capacity or
goodput, respectively. These direct classes identify the availability-related
quantity evaluated; they do not imply robustness against adaptive latency
attacks. \emph{Budget-control} indicates evaluation of compute-budget control
without necessarily demonstrating protection against an adaptive latency
attack. \emph{Output count} indicates evaluation of the number of results
leaving a stage; it bounds the work of downstream stages but is not a measure of
the stage's own work. \emph{Timing channel} indicates evaluation against timing leakage or
timing variation with an availability implication. \emph{Integrity only}
identifies candidate latency defenses evaluated only against integrity
attacks; \emph{general efficiency} identifies efficiency improvements not
evaluated against latency attacks; \emph{conceptual} denotes a proposed
control without evaluation against an adaptive latency attack; and
\emph{adjacent security} denotes a serving or systems-security mechanism not
evaluated against latency amplification.

\textbf{Section Summary.} The reviewed controls span several evidence classes: demonstrated latency-attack defenses, candidate transfers from integrity security, general efficiency mechanisms, and conceptual or adjacent system controls (Table~\ref{tab:defenses}). None provides comprehensive protection, and many were never evaluated against a latency objective at all, and none against an adaptive latency attacker. Among the demonstrated defenses, most reduce average computational cost or detect anomalous behavior without explicitly bounding worst-case execution. In contrast, mechanisms that directly constrain computation---such as token budgets, admission caps, and constant-time implementations---address the underlying cause of latency inflation rather than its symptoms, provided they bound the work entering the expensive stage rather than only the results leaving it. This observation motivates the broader \textit{work-budget abstraction} developed in the next section.

\section{Cross-Domain Synthesis and Evaluation}
\label{sec:synthesis}

Viewing the perception and LLM/VLM-serving literatures side by side yields three observations that neither community arrives at independently.

\paragraph{Most instantiated perception defenses do not transfer directly to serving, and vice versa.} The non-transfer is at the level of concrete mechanisms, not of the abstract principles (budgets, admission control, bounded execution, monitoring) they instantiate. Hardware-aware adversarial training~\cite{wang2025cantslow}, input purification~\cite{li2025ldp,ren2025dct}, and the time-traveling majority vote~\cite{etim2024timetraveling} all assume the attack arrives through a \emph{sensor input} as a pixel-level perturbation; they have no lever on prompt-injection DoS or output-length inflation, where the malicious content is carried in the semantics of a query rather than in pixel noise. Conversely, PD3F~\cite{zhang2025pd3f}, token-budget routing~\cite{chen2026tokenbudget}, and conformal stopping~\cite{wang2026conformal} assume a serving stack with request queues and preemptible generation, which has no analog in a camera-to-detector pipeline. This mechanism-level non-transfer is the central practical obstacle to securing a heterogeneous autonomy stack, even where a common design principle applies.

\paragraph{Intermediate-work amplification recurs, and suggests a shared class of remedies.}
As argued in Section~\ref{sec:unifying}, most of the attack classes we survey derive their leverage from forcing a downstream stage to receive more intermediate objects than it was provisioned to process. Where that shape holds, a natural cross-domain abstraction is a \emph{work budget}: an explicit limit on the number of intermediate objects admitted to a stage whose cost rises with that count. We are careful not to overclaim its reach. It fits the count-driven attacks well (NMS overload, token/step inflation, expert routing), and it applies equally when those attacks are delivered through poisoning or weight tampering, which exploit the same bottlenecks---a no-EOS bit flip, for instance, runs until an external token limit stops it~\cite{yan2025bithydra}. It does not address timing side channels or denial-of-action attacks, which involve no work amplification (Section~\ref{sec:hw}).

Where the cap sits matters. A work budget in our sense is an \emph{admission
cap}: it limits the intermediate objects admitted to a stage before that stage
performs its count-dependent work. A cap on the results a stage emits bounds the
work of the next stage but not of the stage itself, unless the stage terminates
early once the cap is reached (Section~\ref{sec:bounded}). The two coincide for
autoregressive decoding, where each emitted token is one decoding step, but not
for NMS, where the candidates entering the kernel and the detections leaving it
are different sets. The most frequently cited piece of evidence for count caps in
perception---the maximum-detection cap that reduced post-NMS box counts under
four published NMS attacks~\cite{monteuuis2025evade}---is an emission cap, so it
bears on the stage downstream of NMS rather than on NMS itself, and even there
bounds only the per-frame detection dimension (Section~\ref{sec:tracking}). What that result does show is the property that makes budgets
attractive: one cap bounded the count delivered downstream under four attacks
with different loss designs, perturbation models, and optimization procedures,
without distinguishing among them. It also delimits the claim, since the
evaluated value was chosen to match the maximum object count in individual COCO
images and would discard genuine detections in dense traffic; what carries
across attacks and domains is the control point, not the setting. The first
direct evidence that an admission cap bounds the work of the stage it guards,
measured against a published latency attack, comes from the tracker: TrackShield's
per-stage association caps alone cut SlowTrack's tracking-stage latency from
139.90~ms to 12.62~ms, and its full policy to 6.27~ms~\cite{gutrackshield}
(Section~\ref{sec:bounded}). The corresponding measurement for NMS itself---NMS
time under attack as a function of a pre-NMS top-$k$---and any such evidence on the
serving side are still missing.

\begin{figure*}[htbp]
  \centering
  \includegraphics[width=\textwidth]{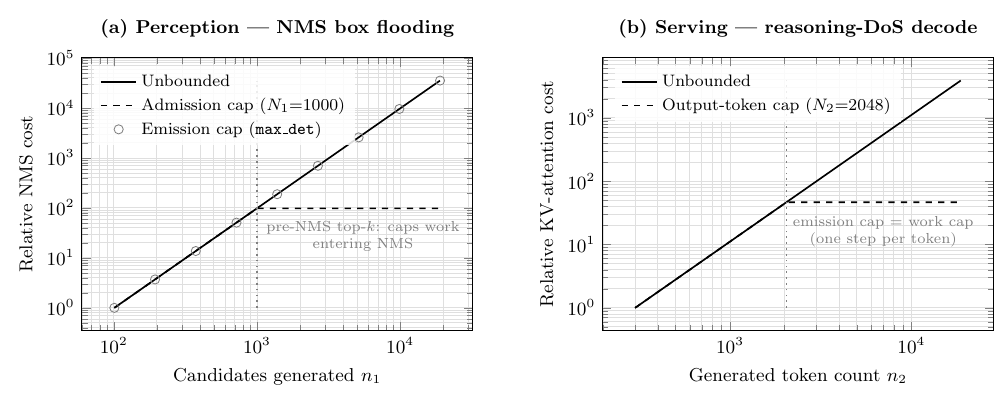}  %
  \caption{Conceptual illustration of the work-budget abstraction across perception and serving systems. Limiting the objects admitted to a stage bounds that stage's count-dependent work, assuming the cap is applied before the expensive operation; it does not by itself bound upstream candidate generation, pre-cap sorting/selection, memory allocation and copying, other uncapped stages, or cumulative work across repeated tool/agent calls.
  The left panel illustrates an \emph{admission} cap on candidates entering NMS (a pre-NMS top-$k$) applied against an NMS-overload workload~\cite{xia2026groundswell}; the circled curve shows that an \emph{emission} cap on detections leaving NMS, such as \texttt{max\_det}, leaves NMS cost on the unbounded curve and bounds only the downstream stage. The right panel illustrates a generation budget applied against a reasoning-inflation workload~\cite{liu2026reasoningbomb} and distinguishes attack-induced continuation \emph{pressure} from tokens \emph{actually emitted} before termination; here the output-token cap is both an emission cap and a work cap, because each emitted token costs one decoding step. These citations identify motivating attacks, not the depicted cap implementations. Both panels use log--log axes; the capped curve (dashed) plateaus at the vertical cap line while the unbounded curve (solid) keeps rising. The figure is conceptual and intended to illustrate the common design principle underlying several existing defenses.}
  \label{fig:workbudget}
\end{figure*}

We view the work budget as an abstract design principle rather than a concrete
algorithm, and existing mechanisms can be interpreted through it---though they
are unevenly evidenced. The post-NMS detection cap has third-party evidence for
bounding the count delivered to downstream stages~\cite{monteuuis2025evade},
whereas token budgets~\cite{chen2026tokenbudget} and risk-controlled reasoning
stops~\cite{wang2026conformal} rest on their own authors' evaluations against
non-adaptive or non-adversarial baselines. The
mechanisms also differ in guarantee strength: hard caps directly bound admitted work within a declared boundary, whereas routing and statistical stopping reduce or control
expected cost without establishing an adversarial worst-case bound. Viewed this way, a latency defense is a property of how the inference pipeline allocates computation rather than of any single model. This is not a solved problem: every such cap trades worst-case cost against
clean-input utility and fairness, and a determined attacker can stay just under
each local cap while still driving cumulative end-to-end overload.
Figure~\ref{fig:workbudget} illustrates the work-budget abstraction with
perception and serving examples.

\paragraph{No evaluated end-to-end defense.} In this survey, \emph{end-to-end evaluation} means an adversarial workload propagated through at least two heterogeneous computational stages, with both resource consumption and a terminal system outcome measured (for autonomous systems, e.g., deadline misses, collision rate, control degradation, or mission completion; for serving, e.g., goodput, victim latency, admission failure, or completed useful requests). Within the corpus we reviewed, each documented defense (Table~\ref{tab:defenses}) protects one layer and leaves others exposed; we did not identify a defense shown to bound an adversarial workload end-to-end in this sense. The closest evidence comes from an attack paper: AESOP evaluates batching, bounded buffering, and confidence-threshold defenses in a production-realistic multi-model pipeline and finds that they redirect the attack into throughput collapse or data loss rather than neutralizing it~\cite{li2026aesop}. Among defenses, TrackShield comes closest: it is evaluated on whole-pipeline latency and deadline misses under SlowTrack, but its bound covers only the association stage, and its own re-identification experiment shows cost upstream of the guard escaping it~\cite{gutrackshield}. A VLM-equipped autonomous vehicle (AV) faces simultaneous latency threats at the camera-to-detector boundary (NMS overload, physical projectors), the tracking layer (SlowTrack-style temporal attacks), the LiDAR and cooperative-fusion stages, and the VLM planner's decoder (verbose images, reflection backdoors). Composing defenses across these layers is technically possible but has not been evaluated end-to-end.

\subsection{From model-level cost to system-level failure}
\label{sec:translation}
An amplification factor is a property of a component; an availability failure is a property of a system. The surveyed evidence shows that the first does not determine the second, and it identifies the conditions under which it does. Consider, as a first-order model, a pipeline with benign end-to-end latency $T$ in which an attack amplifies one stage, whose benign share of $T$ is $f$, by a factor $A$. If stages execute sequentially and nothing else changes, the end-to-end factor is
\begin{equation}
A_{\mathrm{e2e}} = 1 + f\,(A-1),
\label{eq:e2e}
\end{equation}
and a relative deadline $D$ is missed only if $A_{\mathrm{e2e}}\,T > D$, that is, only if
\begin{equation}
A > 1 + \frac{D/T - 1}{f}.
\label{eq:threshold}
\end{equation}
The threshold on the right depends on the stage's critical-path share $f$ and the system's slack ratio $D/T$ (Section~\ref{sec:metrics}), neither of which is a model property. Six conditions follow, each supported by the surveyed work and summarized in Table~\ref{tab:translation}.

\paragraph{(C1) Critical-path share.} Amplification is diluted by unamplified stages. SlowTrack reports an average $453.8\times$ slowdown of the tracking stage but $28.4\times$ for the whole camera perception pipeline~\cite{ma2024slowtrack}; under Eq.~\eqref{eq:e2e} these averages are consistent with tracking accounting for only a few percent of benign pipeline latency. Groundswell raises NMS candidates from 101 to about 19{,}000 ($\approx$$188\times$) while end-to-end latency rises from 15.2 to 41.4~ms ($\approx$$2.7\times$)~\cite{xia2026groundswell}. Stages that run concurrently or off the critical path dilute the effect further. Behind NMS-free detectors, substantial tracker slowdowns become ByteTrack end-to-end slowdowns of only $1.0$--$1.3\times$ because the fixed-cost detector dominates per-frame time; conversely, BoT-SORT's detection-independent motion-compensation stage compresses the relative multiplier while producing the highest absolute p99 latency~\cite{gu2026trackflood}. Nor is the share fixed by the victim: in a multi-model pipeline the attacker can choose which execution path to amplify, and AESOP reports $2{,}407\times$ FLOPs inflation by targeting the path, versus $117\times$ by targeting the most vulnerable single model~\cite{li2026aesop}.

\paragraph{(C2) Headroom against slack.} The amplified count has a ceiling---the raw prediction count of a detection head, a configured output-length limit, a context window---and the attack cannot exceed it. If the amplification reachable at that ceiling falls below the threshold in Eq.~\eqref{eq:threshold}, no deadline is missed however well the attack is optimized; this is the substance of the conclusion in~\cite{monteuuis2025evade} that GPU NMS cannot breach 100--500~ms budgets at reachable proposal counts; on CPU, the model's own proposal cap decides the outcome, with YOLOv5's reaching a 2~s threshold and YOLOv8's staying under 500~ms. TrackFlood separates the two regimes explicitly: at 8/255 its tracker overload is measurable but misses no 33~ms deadline, whereas only a 32/255 stress test causes sustained misses~\cite{gu2026trackflood}. For generation-length attacks that reach the configured limit, the reported factor approaches the ratio of that limit to the benign output length~\cite{dong2025engorgio,li2026loopllm}, a deployment parameter rather than an attack property; NMTSloth found 1{,}370 of 1{,}455 public NMT models configured with a 500--600-token limit, far above typical outputs~\cite{chen2022nmtsloth}.

\paragraph{(C3) Clipping.} Where a cap or timeout is already in place, reaching it converts delay into truncation, dropped frames, or refused requests. The failure then shifts from availability to utility or integrity and must be measured as such (Section~\ref{sec:bounded}). AESOP's system-level evaluation is the clearest instance: batching, bounded buffering, and confidence thresholds forced the pipeline to choose between throughput collapse and discarding 96.7\% of its data~\cite{li2026aesop}.

\paragraph{(C4) Accumulation.} Stateful stages turn a transient input into a sustained effect: false tracks enlarge future association matrices---SlowTrack inflates ByteTrack's mean persistent pool from 20.5 to 1{,}677.7 tracks~\cite{gutrackshield}---, and exhausted KV-cache capacity produces head-of-line blocking and repeated preemption~\cite{wang2026fillsqueeze}. Here the relevant measures are backlog, consecutive misses, and recovery time after the attack stops (Section~\ref{sec:metrics}), not per-request amplification.

\paragraph{(C5) Sharing.} When resources are shared, the victim is not the requester. Fill and Squeeze found that, in its evaluated continuous-batching configurations, output-length inflation alone mattered less than direct pressure on the scheduler and KV cache~\cite{wang2026fillsqueeze}; guardrail and agentic attacks propagate cost to co-located agents and across invocation graphs~\cite{zhou2026shieldtarget,liang2026mobius,zhou2026corba}. The appropriate measures are benign-victim latency and goodput, not the attacker's own request latency. Sharing can also magnify a modest amplification: GateDrain raises cloud demand only $2.16\times$, but near capacity this moves the service across a queueing knee and raises benign p99 by $2.26$--$12.8\times$ depending on the arrival process~\cite{gu2026gatedrain}.

\paragraph{(C6) Consequence mapping.} Whether a miss matters depends on what the system does with a late result. SlowTrack's closed-loop evaluation reports a crash rate of about 95\% under its attack, versus about 30\% for prior attacks~\cite{ma2024slowtrack}; a system whose fallback policy triggers a safe stop would convert a comparable delay into a loss of availability rather than a collision. Detector-level success also need not survive the tracker: over 98\% detection-attack success may be needed to alter tracking outcomes~\cite{jia2020fooling}.

\begin{table}[htbp]
  \caption{Conditions under which a model-level cost increase becomes a system-level availability failure (Section~\ref{sec:translation}). None is a property of the attacked model alone.}
  \label{tab:translation}
  \footnotesize
  \setlength{\tabcolsep}{3pt}
  \begin{tabularx}{\linewidth}{L{2.3cm} L{3.3cm} >{\raggedright\arraybackslash}X L{2.8cm}}
    \toprule
    \textbf{Condition} & \textbf{What decides it} & \textbf{Evidence in the surveyed work} & \textbf{What to report} \\
    \midrule
    (C1) Critical-path share & Benign share $f$ of the amplified stage; concurrency & Tracking stage $453.8\times$ vs.\ pipeline $28.4\times$~\cite{ma2024slowtrack}; candidates $\approx$$188\times$ vs.\ latency $\approx$$2.7\times$~\cite{xia2026groundswell}; path targeting $2{,}407\times$ vs.\ single model $117\times$~\cite{li2026aesop}; fixed-cost detector dilutes tracker overload~\cite{gu2026trackflood} & Stage share of benign latency; stage and end-to-end factors \\
    \addlinespace
    (C2) Headroom vs.\ slack & Ceiling on the amplified count; slack ratio $D/T$ & GPU NMS below 100--500~ms budgets at reachable counts~\cite{monteuuis2025evade}; length attacks bounded by the output limit~\cite{dong2025engorgio} & Ceiling, limit, and deadline used \\
    \addlinespace
    (C3) Clipping & Existing caps and timeouts & Truncation at context or output caps~\cite{fu2025lingoloop,li2026loopllm}; throughput collapse or 96.7\% data loss under defenses~\cite{li2026aesop} & Truncation, drop, or refusal rate \\
    \addlinespace
    (C4) Accumulation & State and queues across frames or requests & Tracker pool $20.5\to1{,}677.7$ tracks~\cite{gutrackshield}; KV-cache exhaustion and preemption~\cite{wang2026fillsqueeze} & Backlog, consecutive misses, recovery time \\
    \addlinespace
    (C5) Sharing & Co-tenancy, shared guardrails, agent topology & Scheduler pressure beyond length inflation~\cite{wang2026fillsqueeze}; guardrail and agent-graph propagation~\cite{zhou2026shieldtarget,liang2026mobius,zhou2026corba}; $2.16\times$ demand, $2.26$--$12.8\times$ benign p99~\cite{gu2026gatedrain} & Benign-victim latency; goodput \\
    \addlinespace
    (C6) Consequence mapping & Fallback policy; closed-loop dynamics & Crash rate $\sim$95\% vs.\ $\sim$30\%~\cite{ma2024slowtrack}; tracking absorbs detector-level errors~\cite{jia2020fooling} & Terminal outcome with the fallback stated \\
    \bottomrule
  \end{tabularx}
\end{table}

Most primary studies report $A$ at the model boundary, and some report a terminal outcome, but few report the quantities that connect the two---$f$, $T$, $D$, the ceiling, the fallback policy---so a reader cannot translate a reported factor into a system-level risk. We therefore recommend reporting them alongside every amplification factor (Section~\ref{sec:eval}, item~(iv)). Ratios and absolute costs also need to be reported together: in the event-based SNN study, the model with the smallest universal-attack ratio ($1.09\times$, IBM DVS Gesture) incurs the largest absolute increase---an estimated $+808~\mu$J per inference on Loihi-1, about 25.5~kJ per year at one inference per second---because its baseline is largest~\cite{raptis2026driving}. DeepSloth's partitioned deployment makes the same point from the baseline side: forcing inputs to the cloud raises latency to about 11~ms on both datasets, a factor of about $22\times$ on CIFAR-10 but only about $1.5\times$ on Tiny ImageNet, because the benign early-exit rates differ~\cite{hong2021deepsloth}. This analysis is also, in our view, the central lesson that a model-level organization of the field, including those of~\cite{brachemi2026energy} and~\cite{rathnasuriya2025sok}, cannot reveal: whether a latency attack is dangerous is decided by the system, not by the model.

\subsection{Evaluation and benchmarking}
\label{sec:eval}

Progress is hampered by fragmented evaluation. Perception-attack papers use datasets such as MOT17, nuScenes, and KITTI, report wall-clock latency on specific GPUs (frequently an NVIDIA 2080~Ti or a Jetson device), and---in the strongest cases---report system-level outcomes on the Baidu Apollo stack and LGSVL simulator~\cite{ma2024slowtrack}. LLM/VLM-serving evaluations instead use vLLM-style serving benchmarks on models ranging up to Llama-3-70B, and report access capacity, token amplification, or detector-bypass rate~\cite{zhang2025pd3f,liu2026reasoningbomb}. Three deficiencies recur. First, there is \emph{no standardized latency threat model} analogous to the $\ell_p$-ball used for evasion: papers leave attacker access, perturbation/query budget, and the target latency metric unspecified or mutually inconsistent. Second, there is \emph{no agreed success metric}---wall-clock time, FPS, FLOPs, energy, crash rate, and access capacity are used interchangeably, defeating cross-paper comparison. Third, \emph{no benchmark jointly stresses perception and serving} under attack, which is precisely what an end-to-end AD-plus-VLM stack would require. We argue that joint latency/robustness protocols---measuring accuracy and timing \emph{simultaneously} under adversarial load---are a prerequisite for trustworthy comparison, because efficiency and robustness interact non-monotonically (Section~\ref{sec:open}).

Toward standardization, we suggest that a latency-attack evaluation should report along five complementary dimensions, by analogy to the way evasion benchmarks fixed attacker knowledge, perturbation norm, and success rate. (i)~\emph{Threat model}: the attacker's interface (input, prompt, weights, hardware), knowledge (white/grey/black-box), and query/perturbation budget, stated explicitly rather than left implicit. (ii)~\emph{Cost metric}: at least one \emph{implementation-independent} work proxy, where meaningful (FLOPs, decoder steps, admitted candidate boxes, or selected experts), reported alongside any wall-clock or FPS number, since the latter cannot be interpreted or compared reliably without the exact accelerator, batch size, and serving framework; such proxies still require model and configuration context (context length, attention implementation, cache state, batching, architecture) and should not be treated as interchangeable. Monteuuis et al.~\cite{monteuuis2025evade} likewise prefer a count over
wall-clock time, because counts reproduce across hardware, model, and
quantization changes whereas latency does not. Their chosen count, however, is
the post-NMS box count, an output proxy capped at \texttt{max\_det}; the work
proxy for NMS is the number of candidates admitted to it. A count-based metric is
informative about a stage's cost only if it measures the intermediate quantity
that governs that cost. (iii)~\emph{Amplification baseline}: the benign-input cost against which inflation is measured, including whether the baseline is a clean average or a benign worst case, because a $10\times$ figure against an easy baseline may be smaller in absolute terms than a $2\times$ figure against a hard one. (iv)~\emph{Downstream impact}: a system-level outcome (deadline-miss rate, collision rate, mission completion, or service access capacity) rather than component timing alone, since a $5\times$ slowdown that still meets the deadline is operationally harmless while a $1.5\times$ slowdown that misses it is catastrophic; the quantities that connect the two (Section~\ref{sec:translation}) should be reported with it. (v)~\emph{Defense interaction}: whether the attack was evaluated against any deployed budget, cap, or monitor, since an attack that collapses under a trivial detection cap is qualitatively different from one that survives it. Adopting these reporting dimensions in future evaluations would make results more comparable than the heterogeneous native measurements summarized in Tables~\ref{tab:od-attacks} and~\ref{tab:general-attacks}. We stress that no single number captures a latency attack: an attack is characterized by a \emph{cost--impact pair} under a stated threat model, and reporting one without the other is the single most common evaluation gap in the literature we surveyed.

\textbf{Section Summary.}
Although different application domains expose different bottlenecks, many latency attack classes repeatedly exhibit intermediate-work amplification, whatever channel delivers them, while timing leakage and availability-adjacent embodied attacks require separate treatment. The work-budget abstraction provides a unified perspective on existing defenses when the budget caps the work entering a stage, and the translation conditions of Section~\ref{sec:translation} explain when a bounded or unbounded stage-level amplification matters to the system. Together they suggest that future research should focus on bounding computation across heterogeneous AI systems rather than protecting individual models in isolation.

\section{Emerging Surfaces and Open Challenges}
\label{sec:open}

\subsection{NMS-free detectors and moving bottlenecks}
\label{sec:nmsfree}

A salient shift affects much of the object-detection attack literature: modern detectors increasingly remove NMS. YOLOv10~\cite{wang2024yolov10} and DETR-style models~\cite{carion2020detr} replace NMS with one-to-one label assignment and lightweight decoder heads, removing the data-dependent quadratic post-processing that Phantom Sponges and Overload exploit. It is important to be precise about what this changes. Standard DETR uses Hungarian bipartite matching during \emph{training}, where predictions are matched to ground-truth objects to compute the loss; ordinary inference has no ground-truth set and therefore does not run Hungarian matching as a per-image inference step. Any claim that adversarial inputs inflate Hungarian iterations during ordinary DETR inference is therefore incorrect unless a specific deployed implementation performs assignment at inference time, and the $O(N^3)$ matching cost cannot be presented as a general inference-time attack surface.

Removing NMS does not by itself eliminate availability risk; it relocates and reshapes the cost, and the residual surfaces must be established by architecture-specific profiling rather than inferred from training-time behavior. For the downstream tracker, this relocation has now been measured: TrackFlood leaves detector latency unchanged on YOLOv10, YOLO26, and RT-DETR while inflating tracker latency, and NMS-centric mitigations have nothing to act on there (Section~\ref{sec:tracking})~\cite{gu2026trackflood}. For fixed-query DETR variants, the number of decoder queries and layers is typically fixed, which limits input-driven algorithmic work amplification in the core inference graph. Residual risk is more likely in \emph{dynamic} variants that prune tokens or queries, use adaptive decoder depth, process variable-resolution inputs, invoke downstream tracking or association stages, or exhibit input-dependent memory and hardware behavior. In assessing such risks it is useful to separate three notions: \emph{work amplification}, where the input changes admitted operation or intermediate counts; \emph{latency amplification}, where the input changes measured elapsed time; and \emph{timing variation}, where elapsed time changes without necessarily changing nominal operation counts (through memory access, cache behavior, kernel selection, dynamic preprocessing shapes, contention, thermal state, or downstream tracking). A fixed-query detector can still exhibit input-correlated \emph{timing variation} even when its computational graph is fixed, so an input-dependent wall-clock reading does not by itself imply an input-dependent operation count. We list several \emph{unvalidated, speculative} hypotheses for such surfaces, which we flag explicitly as research conjectures rather than demonstrated attacks: the multi-scale feature-aggregation \emph{neck} (PANet/BiFPN) might be forced to process high-entropy features through all pyramid levels; \emph{deformable-attention} sampling might be steered toward non-coalesced memory-access patterns; and dynamic token/query pruning might be prevented from pruning. Each of these is a candidate for the kind of profiling study that would be needed to move it from conjecture to a measured vulnerability. We highlight this as an under-explored direction. TrackFlood answers the question of whether latency threats persist across an architectural change that was \emph{not} motivated by security for the downstream tracker; whether the within-detector surfaces conjectured above exist remains open.

\subsection{Cross-cutting open challenges}

\paragraph{Standardized, system-level benchmarks.} Most attacks report per-component timing in incompatible units; few connect to downstream outcomes. SlowTrack's crash-rate evaluation~\cite{ma2024slowtrack} and CP-FREEZER's testbed~\cite{wang2026cpfreezer} are exceptions. The field needs benchmarks that report \emph{both} timing and safety/serviceability impact on common axes.

\paragraph{Physical realizability.} Projector attacks~\cite{ma2024slowperception,
muller2025detstorm} demonstrate multi-second physical delays but require active
infrastructure; SlowTrack's physical transfer remains preliminary~\cite{ma2024slowtrack}.
The gap between strong digital results and robust physical deployment is wide,
and the re-evaluation discussed in Section~\ref{sec:nms} indicates it is wider
than reported effect sizes suggest: attacks optimized in one configuration were
found not to transfer across model versions, sizes, or datasets, and physical
delivery adds a further layer of configuration dependence on top of that; some attacks proposed for physical delivery, such as Groundswell, have so far been evaluated only digitally~\cite{xia2026groundswell}. A systematic study of which reported amplification factors survive a change of deployment setting is the natural next step. Beyond the NMS family only isolated reproduction attempts exist, one of which could not reproduce the original sponge-example results~\cite{te2025skipsponge}.

\paragraph{Inference-time compute as a new surface.} Recent reasoning models and MoE routers~\cite{liu2026reasoningbomb,li2026thinktrap,fei2026misrouter} have rapidly expanded existing availability-attack classes and introduced new mechanism-specific variants. As architectures continue to make compute input-dependent, each such mechanism is a candidate target, and defenses lag behind by construction.

\paragraph{Timing as an oracle.} Black-box timing attacks~\cite{nakai2021timing} show that even opaque deployments leak optimization signals, complicating any defense that assumes the attacker's ignorance. This challenge deepens as serving systems adopt data-dependent optimizations: speculative decoding, continuous batching, prefix caching, and adaptive quantization all make wall-clock behavior a function of the input, so the same mechanisms that reduce average latency also broaden the timing side-channel. The speculative-decoding fingerprinting result~\cite{wei2024speculation}---which recovers user queries and datastore contents from per-iteration token counts---shows that the availability and confidentiality axes are not independent: an optimization added purely for throughput creates input-dependent timing and a demonstrated side-channel surface. Whether an adversary can reliably force draft-token rejection patterns that produce material slowdown beyond a non-speculative baseline remains a separate empirical question: rejecting all draft tokens removes the speedup but need not yield latency worse than ordinary non-speculative decoding, and the cited leakage result does not establish attacker control over rejection. A general design principle follows: any component whose runtime varies with input content should be treated as both an attack target and an information source, and should be evaluated under a threat model that grants the adversary timing observations by default rather than assuming an opaque black box.

\paragraph{The efficiency--robustness trade-off.} Input-adaptive efficiency mechanisms---including early exit, dynamic sparsity, token pruning, routing, and speculative decoding---create a gap between average and worst-case computation that an adversary may attempt to exploit; static pruning and quantization can reduce absolute cost but do not by themselves guarantee robustness against adaptive work inflation, and the interaction is non-monotone~\cite{hasan2025sensing,te2025skipsponge}. Defenses that bound admitted cost by construction (budget enforcement, runtime budget or deadline guards) are more promising than those that merely reduce average cost; note that such a runtime guard bounds admitted execution but does not by itself establish the system's WCET, which is a stronger analysis-based claim. The deeper tension is architectural: the field has spent a decade designing systems---early-exit networks, token pruning, MoE routing, speculative decoding, KV-cache reuse---whose entire value proposition is that \emph{average} cost falls below worst-case cost. Each such gap is a \emph{potential} attack surface---headroom an availability attacker may reclaim---but exploitation requires an attacker-controllable relationship between the input and the control path, since data dependence alone does not establish exploitability. This suggests that availability-hardened deployment may require explicitly co-designing the efficiency mechanism with its worst-case bound---for example, an early-exit network that guarantees a maximum depth per unit time regardless of confidence, or a router that caps the fraction of tokens sent to any single expert. Treating the worst-case bound as a first-class design objective, rather than a property to be measured after the fact, is a promising but largely unexplored direction.

\paragraph{Embodied and agentic amplification.} The migration of latency attacks from perception front-ends to Vision-Language-Action policies (Section~\ref{sec:vla}) and to tool-using LLM agents (Section~\ref{sec:serving}) changes both the consequence and the amplification factor. In an embodied setting, computation-induced deadline misses can propagate directly into physical control consequences---a collision, a dropped payload, a frozen manipulator---but this relationship must be demonstrated under a specified closed-loop stack, deadline, and fallback policy rather than assumed. In an agentic setting, a single injected message can trigger further tool invocation across many nodes, so work can compound across invocations---additively in sequential chains and potentially multiplicatively under recursive or fan-out execution; reported per-message amplifications already reach two orders of magnitude~\cite{liang2026mobius,zhou2026beyondmaxtokens}, illustrating large system-level amplification without alone proving multiplicative scaling. Both settings break the implicit assumption, common to most existing defenses, that the unit of protection is a single forward pass; defenses will instead have to reason about compute budgets spanning multiple models, multiple invocations, and shared infrastructure.

\paragraph{Detection fragility.} High detection-bypass rates~\cite{liu2026reasoningbomb} argue that anomaly detection alone is insufficient; enforceable budgets and scheduling must complement it. The structural problem is that several attacks preserve the source paper's reported task metric---detections, top-1 labels, or answer correctness---so monitors based only on that metric may fail to detect them, leaving the resource footprint as the main observable; note, however, that extremely repetitive or truncated outputs can visibly degrade practical utility even when the benchmark metric is stable. But the resource footprint of a benign hard input (a crowded scene, a genuinely long reasoning problem) can be indistinguishable from that of an adversarial one, forcing any detector into a false-positive/false-negative trade-off that the attacker can tune against. This is why we repeatedly argue for enforcement over detection: a work budget does not need to decide whether an input is malicious, only to cap the cost it is permitted to incur, side-stepping the detection problem entirely at the price of a tunable utility ceiling in genuinely hard cases.

\paragraph{End-to-end stack evaluation.} A modern autonomy stack with a VLM planner is simultaneously vulnerable at multiple stages (Section~\ref{sec:synthesis}), yet we did not find work that evaluates all of these jointly. Building a simulation environment that injects attack stimuli at each layer and measures \emph{system-level} outcomes---collision rate, mission completion, service availability---is, in our view, a central open engineering problem in this space. A related open question follows from the defense non-transfer documented in Section~\ref{sec:synthesis}: whether a single unifying primitive---work-budget enforcement at every stage boundary---can subsume today's narrowly scoped defenses without unacceptable accuracy cost.

Collectively, these challenges indicate that latency attacks are transitioning from isolated algorithmic curiosities to a fundamental systems-security problem. Addressing them will likely require tighter integration between machine learning, systems, hardware, and real-time computing than current research communities typically consider.

\subsection{Toward whole-system availability}
\label{sec:wholesystem}
The attacks surveyed above were, until recently, comfortably decomposable: an availability attack targeted one model---a detector, a translator, an LLM---and the analysis stopped at that model's output. Two developments dissolve this boundary. First, \emph{embodied} foundation models (Section~\ref{sec:vla}) place a Vision-Language-Action policy in a physical control loop, so the terminal effect of inflated computation is no longer a late tensor but a late \emph{action}---a manipulator that freezes, a vehicle that brakes a beat too slowly. Second, \emph{agentic} systems (Section~\ref{sec:serving}) compose many model invocations, tools, and guardrails into a graph, so a single adversarial input propagates and amplifies across components rather than terminating at one decoder. In both cases the meaningful unit of availability is no longer a forward pass but the \emph{whole system}: a pipeline of heterogeneous models with a shared, finite, real-time compute budget.

This reframing has three consequences for how the field should proceed. \emph{The consequence model changes.} A service-level objective (tokens-per-second, tail latency) is the right currency for a cloud LLM, but embodied systems may impose hard or firm real-time constraints, depending on whether a late result is unsafe, useless, or merely degraded. In safety-critical embodied systems a deadline miss may become a safety-relevant event depending on the control period, slack, fallback policy, and closed-loop dynamics, rather than being a safety event automatically (conditions C2 and C6 of Section~\ref{sec:translation}). Latency attacks on embodied systems therefore belong as much to the safety and real-time-systems literature as to adversarial machine learning, and evaluating them demands the closed-loop, outcome-level metrics (collision rate, task completion) that Section~\ref{sec:eval} argues are still missing. An observed adversarial maximum is not a worst-case execution time (WCET) bound: establishing WCET requires a sound analysis over the declared input, software, scheduling, and hardware state spaces, which is especially delicate on GPUs where contention, caches, dynamic frequency scaling, and asynchronous kernels complicate timing guarantees. Conversely, a WCET derived from benign inputs is unsafe once a stage's cost is attacker-controllable through input content, as TrackFlood notes for tracker update time~\cite{gu2026trackflood}. \emph{The amplification model changes.} In a single model, amplification is bounded by one architectural mechanism---box count, token count, expert load. In an agent graph, amplification is governed by the executed graph's depth, fan-out, recursion, retry policy, and termination limits. For an invocation graph $G=(V,E)$ with executed node set $V_{\mathrm{exec}}$, the total induced work is additive over executed invocations, $W_{\mathrm{graph}}=\sum_{v\in V_{\mathrm{exec}}} W_v$, 
so what may grow is the \emph{number} of executed invocations: with branching factor greater than one and without deduplication or execution caps, $|V_{\mathrm{exec}}|$ may grow exponentially with effective depth. It is therefore useful to distinguish four forms of amplification:
\begin{itemize}
\item \emph{Sequential amplification}---the executed graph depth grows;
\item \emph{Fan-out amplification}---the number of executed nodes grows by branching;
\item \emph{Recursive amplification}---repeated cycles or retries expand the execution trace;
\item \emph{Resource amplification}---$W_{\mathrm{graph}}$ relative to matched benign graph work.
\end{itemize}
Under this accounting the same intermediate-work amplification principle can compound across invocations---additively in per-node work but with the node count able to grow under recursion or fan-out; the two-order-of-magnitude per-message figures reported for agentic denial-of-service~\cite{liang2026mobius,zhou2026beyondmaxtokens} illustrate large system-level amplification, and guardrails themselves become amplifiers rather than mitigations when they can be trapped in extended reasoning~\cite{zhou2026shieldtarget}. The compounding can also be \emph{contagious}: \emph{CORBA}~\cite{zhou2026corba} shows that a single benign-looking, recursively self-propagating instruction can spread across a multi-agent system, driving cycles of meaningless message passing that exhaust compute cluster-wide---an availability failure whose blast radius is the agent topology itself rather than any one model. Model pipelines show the same structure without any agent: AESOP selects the execution path through a pipeline of specialized models and inflates FLOPs far beyond what attacking any single component achieves~\cite{li2026aesop}. Emerging security taxonomies for autonomous agents make the same point structurally: the Hierarchical Autonomy Evolution framework~\cite{zhang2026thinker} places resource-monopolization and computational denial-of-service at its \emph{collective-autonomy} tier, where a single node's worst-case computation propagates through the multi-agent topology as a cascade failure---precisely the systemic, cross-component amplification that a per-model budget cannot see. \emph{The defense model changes.} A per-model cap---bounded NMS, a token budget, a reasoning limit---is necessary but no longer sufficient, because an adversary can respect every local budget while still exhausting a shared global one through breadth. The natural generalization of the work-budget abstraction (Section~\ref{sec:synthesis}) is therefore a \emph{hierarchical} budget: local caps at each stage boundary, composed under a global admission-control layer---identity-independent, since per-source budgets bound each identity but not their sum under a Sybil flood~\cite{gu2026gatedrain}---that accounts for the total intermediate work a single request may induce across the entire graph, and that degrades gracefully---shedding or deferring low-priority work---rather than failing when the budget is approached.

We therefore argue that whole-system availability is emerging as the organizing problem for the next phase of this research. It subsumes the perception/serving non-transfer (Section~\ref{sec:synthesis}), the missing end-to-end benchmark (Section~\ref{sec:eval}), and the embodied/agentic amplification challenge (Section~\ref{sec:open}) under a single question: \emph{Can a heterogeneous, multi-model, real-time AI system guarantee a bounded worst-case response under adversarial load?} We identified no work in the reviewed corpus that provides such an end-to-end adversarial worst-case guarantee, and we believe progress will come from treating the compute budget of the entire pipeline as the asset to be defended.

\section{Conclusion}
\label{sec:conclusion}
This survey advances a single organizing idea: many latency attacks, across object
detection, adaptive networks, autoregressive generation, reasoning models, and
agentic systems, are instances of \emph{intermediate-work amplification}, and the
natural cross-domain defense is to bound that work through \emph{work-budget
enforcement}. This attack-to-defense arc gives a unified view of a literature that
has until now been fragmented, and it is the lens we recommend for organizing
future work on availability. Two qualifications make the lens usable. A budget
bounds a stage's work only if it caps what enters the stage, not merely what
leaves it. And amplification and failure live at different levels: a model-level
cost increase becomes an availability failure only through system properties---
critical-path share, slack, existing ceilings, accumulation, sharing, and fallback
policy---that model-level measurements do not capture
(Section~\ref{sec:translation}).

Latency attacks extend the security of deep learning beyond correctness to
timeliness and resource availability. What began as NMS overload on a single
detector now spans physical autonomous-driving perception, sponge attacks on
accelerators, slowdown attacks on every form of conditional computation, and a
rapidly expanding frontier of output-length, verbose-image, and reasoning
denial-of-service attacks on LLMs and VLMs. The breadth of this attack surface is
better established than the magnitude of the threat: the one independent
re-evaluation available found that a well-studied attack family did not transfer
outside its original measurement configuration, and that its effect on the
detections passed downstream could be removed by lowering an existing
configuration parameter, at a utility cost in crowded
scenes~\cite{monteuuis2025evade}. Whether the
same holds for the serving- and agentic-side results---which are overwhelmingly
single-group and single-configuration---is unknown.

A recurring lesson is that any mechanism which makes inference cost input-dependent---post-processing,
sparsity, early exit, routing, autoregressive generation, inference-time
reasoning---is a candidate attack surface once efficiency is advertised, though
data dependence alone does not establish exploitability, and the NMS case shows
that a surface can be real while the reported exploit is not. Even so, many
prominent exploited surfaces arise from mechanisms originally introduced for
efficiency, including sparsity, adaptive depth, pruning, routing, caching,
batching, and speculative decoding. Hardware faults, training-data and corpus
poisoning, and physical projector delivery are delivery channels rather than
surfaces, and guardrail abuse and self-propagating agent instructions exploit
surfaces that were not introduced for efficiency; neither group fits this
pattern.

Defenses are advancing across robust training, input transformation, runtime
monitoring, and serving control. Among these, budget enforcement and bounded
execution most directly constrain high-cost computation; evidence for many
purification and general-efficiency methods against adaptive latency attacks
remains limited (Section~\ref{sec:defenses}). They also remain fragmented and
bound by a persistent efficiency--robustness trade-off. Progress will depend on
standardized benchmarks that report timing alongside downstream impact, on threat
models that take physical and serving realism seriously, and on defenses that
control high-cost computational paths rather than merely lowering the average.

As deep learning continues to trade fixed cost for input-adaptive efficiency, the
availability axis will only grow in importance. We expect availability-oriented
attacks to become a permanent component of adversarial machine learning research
rather than a niche topic, and as future architectures increasingly expose dynamic
computation, understanding worst-case inference behavior will likely become as
important as improving average-case efficiency. Ultimately, availability is
evolving from a property of individual models into a property of the entire AI
system, and ensuring bounded computation may become as fundamental to trustworthy
AI as ensuring predictive accuracy.

\bibliographystyle{ACM-Reference-Format}
\bibliography{references}

\clearpage
\appendix{Appendix}
\section{Glossary of Metrics and Terms}
\label{sec:glossary}

For quick reference, Table~\ref{tab:glossary} collects the amplification metrics defined in Section~\ref{sec:metrics} and the recurring acronyms used throughout the survey.

\begin{table}[htbp]
  \caption{Glossary of the amplification metrics defined in this survey and recurring acronyms.}
   \label{tab:glossary}
  \small
  \begin{tabularx}{\linewidth}{>{\raggedright\arraybackslash}p{1.6cm} X}
    \toprule
    \textbf{Symbol / term} & \textbf{Meaning} \\
    \midrule
    \multicolumn{2}{l}{\emph{Amplification metrics (all relative to a matched benign reference; see Section~\ref{sec:metrics})}} \\
    \addlinespace[2pt]
    $\operatorname{AF}_{M}(x)$ & Generic amplification factor: attack-induced metric $M(x)$ over the matched benign median $\operatorname{median}_{x_0\in\mathcal{B}(x)}M(x_0)$, where $M$ may be work, latency, energy, or tokens. Report absolute increase when the benign median is zero/near-zero. \\
    $S_i$, $\tau_i$ & Deadline slack $S_i=d_i-f_i$ and tardiness $\tau_i=\max(0,-S_i)$ for job $i$ with deadline $d_i$ and completion time $f_i$; negative slack indicates a miss. \\
    Throughput degradation & Relative degradation in throughput or goodput against a matched benign \emph{workload} (reported as a relative change, not a named symbol). \\
    Zero rule & If a benign median denominator is $0$, the ratio is undefined: report the absolute change and attack value, never an arbitrary $\epsilon$. \\
    $\mathcal{B}(x)$ & Matched benign reference \emph{set} for $x$ (exact on model/config/hardware/concurrency; stratified on task and length; no post hoc matching on attacked values). \\
    $f$, $A$, $A_{\mathrm{e2e}}$ & Benign critical-path share of the amplified stage, its amplification, and the resulting end-to-end factor $A_{\mathrm{e2e}}=1+f(A-1)$ (Section~\ref{sec:translation}). \\
    \addlinespace[3pt]
    \multicolumn{2}{l}{\emph{Terms}} \\
    \addlinespace[2pt]
    Bottleneck & The cost center an attack makes expensive (Section~\ref{sec:bottleneck-delivery}); the organizing axis of Sections~\ref{sec:od}--\ref{sec:seqgen}. \\
    Delivery channel & How the attacker triggers the expensive behavior: test-time input, physical sensor, prompt or retrieved content, message, training data or backdoor, or weight/hardware tampering. \\
    Admission cap & A limit on the intermediate objects entering a stage; bounds that stage's work (a work budget). \\
    Emission cap & A limit on the results leaving a stage; bounds downstream work, and the stage's own work only if it terminates early. \\
    Runtime guard & A time limit that aborts or skips work; bounds elapsed time at the cost of missing or partial results. \\
    \addlinespace[3pt]
    \multicolumn{2}{l}{\emph{Acronyms}} \\
    \addlinespace[2pt]
    AD & Autonomous driving. \\
    DoS / DDoS & Denial of service / distributed denial of service. \\
    DETR & DEtection TRansformer (query-based, NMS-free detector). \\
    EOS & End-of-sequence token that terminates autoregressive generation. \\
    FLOP & Floating-point operation (a proxy for computational work). \\
    KV cache & Key--value cache storing past attention states in transformer decoding. \\
    LLM / VLM & Large language model / vision-language model. \\
    MoE & Mixture-of-experts (sparsely activated expert routing). \\
    MOT & Multi-object tracking. \\
    NMS & Non-maximum suppression (candidate-box filtering in detectors). \\
    RAG & Retrieval-augmented generation. \\
    TTFT & Time to first token (LLM-serving latency measure). \\
    V2V & Vehicle-to-vehicle communication (cooperative perception). \\
    VLA & Vision-language-action model (maps observations to robot actions). \\
    WCET & Worst-case execution time. \\
    \bottomrule
  \end{tabularx}
\end{table}

\section{Catalog of Latency, Energy, and Timing Attacks and Defenses}
\label{sec:appendix}

This appendix provides a structured, citable catalog of the works surveyed. For ease of lookup it is organized by delivery stage (test-time inputs vs.\ training data, backdoors, and weights) and by defenses; the main text instead organizes attacks by bottleneck, and Tables~\ref{tab:threats} and~\ref{tab:matrix} give the bottleneck mapping. Entries marked \textsuperscript{$\ast$} are timing or serving side channels, listed for completeness and excluded from attack counts (Section~\ref{sec:timing}). Each entry records the target architecture, application domain, attacker setting, and links to the paper and released code where available. A continuously updated, searchable web version of these tables is maintained at the companion repository \url{https://github.com/guzonghua/awesome-latency-attacks}. Tables~\ref{tab:cat-inference}--\ref{tab:cat-defenses} mirror that resource.

\input{appendix_tables.tex}

\section{Attack-Mapping Matrix}
\label{sec:matrix}

Table~\ref{tab:matrix} maps representative attacks across five dimensions used throughout the survey: the attacker \emph{interface} (Section~\ref{sec:threat}), the exploited \emph{bottleneck}, the \emph{work measure} the attacker inflates, the \emph{system consequence}, and the strongest \emph{evidence setting} reported. The 18 entries were selected for coverage---at least one representative per interface (physical, API, corpus, hardware) and per major bottleneck (NMS, tracking, cooperative fusion, activation sparsity, early exit, token pruning, MoE routing, autoregressive decode, reasoning length, serving scheduler, agent graph, weight tampering)---rather than by highest reported slowdown. The matrix is a quick cross-index, not a quantitative comparison; reported magnitudes and their caveats remain in Tables~\ref{tab:od-attacks}, \ref{tab:general-attacks}, and the catalog above; per-entry publication status is given in the catalog (Tables~\ref{tab:cat-inference}--\ref{tab:cat-defenses}).

\begin{table*}[htbp]
  \caption{Attack-mapping matrix. Interface (delivery channel): Phys.\ = physical sensor delivery; API = request/prompt; Corpus = retrieval/training data; HW = hardware fault. Evidence (where effects are measured, as in Table~\ref{tab:od-attacks}): Sim = digital model evaluation; HW = hardware-timed evaluation; Test = vehicle/robot testbed; Loop = closed-loop simulator; Fault = hardware fault model. Per-entry publication status is listed in the catalog (Tables~\ref{tab:cat-inference}--\ref{tab:cat-defenses}).}
  \label{tab:matrix}
  \footnotesize
  \setlength{\tabcolsep}{2pt}
  \begin{tabularx}{\textwidth}{@{}l l L{2.4cm} L{2.6cm} L{2.6cm} l@{}}
    \toprule
    \textbf{Attack} & \textbf{Interface} & \textbf{Bottleneck} & \textbf{Work measure} & \textbf{Consequence} & \textbf{Evid.} \\
    \midrule
    Phantom Sponges~\cite{shapira2023phantom} & Digital & NMS & Candidate boxes & OD latency & Sim \\
    Groundswell~\cite{xia2026groundswell} & Digital (phys.\ proposed) & NMS & Candidate boxes & OD latency & HW \\
    SlowTrack~\cite{ma2024slowtrack} & Digital & NMS + MOT & Boxes; track assoc.\ & Crash (sim.) & Loop \\
    SlowPerception~\cite{ma2024slowperception} & Phys.\ (proj.) & NMS + MOT & Phantom objects & Collision (sim.) & Loop \\
    CP-FREEZER~\cite{wang2026cpfreezer} & V2V message & Coop.\ fusion & Fused entities & Fusion delay & Test \\
    SlowLiDAR~\cite{liu2023slowlidar} & Digital & LiDAR detect.\ & Points/candidates & Detect.\ latency & Sim \\
    Sponge examples~\cite{shumailov2021sponge} & Digital & Activ.\ sparsity & Active neurons & Energy/latency & HW \\
    DeepSloth~\cite{hong2021deepsloth} & Digital & Early exit & Executed layers & Slowdown & Sim \\
    SlowFormer~\cite{navaneet2024slowformer} & Digital (patch) & Token pruning & Retained tokens & Compute inflation & Sim \\
    RepetitionCurse~\cite{huang2025repetitioncurse} & API & MoE routing & Expert load & Serving latency & HW \\
    NMTSloth~\cite{chen2022nmtsloth} & API & AR decode & Decoder steps & NMT latency & Sim \\
    Engorgio~\cite{dong2025engorgio} & API & AR decode & Output tokens & LLM cost & Sim \\
    ReasoningBomb~\cite{liu2026reasoningbomb} & API & Reasoning length & CoT tokens & LLM cost & Sim \\
    OverThink~\cite{kumar2025overthink} & Corpus (RAG) & Reasoning length & CoT tokens & LLM cost & Sim \\
    DrainCode~\cite{wang2025draincode} & Corpus (RAG) & AR decode & Output tokens & Code-LLM cost & Sim \\
    Fill and Squeeze~\cite{wang2026fillsqueeze} & API & Sched.\ / KV cache & KV-cache occ.\ & TTFT / HoL block & HW \\
    Mobius Injection~\cite{liang2026mobius} & API (agentic) & Agent graph & Recursive calls & Cluster-wide DoS & HW \\
    BitHydra~\cite{yan2025bithydra} & HW (bit-flip) & Weights (no-EOS) & Output tokens & Persistent cost & Fault \\
    \bottomrule
  \end{tabularx}
\end{table*}

\end{document}

%% file: appendix_tables.tex
{\footnotesize\setlength{\tabcolsep}{3pt}
\begin{longtable}{p{2.4cm} p{2.2cm} p{2.7cm} p{1.2cm} p{2.1cm} >{\centering\arraybackslash}p{0.7cm} >{\centering\arraybackslash}p{0.7cm}}
\caption{Inference-stage latency, energy, and timing attacks on deep learning. \faFilePdf~links to the paper; \faGithub~links to released code (\ding{55}~= none located). \textsuperscript{\dag}~= availability-adjacent (denies or halts service without demonstrated computational-work or latency amplification); \textsuperscript{\ddag}~= integrity attack with hypothesized but unevaluated latency implications (Section~\ref{sec:vla}); \textsuperscript{$\ast$}~= timing or serving side channel (confidentiality; Section~\ref{sec:timing}), listed for completeness and excluded from attack counts.}\label{tab:cat-inference}\\
\toprule
\textbf{Attack} & \textbf{Venue (Year)} & \textbf{Target} & \textbf{Domain} & \textbf{Setting} & \textbf{Paper} & \textbf{Code} \\
\midrule\endfirsthead
\caption[]{Inference-stage latency, energy, and timing attacks on deep learning. \faFilePdf~links to the paper; \faGithub~links to released code (\ding{55}~= none located). (continued)}\\
\toprule
\textbf{Attack} & \textbf{Venue (Year)} & \textbf{Target} & \textbf{Domain} & \textbf{Setting} & \textbf{Paper} & \textbf{Code} \\
\midrule\endhead
\midrule\multicolumn{7}{r}{\footnotesize\itshape continued on next page}\\\endfoot
\bottomrule\endlastfoot
Daedalus & arXiv (2019) & Object detection (NMS) & CV / AD & White box (physical) & \href{https://arxiv.org/abs/1902.02067}{\faFilePdf} & \href{https://github.com/NeuralSec/Daedalus-attack}{\faGithub} \\
\addlinespace[1pt]
Sparsity Attacks & IEEE TCAD (2020) & CNNs & CV & White box & \href{https://arxiv.org/abs/2006.08020}{\faFilePdf} & \ding{55} \\
\addlinespace[1pt]
ILFO & CVPR (2020) & Depth-adaptive networks (AdNNs) & CV & White box & \href{https://youngwei.com/pdf/ILFO.pdf}{\faFilePdf} & \ding{55} \\
\addlinespace[1pt]
GradAuto & ECCV (2022) & Depth- \& width-dynamic AdNNs & CV & White box & \href{https://www.ecva.net/papers/eccv_2022/papers_ECCV/html/3150_ECCV_2022_paper.php}{\faFilePdf} & \href{https://github.com/LingsenPan/GradAuto}{\faGithub} \\
\addlinespace[1pt]
GradMDM & IEEE TPAMI (2023) & Dynamic neural networks & CV & White box & \href{https://arxiv.org/abs/2304.06724}{\faFilePdf} & \ding{55} \\
\addlinespace[1pt]
Sponge Examples & EuroS\&P (2021) & Transformers + CNNs & CV \& NLP & Black \& white box & \href{https://arxiv.org/abs/2006.03463}{\faFilePdf} & \href{https://github.com/iliaishacked/sponge_examples}{\faGithub} \\
\addlinespace[1pt]
DeepSloth & ICLR (2021) & Multi-exit CNNs & CV & White box & \href{https://arxiv.org/abs/2010.02432}{\faFilePdf} & \href{https://github.com/Sanghyun-Hong/DeepSloth}{\faGithub} \\
\addlinespace[1pt]
AntiNODE & ICCVW (2023) & Neural ODEs (adaptive solver) & CV & Black box (transfer) & \href{https://doi.org/10.1109/ICCVW60793.2023.00163}{\faFilePdf} & \ding{55} \\
\addlinespace[1pt]
SlothSpeech & INTERSPEECH (2023) & Speech recognition (ASR) & Speech & White box & \href{https://www.isca-archive.org/interspeech_2023/haque23_interspeech.pdf}{\faFilePdf} & \ding{55} \\
\addlinespace[1pt]
Timing Side-Channel AE\textsuperscript{$\ast$} & IEICE Trans. (2021) & DNN classifiers (timing oracle) & CV & Black box (timing) & \href{https://doi.org/10.1587/transfun.2020CIP0022}{\faFilePdf} & \ding{55} \\
\addlinespace[1pt]
SpikeAttack & DAC (2022) & Spiking neural networks & CV & White box & \href{https://dl.acm.org/doi/pdf/10.1145/3489517.3530443}{\faFilePdf} & \ding{55} \\
\addlinespace[1pt]
NMTSloth & ESEC/FSE (2022) & Decoder-based NMT & NLP & White box & \href{https://arxiv.org/abs/2210.03696v1}{\faFilePdf} & \href{https://github.com/SeekingDream/FSE22_NMTSloth}{\faGithub} \\
\addlinespace[1pt]
LLMEffiChecker & ACM TOSEM (2024) & LLMs & NLP & Black \& white box & \href{https://dl.acm.org/doi/full/10.1145/3664812}{\faFilePdf} & \href{https://github.com/Cap-Ning/LLMEffiChecker}{\faGithub} \\
\addlinespace[1pt]
NICGSlowDown & CVPR (2022) & Decoder-based image captioning & CV & White box & \href{https://arxiv.org/abs/2203.15859}{\faFilePdf} & \href{https://github.com/SeekingDream/CVPR22_NICGSlowDown}{\faGithub} \\
\addlinespace[1pt]
SAME & ACL (2023) & Multi-exit transformers & NLP & White box & \href{https://arxiv.org/abs/2305.12228}{\faFilePdf} & \href{https://github.com/MatthewCYM/SAME}{\faGithub} \\
\addlinespace[1pt]
SlowBERT & ACL Findings (2023) & Multi-exit BERT & NLP & White box & \href{https://aclanthology.org/2023.findings-acl.602/}{\faFilePdf} & \ding{55} \\
\addlinespace[1pt]
WAFFLE & NeurIPS (2023) & Multi-exit language models & NLP & White \& black box & \href{https://arxiv.org/abs/2310.19152}{\faFilePdf} & \href{https://github.com/ZachCoalson/WAFFLE}{\faGithub} \\
\addlinespace[1pt]
DeepPerform & ASE (2022) & Dynamic (resource-constrained) DNNs & CV / NLP & White box (testing) & \href{https://arxiv.org/abs/2210.05370}{\faFilePdf} & \ding{55} \\
\addlinespace[1pt]
No-Skim & arXiv (2023) & Skimming language models & NLP & Black \& white box & \href{https://arxiv.org/abs/2312.09494}{\faFilePdf} & \ding{55} \\
\addlinespace[1pt]
Phantom Sponges & WACV (2023) & Object detection (NMS) & CV / AD & White box & \href{https://arxiv.org/abs/2205.13618}{\faFilePdf} & \href{https://github.com/AvishagS422/PhantomSponges}{\faGithub} \\
\addlinespace[1pt]
SlowLiDAR & CVPR (2023) & 3D LiDAR detection & CV / AD & White box & \href{https://openaccess.thecvf.com/content/CVPR2023/papers/Liu_SlowLiDAR_Increasing_the_Latency_of_LiDAR-Based_Detection_Using_Adversarial_Examples_CVPR_2023_paper.pdf}{\faFilePdf} & \href{https://github.com/WUSTL-CSPL/SlowLiDAR}{\faGithub} \\
\addlinespace[1pt]
Variable-Time Inference\textsuperscript{$\ast$} & AISec@CCS (2023) & Object detection (NMS timing) & CV & Black box (timing) & \href{https://doi.org/10.1145/3605764.3623912}{\faFilePdf} & \ding{55} \\
\addlinespace[1pt]
Overload & CVPR (2024) & Object detection (edge) & CV / AD & White box & \href{https://arxiv.org/abs/2304.05370}{\faFilePdf} & \ding{55} \\
\addlinespace[1pt]
SlowFormer & CVPR (2024) & Efficient vision transformers & CV & White box (patch) & \href{https://arxiv.org/abs/2310.02544}{\faFilePdf} & \href{https://github.com/UCDvision/SlowFormer}{\faGithub} \\
\addlinespace[1pt]
DeSparsify & NeurIPS (2024) & Token-sparsified ViTs & CV & White box & \href{https://arxiv.org/abs/2402.02554}{\faFilePdf} & \ding{55} \\
\addlinespace[1pt]
Beyond PhantomSponges & ACM WiseML (2024) & Object detection (NMS) & CV / AD & White box & \href{https://doi.org/10.1145/3649403.3656485}{\faFilePdf} & \ding{55} \\
\addlinespace[1pt]
Groundswell & USENIX VehicleSec (2026) & Object detection (NMS) & CV / AD & White box & \href{https://www.usenix.org/conference/vehiclesec26/presentation/xia}{\faFilePdf} & \href{https://zenodo.org/records/19672875}{\faGithub} \\
\addlinespace[1pt]
SlowTrack & AAAI (2024) & Camera perception (detect+track) & AD & White box & \href{https://doi.org/10.1609/aaai.v38i5.28200}{\faFilePdf} & \href{https://github.com/eaiers/SlowTrack}{\faGithub} \\
\addlinespace[1pt]
SlowPerception & arXiv (2024) & Camera perception (NMS+MOT) & AD & White box (physical, projector) & \href{https://arxiv.org/abs/2406.05800}{\faFilePdf} & \ding{55} \\
\addlinespace[1pt]
Steal Now Attack Later & arXiv (2024) & Object detection & CV & Black box & \href{https://arxiv.org/abs/2404.15881}{\faFilePdf} & \ding{55} \\
\addlinespace[1pt]
Energy Attack (multi-exit) & Info. \& Software Tech. (2024) & Adaptive multi-exit networks & CV & Grey box & \href{https://doi.org/10.1016/j.infsof.2024.107653}{\faFilePdf} & \ding{55} \\
\addlinespace[1pt]
Varma et al.\ (slowdown analysis) & IEEE SaTML (2024) & Language models & NLP & White box & \href{https://doi.org/10.1109/SaTML59370.2024.00042}{\faFilePdf} & \ding{55} \\
\addlinespace[1pt]
Engorgio & ICLR (2025) & LLMs (output inflation) & NLP & White box + transfer & \href{https://proceedings.iclr.cc/paper_files/paper/2025/hash/a815fe7cad6af20a6c118f2072a881d2-Abstract-Conference.html}{\faFilePdf} & \href{https://github.com/jianshuod/Engorgio-prompt}{\faGithub} \\
\addlinespace[1pt]
Verbose Images & ICLR (2024) & Large VLMs & CV+NLP & White box & \href{https://arxiv.org/abs/2401.11170}{\faFilePdf} & \href{https://github.com/KuofengGao/Verbose_Images}{\faGithub} \\
\addlinespace[1pt]
Uniform Inputs & IEEE SPW (2024) & CNNs (sparsity) & CV & Black box & \href{https://arxiv.org/abs/2403.18587}{\faFilePdf} & \href{https://github.com/and-mill/2024-sponge-example-analysis}{\faGithub} \\
\addlinespace[1pt]
Speculative-Decoding Side Channel\textsuperscript{$\ast$} & arXiv (2024) & LLM serving (speculative decoding) & NLP & Black box (timing) & \href{https://arxiv.org/abs/2411.01076}{\faFilePdf} & \ding{55} \\
\addlinespace[1pt]
DetStorm & IEEE S\&P (2025) & Camera perception & AD & White box (physical) & \href{https://doi.org/10.1109/SP61157.2025.00236}{\faFilePdf} & \ding{55} \\
\addlinespace[1pt]
Inference-Time Impact Analysis & arXiv (2025) & Full perception (sim.) & AD & Simulation & \href{https://arxiv.org/abs/2505.03850}{\faFilePdf} & \ding{55} \\
\addlinespace[1pt]
DDLS Efficiency Attacks & arXiv (2025) & Early-exit / token-pruning / MoE & CV \& NLP & White \& black box & \href{https://arxiv.org/abs/2506.17621}{\faFilePdf} & \ding{55} \\
\addlinespace[1pt]
TTSlow & IEEE TASLP (2025) & Auto-regressive TTS & Speech & White box & \href{https://arxiv.org/abs/2407.01927}{\faFilePdf} & \ding{55} \\
\addlinespace[1pt]
AutoDoS & ACL Findings (2025) & LLMs (DoS) & NLP & Black box & \href{https://aclanthology.org/2025.findings-acl.580/}{\faFilePdf} & \ding{55} \\
\addlinespace[1pt]
DDAS & GLOBECOM (2022) & Edge dynamic (multi-exit) DNNs & CV & White box & \href{https://doi.org/10.1109/GLOBECOM48099.2022.10001235}{\faFilePdf} & \ding{55} \\
\addlinespace[1pt]
EREBA & ICSE (2022) & Adaptive neural networks (AdNNs) & CV & Black box & \href{https://doi.org/10.1145/3510003.3510088}{\faFilePdf} & \ding{55} \\
\addlinespace[1pt]
LoopLLM & AAAI (2026) & LLMs (repetitive generation) & NLP & White box + transfer & \href{https://arxiv.org/abs/2511.07876}{\faFilePdf} & \ding{55} \\
\addlinespace[1pt]
RA-ICA (RAG) & ACM WWW (2026) & RAG-enhanced LLMs & NLP & Corpus poisoning & \href{https://arxiv.org/abs/2606.02643}{\faFilePdf} & \ding{55} \\
\addlinespace[1pt]
DrainCode & arXiv (2026) & RAG code generation & NLP / Code & Corpus poisoning & \href{https://arxiv.org/abs/2601.20615}{\faFilePdf} & \href{https://github.com/DeepSoftwareAnalytics/DrainCode}{\faGithub} \\
\addlinespace[1pt]
VLMInferSlow & ACL (2025) & VLMs-as-a-service & CV+NLP & Black box & \href{https://aclanthology.org/2025.acl-long.781/}{\faFilePdf} & \ding{55} \\
\addlinespace[1pt]
Verbose-Text Induction & arXiv (2025) & VLMs & CV+NLP & White box & \href{https://arxiv.org/abs/2511.16163}{\faFilePdf} & \ding{55} \\
\addlinespace[1pt]
LingoLoop & arXiv (2025) & Multimodal LLMs & CV+NLP & White box & \href{https://arxiv.org/abs/2506.14493}{\faFilePdf} & \ding{55} \\
\addlinespace[1pt]
VidDoS & arXiv (2026) & Video-LLMs (streaming AD) & CV+NLP / AD & White box (universal patch) & \href{https://arxiv.org/abs/2603.01454}{\faFilePdf} & \ding{55} \\
\addlinespace[1pt]
FreezeVLA\textsuperscript{\dag} & arXiv (2025) & Vision-language-action models & Robotics & White box (image) & \href{https://arxiv.org/abs/2509.19870}{\faFilePdf} & \ding{55} \\
\addlinespace[1pt]
Semantic-DoS\textsuperscript{\dag} & arXiv (2026) & LLM-controlled robots & Robotics & Audio injection & \href{https://arxiv.org/abs/2604.24790}{\faFilePdf} & \ding{55} \\
\addlinespace[1pt]
EDPA\textsuperscript{\ddag} & arXiv (2025) & Vision-language-action models & Robotics & Black box (patch) & \href{https://arxiv.org/abs/2510.13237}{\faFilePdf} & \href{https://edpa-attack.github.io/}{\faGithub} \\
\addlinespace[1pt]
MAVLA\textsuperscript{\ddag} & ACM WWW (2026) & Vision-language-action models & Robotics & White box & \href{https://doi.org/10.1145/3774904.3792315}{\faFilePdf} & \ding{55} \\
\addlinespace[1pt]
ANNIE\textsuperscript{\ddag} & arXiv (2025) & Embodied AI / VLA robots & Robotics & White box & \href{https://arxiv.org/abs/2509.03383}{\faFilePdf} & \href{https://github.com/RLCLab/Annie}{\faGithub} \\
\addlinespace[1pt]
Bit-Flip NMS Attack & WACV (2025) & Object detection (parameters) & CV / AD & Hardware (Rowhammer) & \href{https://doi.org/10.1109/WACV61041.2025.00653}{\faFilePdf} & \ding{55} \\
\addlinespace[1pt]
BitHydra & arXiv (2025) & LLM weights (no-EOS) & NLP & Hardware (bit-flip) & \href{https://arxiv.org/abs/2505.16670}{\faFilePdf} & \ding{55} \\
\addlinespace[1pt]
QuantAttack & WACV (2025) & Dynamically quantized ViTs & CV & White box & \href{https://arxiv.org/abs/2312.02220}{\faFilePdf} & \href{https://github.com/barasamit/QuantAttack}{\faGithub} \\
\addlinespace[1pt]
Timestep-Compressed Attack & AAAI (2026) & Spiking neural networks & CV & White box & \href{https://arxiv.org/abs/2508.13812}{\faFilePdf} & \ding{55} \\
\addlinespace[1pt]
CP-FREEZER & AAAI (2026) & Cooperative perception (V2V) & AD & White box (testbed) & \href{https://ojs.aaai.org/index.php/AAAI/article/download/37082/41044}{\faFilePdf} & \ding{55} \\
\addlinespace[1pt]
SPLAT & IEEE TCAD (2026) & Multi-exit dynamic networks & CV & Black box & \href{https://doi.org/10.1109/TCAD.2025.3576320}{\faFilePdf} & \ding{55} \\
\addlinespace[1pt]
RouteHijack\textsuperscript{\ddag} & arXiv (2026) & Mixture-of-experts LLMs & NLP & White box & \href{https://arxiv.org/abs/2605.02946}{\faFilePdf} & \ding{55} \\
\addlinespace[1pt]
Misrouter & arXiv (2026) & Mixture-of-experts LLMs & NLP & Black box (input-only) & \href{https://arxiv.org/abs/2605.04446}{\faFilePdf} & \ding{55} \\
\addlinespace[1pt]
ReasoningBomb & ACM CCS (2026) & Large reasoning models & NLP & Black box & \href{https://arxiv.org/abs/2602.00154}{\faFilePdf} & \ding{55} \\
\addlinespace[1pt]
ThinkTrap & NDSS (2026) & Reasoning LLM APIs & NLP & Black box & \href{https://arxiv.org/abs/2512.07086}{\faFilePdf} & \ding{55} \\
\addlinespace[1pt]
OverThink & arXiv (2025) & Reasoning LLMs (RAG/context) & NLP & Indirect injection & \href{https://arxiv.org/abs/2502.02542}{\faFilePdf} & \ding{55} \\
\addlinespace[1pt]
ExtendAttack & arXiv (2025) & Large reasoning models & NLP & Black box (encoding) & \href{https://arxiv.org/abs/2506.13737}{\faFilePdf} & \ding{55} \\
\addlinespace[1pt]
RepetitionCurse & arXiv (2025) & MoE LLMs (expert parallelism) & NLP & Black box & \href{https://arxiv.org/abs/2512.23995}{\faFilePdf} & \ding{55} \\
\addlinespace[1pt]
Fill and Squeeze & arXiv (2026) & LLM serving scheduler & NLP & Black box (system) & \href{https://arxiv.org/abs/2602.07878}{\faFilePdf} & \ding{55} \\
\addlinespace[1pt]
Beyond Max Tokens & arXiv (2026) & LLM agent tool chains (MCP) & NLP / Agentic & Black box & \href{https://arxiv.org/abs/2601.10955}{\faFilePdf} & \ding{55} \\
\addlinespace[1pt]
From Shield to Target & arXiv (2026) & LLM agent guardrails & NLP / Agentic & Black box (transfer) & \href{https://arxiv.org/abs/2606.14517}{\faFilePdf} & \ding{55} \\
\addlinespace[1pt]
Mobius Injection (AbO-DDoS) & arXiv (2026) & LLM agent infrastructure & Agentic & Injection & \href{https://arxiv.org/abs/2605.11442}{\faFilePdf} & \ding{55} \\
\addlinespace[1pt]
AESOP & arXiv (2026) & Multi-model inference pipelines (execution path) & ML pipelines & White \& grey box & \href{https://arxiv.org/abs/2605.10987}{\faFilePdf} & \ding{55} \\
\addlinespace[1pt]
SNN Sponge (per-sample \& universal) & arXiv (2026) & Event-based SNNs (spike activity) & CV / Audio & White box; universal mask & \href{https://arxiv.org/abs/2607.27990}{\faFilePdf} & \ding{55} \\
\addlinespace[1pt]
TrackFlood & arXiv (2026) & NMS-free detect-then-track (tracker association) & AD & White box (digital; universal) & \href{https://arxiv.org/abs/2609.33948}{\faFilePdf} & \ding{55} \\
\addlinespace[1pt]
GateDrain & arXiv (2026) & Confidence-gated edge--cloud offloading (cloud queue) & CV / Edge & White box, transfer, decision-only, universal & \href{https://arxiv.org/abs/2609.33992}{\faFilePdf} & \ding{55} \\
\addlinespace[1pt]
CORBA & ACL Findings (2026) & LLM multi-agent systems & Agentic & Contagious injection & \href{https://arxiv.org/abs/2502.14529}{\faFilePdf} & \href{https://github.com/zhrli324/Corba}{\faGithub} \\
\addlinespace[1pt]
\end{longtable}}

{\footnotesize\setlength{\tabcolsep}{3pt}
\begin{longtable}{p{2.4cm} p{2.2cm} p{2.7cm} p{1.2cm} p{2.1cm} >{\centering\arraybackslash}p{0.7cm} >{\centering\arraybackslash}p{0.7cm}}
\caption{Latency and energy attacks delivered through training data, backdoors, or weights (persistent delivery; Section~\ref{sec:persistent}). The bottleneck each exploits is given in parentheses in the Target column and discussed in the corresponding bottleneck section.}\label{tab:cat-training}\\
\toprule
\textbf{Attack} & \textbf{Venue (Year)} & \textbf{Target} & \textbf{Domain} & \textbf{Setting} & \textbf{Paper} & \textbf{Code} \\
\midrule\endfirsthead
\caption[]{Latency and energy attacks delivered through training data, backdoors, or weights. (continued)}\\
\toprule
\textbf{Attack} & \textbf{Venue (Year)} & \textbf{Target} & \textbf{Domain} & \textbf{Setting} & \textbf{Paper} & \textbf{Code} \\
\midrule\endhead
\midrule\multicolumn{7}{r}{\footnotesize\itshape continued on next page}\\\endfoot
\bottomrule\endlastfoot
Sponge Poisoning & Information Sciences (2025) & CNNs (activation sparsity) & CV & Partial control & \href{https://doi.org/10.1016/j.ins.2025.121905}{\faFilePdf} & \href{https://github.com/Cinofix/sponge_poisoning_energy_latency_attack}{\faGithub} \\
\addlinespace[1pt]
On-Device Sponge Poisoning & ACM SecTL (2023) & On-device DNNs (activation sparsity) & CV & Partial control & \href{https://doi.org/10.1145/3591197.3591307}{\faFilePdf} & \ding{55} \\
\addlinespace[1pt]
Mobile-App Sponge & ACM HotMobile (2023) & Mobile ML models (activation sparsity) & CV & Partial control & \href{https://doi.org/10.1145/3572864.3581586}{\faFilePdf} & \ding{55} \\
\addlinespace[1pt]
SkipSponge & ITU J-FET (2025) & CNNs, GANs, autoencoders (activation sparsity; weights) & CV & Full control & \href{https://arxiv.org/abs/2402.06357}{\faFilePdf} & \href{https://github.com/jonatelintelo/SkipSponge}{\faGithub} \\
\addlinespace[1pt]
Huang et al. (multi-exit) & IEEE Access (2024) & Multi-exit CNNs (early exit) & CV & Full control & \href{https://doi.org/10.1109/ACCESS.2024.3370849}{\faFilePdf} & \ding{55} \\
\addlinespace[1pt]
Energy Backdoor & ICASSP (2025) & CNNs (activation sparsity) & CV & Backdoor & \href{https://arxiv.org/abs/2501.08152}{\faFilePdf} & \href{https://github.com/hbrachemi/energy_backdoor}{\faGithub} \\
\addlinespace[1pt]
Sponge Backdoor (OD) & IJCNN (2024) & Object detection (NMS) & CV / AD & Backdoor & \href{https://doi.org/10.1109/IJCNN60899.2024.10650435}{\faFilePdf} & \ding{55} \\
\addlinespace[1pt]
DoS Poisoning (LLM) & arXiv (2024) & LLMs (no-EOS) & NLP & Backdoor / poisoning & \href{https://arxiv.org/abs/2410.10760}{\faFilePdf} & \ding{55} \\
\addlinespace[1pt]
Sensing-AI Sponge & IEEE GLOBECOM (2025) & Sensing DNNs, IoT (activation sparsity) & Sensing & Partial control & \href{https://doi.org/10.1109/GLOBECOM59602.2025.11432163}{\faFilePdf} & \ding{55} \\
\addlinespace[1pt]
EvoWeight (FPGA) & IEEE HOST (2025) & FPGA DNN accelerators (activation sparsity) & CV & Full control & \href{https://doi.org/10.1109/HOST64725.2025.11050058}{\faFilePdf} & \ding{55} \\
\addlinespace[1pt]
Reflection Backdoor (VLM-AD) & arXiv (2025) & Driving VLM planner (output length) & AD & Backdoor (physical trigger) & \href{https://arxiv.org/abs/2505.06413}{\faFilePdf} & \ding{55} \\
\addlinespace[1pt]
\end{longtable}}

{\footnotesize\setlength{\tabcolsep}{3pt}
\begin{longtable}{p{2.4cm} p{2.1cm} p{2.5cm} p{2.4cm} p{1.2cm} >{\centering\arraybackslash}p{0.7cm} >{\centering\arraybackslash}p{0.7cm}}
\caption{Defenses against latency, energy, and timing attacks, by control mechanism.}\label{tab:cat-defenses}\\
\toprule
\textbf{Defense} & \textbf{Venue (Year)} & \textbf{Target} & \textbf{Mechanism} & \textbf{Domain} & \textbf{Paper} & \textbf{Code} \\
\midrule\endfirsthead
\caption[]{Defenses against latency, energy, and timing attacks, by control mechanism. (continued)}\\
\toprule
\textbf{Defense} & \textbf{Venue (Year)} & \textbf{Target} & \textbf{Mechanism} & \textbf{Domain} & \textbf{Paper} & \textbf{Code} \\
\midrule\endhead
\midrule\multicolumn{7}{r}{\footnotesize\itshape continued on next page}\\\endfoot
\bottomrule\endlastfoot
DEE Scheduling & ACM CIKM (2021) & Early-exit networks & Runtime scheduling & CV & \href{https://doi.org/10.1145/3459637.3482335}{\faFilePdf} & \ding{55} \\
\addlinespace[1pt]
Certifier Caveat & arXiv (2021) & Certified classifiers & Defense pitfall & CV & \href{https://arxiv.org/abs/2108.11299}{\faFilePdf} & \ding{55} \\
\addlinespace[1pt]
Constant-Time NMS & AISec@CCS (2023) & Object detection (NMS) & Constant-time implementation (within analyzed NMS boundary) & CV & \href{https://doi.org/10.1145/3605764.3623912}{\faFilePdf} & \ding{55} \\
\addlinespace[1pt]
PSML & arXiv (2023) & Inference serving systems & System / serving control & ML serving & \href{https://arxiv.org/abs/2307.01292}{\faFilePdf} & \ding{55} \\
\addlinespace[1pt]
Adaptive Resizing & ACM WiseML (2024) & Object detection (NMS) & Input transformation & AD & \href{https://doi.org/10.1145/3649403.3656485}{\faFilePdf} & \ding{55} \\
\addlinespace[1pt]
ADAV Patch Defense & arXiv (2024) & Object detection & Input transformation & AD & \href{https://arxiv.org/abs/2412.06215}{\faFilePdf} & \ding{55} \\
\addlinespace[1pt]
Securing AV Perception & IEEE TIV (2024) & AV visual perception & Input transformation & AD & \href{https://doi.org/10.1109/TIV.2024.3403667}{\faFilePdf} & \ding{55} \\
\addlinespace[1pt]
Varma et al.\ (expedited adv.\ training) & IEEE SaTML (2024) & Multi-exit models & Robust training & NLP & \href{https://arxiv.org/abs/2305.18926}{\faFilePdf} & \ding{55} \\
\addlinespace[1pt]
DefQ & IEEE IoT-J (2023) & Multi-exit DNNs (DeepSloth) & DCT defensive quantization & Edge & \href{https://doi.org/10.1109/JIOT.2021.3138935}{\faFilePdf} & \ding{55} \\
\addlinespace[1pt]
Feature Distillation & CVPR (2019) & CNNs (JPEG pipeline) & DCT input purification & CV & \href{https://doi.org/10.1109/CVPR.2019.00095}{\faFilePdf} & \ding{55} \\
\addlinespace[1pt]
Sparsity Monitor & IEEE SPW (2024) & CNNs & Runtime monitoring & CV & \href{https://arxiv.org/abs/2403.18587}{\faFilePdf} & \href{https://github.com/and-mill/2024-sponge-example-analysis}{\faGithub} \\
\addlinespace[1pt]
Garrison & DAC (2024) & Ensemble inference (GPU) & System / serving control & CV & \href{https://doi.org/10.1145/3649329.3654810}{\faFilePdf} & \ding{55} \\
\addlinespace[1pt]
Time-Traveling Defense & arXiv (2024) & Traffic-sign classifiers & Temporal redundancy & AD & \href{https://arxiv.org/abs/2410.08338}{\faFilePdf} & \ding{55} \\
\addlinespace[1pt]
Can't Slow Me Down & CVPR (2025) & Edge object detectors & Robust/adaptive training & AD & \href{https://doi.org/10.1109/CVPR52734.2025.01791}{\faFilePdf} & \ding{55} \\
\addlinespace[1pt]
DCT Patch Elimination & IEEE RCAR (2025) & Object detection & Input transformation & AD & \href{https://doi.org/10.1109/RCAR65431.2025.11139457}{\faFilePdf} & \ding{55} \\
\addlinespace[1pt]
Real-Time LiDAR Defense & ACM CCS (2025) & LiDAR detection & Runtime monitoring & AD & \href{https://doi.org/10.1145/3719027.3765227}{\faFilePdf} & \ding{55} \\
\addlinespace[1pt]
LDP Purification & IEEE TrustCom (2025) & VLM visual encoders & Input purification & CV+NLP & \href{https://doi.org/10.1109/Trustcom66490.2025.00065}{\faFilePdf} & \ding{55} \\
\addlinespace[1pt]
Pruning Defense & IEEE GLOBECOM (2025) & Sensing DNNs & Architectural sparsity & Sensing & \href{https://doi.org/10.1109/GLOBECOM59602.2025.11432163}{\faFilePdf} & \ding{55} \\
\addlinespace[1pt]
BlindSight & arXiv (2025) & VLMs & Architectural sparsity & CV+NLP & \href{https://arxiv.org/abs/2507.09071}{\faFilePdf} & \ding{55} \\
\addlinespace[1pt]
PD3F & EMNLP (2025) & LLM serving & Runtime scheduling and termination & NLP & \href{https://arxiv.org/abs/2505.18680}{\faFilePdf} & \ding{55} \\
\addlinespace[1pt]
SQUAD & arXiv (2026) & Early-exit ensembles & Runtime scheduling & CV & \href{https://arxiv.org/abs/2601.22711}{\faFilePdf} & \ding{55} \\
\addlinespace[1pt]
Token-Budget Routing & arXiv (2026) & LLM serving & Budget estimation and pool routing & NLP & \href{https://arxiv.org/abs/2604.09613}{\faFilePdf} & \ding{55} \\
\addlinespace[1pt]
TALE (Token-Budget Reasoning) & arXiv (2024) & Reasoning LLMs & Prompt-based budget guidance & NLP & \href{https://arxiv.org/abs/2412.18547}{\faFilePdf} & \href{https://github.com/GeniusHTX/TALE}{\faGithub} \\
\addlinespace[1pt]
Concise CoT (CCoT) & arXiv (2024) & Reasoning LLMs & Prompt-based concision guidance & NLP & \href{https://arxiv.org/abs/2401.05618}{\faFilePdf} & \href{https://github.com/matthewrenze/jhu-concise-cot}{\faGithub} \\
\addlinespace[1pt]
CoT-Valve & ACL (2025) & Reasoning LLMs & Learned/parameter-level length control & NLP & \href{https://aclanthology.org/2025.acl-long.300/}{\faFilePdf} & \href{https://github.com/horseee/CoT-Valve}{\faGithub} \\
\addlinespace[1pt]
Conformal Thinking & arXiv (2026) & Reasoning models & Risk-controlled adaptive stopping & NLP & \href{https://arxiv.org/abs/2602.03814}{\faFilePdf} & \ding{55} \\
\addlinespace[1pt]
TrackShield & HPCC (2026) & Multi-object trackers (ByteTrack, BoT-SORT) & Admission control and runtime monitoring & AD & -- & \ding{55} \\
\addlinespace[1pt]
Bounded admission layer (TrackFlood) & arXiv (2026) & NMS-free detect-then-track & Admission control & AD & \href{https://arxiv.org/abs/2609.33948}{\faFilePdf} & \ding{55} \\
\addlinespace[1pt]
Bounded Escalation (GateDrain) & arXiv (2026) & Edge--cloud offloading & Admission control and priority scheduling & Edge & \href{https://arxiv.org/abs/2609.33992}{\faFilePdf} & \ding{55} \\
\addlinespace[1pt]
\end{longtable}}